\documentclass[pdftex]{pasj01}

\usepackage{url}

\begin{document} 
\Received{}
\Accepted{}

\title{The bright-star masks for the HSC-SSP survey}

\author{Jean \textsc{Coupon}\altaffilmark{1}%
}

\altaffiltext{1}{Department of Astronomy, University of Geneva, ch. d'\'Ecogia 16, 1290
Versoix, Switzerland}
\email{jean.coupon@unige.ch}
\author{Nicole \textsc{Czakon}\altaffilmark{2}}
\altaffiltext{2}{Institute of Astronomy and Astrophysics, Academia Sinica, PO Box 23-141, Taipei 10617, Taiwan}
\author{James \textsc{Bosch}\altaffilmark{3}}
\altaffiltext{3}{Department of Astrophysical Sciences, 4 Ivy Lane, Princeton, NJ 08544, USA}
\author{Yutaka \textsc{Komiyama}\altaffilmark{4,5}}
\altaffiltext{4}{National Astronomical Observatory of Japan, 
2-21-1 Osawa, Mitaka, Tokyo 181-8588, Japan}
\altaffiltext{5}{Department of Astronomy, School of Science, 
SOKENDAI (The Graduate University for Advanced Studies), 
Mitaka,  Tokyo, 181-8588, Japan}
\author{Elinor \textsc{Medezinski}\altaffilmark{3}}
\author{Satoshi \textsc{Miyazaki}\altaffilmark{4,5}}
\author{Masamune \textsc{Oguri}\altaffilmark{6,7,8}}
\altaffiltext{6}{Department of Physics, University of Tokyo, 7-3-1 Hongo, Bunkyo-ku, Tokyo 113-0033, Japan}
\altaffiltext{7}{Kavli Institute for the Physics and Mathematics of the Universe (Kavli IPMU, WPI), University of Tokyo, Chiba 277-8583, Japan}
\altaffiltext{8}{Research Center for the Early Universe, The University of Tokyo, 7-3-1 Hongo, Bunkyo-ku, Tokyo 113-0033, Japan}



\maketitle

\begin{abstract}
We present the procedure to build and validate the bright-star masks for the Hyper-Suprime-Cam Strategic Subaru Proposal (HSC-SSP) survey. To identify and mask the saturated stars in the full HSC-SSP footprint, we rely on the Gaia and Tycho-2 star catalogues. We first assemble a pure star catalogue down to $G_{\rm Gaia} < 18$ after removing $\sim$1.5\% of sources that appear extended in the Sloan Digital Sky Survey (SDSS). We perform visual inspection on the early data from the S16A internal release of HSC-SSP, finding that our star catalogue is 99.2\% pure down to $G_{\rm Gaia} < 18$. Second, we build the mask regions in an automated way using stacked detected source measurements around bright stars binned per $G_{\rm Gaia}$ magnitude. Finally, we validate those masks from visual inspection and comparison with the literature of galaxy number counts and angular two-point correlation functions. This version (Arcturus) supersedes the previous version (Sirius) used in the S16A internal and DR1 public releases. We publicly release the full masks and tools to flag objects in the entire footprint of the planned HSC-SSP observations at this address: \texttt{ftp://obsftp.unige.ch/pub/coupon/brightStarMasks/HSC-SSP/}.
\end{abstract}

\section{Introduction}
\label{sec:intro}

Extragalactic studies rely extensively on multi-wavelength imaging data to measure galaxy photometric redshifts, physical properties and shapes, used in a broad range of probes to explore the cosmology and the formation and evolution of galaxies (see e.g. \cite{Heymans:2012,Magnier:2013,dejong:2015,Albareti:2016,Lang:2016,DES:2016,Aihara:2017}). These probes, such as weak gravitational lensing (\cite{Kuijken:2015} and references therein) and galaxy clustering (\cite{Crocce:2016} and references therein), require high-precision data reduction and measurement algorithms, whose development and validation are driven by the increasing statistical power of the surveys and the increasing relative importance of the various systematics.

Among the systematics affecting the images are the multiple instrumental artefacts that cause spurious detections (satellite/plane trails, diffraction spikes, fast-moving objects, etc.) and real source contamination or screening (bright stars, ghosts, cross talks, etc.). All of these effects need automated processes to deal with the large amount of data generated in wide angular field surveys. But if many artefacts can be curated during data processing \citep{Bertin:1996,Ivezic:2004,Magnier:2004,Erben:2009,Schirmer:2013,Juric:2015}, some others cannot. This is typically the case for the bright stars that saturate on the CCD images. The challenge with bright stars is that the pixels systematically saturate on each individual exposure (so one cannot use co-addition and sigma-clipping techniques to exclude them) and that the brightness information is lost beyond the saturation limit (making it difficult to assess the star properties). Hence, a strategy commonly adopted in the literature is to use external star catalogues and build photometric masks around the bright stars, at the expense of an inevitable loss in area that ranges between 10\% and 20\% \citep{Coupon:2012,Heymans:2012}. For example, the Terapix group was among the first to develop automated tools to mask the bright stars in the Canada-France-Hawaii Telescope Legacy Survey\footnote{\url{http://www.cfht.hawaii.edu/Science/CFHTLS/}} (CFHTLS), later refined by the VIPERS team \citep{Guzzo:2014,Granett:2012} to improve their spectroscopic target selection. Almost all of the modern wide-field surveys such as the Sloan Digital Sky Survey (SDSS, \cite{York:2000aa}), panSTARRS \citep{Chambers:2016}, KiDS \citep{dejong:2013}, or the Dark Energy Survey (DES, \cite{DES:2005}), now use similar strategies. In particular the masks produced by the \texttt{THELI} pipeline \citep{Schirmer:2013,Erben:2009,Erben:2013}, used in the CFHTLenS data, were validated by a systematic visual inspection by the CFHTLenS team \citep{Heymans:2012}, making sure that the stringent requirements for cosmic shear studies \citep{Kilbinger:2013} were met.

For very sensitive CCD cameras like the recently installed Hyper-Suprime-Cam (HSC, \cite{Miyazaki:2012}), mounted on the 8-m Subaru telescope at Mauna Kea in Hawaii, the additional challenge when building a bright-star sample is to find a sufficiently deep and pure star catalogue to cover the entire magnitude range over which stars saturate, but without masking any extended source of interest. Indeed, typical exposures of a few minutes with HSC saturate stars as faint as the 19th (AB) magnitude, and sometimes even fainter when the seeing reaches sharp values under \timeform{0.6''}. 

In this work we describe the procedure to automatically build and validate the bright-star masks for the broad-band filters ($grizY$) of the Hyper-Suprime-Cam Strategic Subaru Proposal (HSC-SSP, \cite{HSC:2017}) survey. HSC-SSP is an extragalactic survey whose aim is to observe $1\,400$~deg$^2$ down to $grizY\sim25-26$ (as well as a number of smaller and deeper fields). To do so, we use the data from the Gaia DR1 \citep{Gaia-Collaboration:2016aa}, Tycho-2 \citep{Hog:2000aa,Pickles:2010aa}, and SDSS DR13 \citep{Albareti:2016} catalogues, to build a complete and pure star catalogue down to the typical HSC saturation limit. We first describe the problem in Section~\ref{sec:pb}, we then describe how we gather a sample of bright stars in Section~\ref{sec:catalogue} and how we build the masks in Section~\ref{sec:masks}. In a last Section~\ref{sec:valid} we estimate the purity and completeness of the bright stars masks, quantify the improvement on a number of observables, and describe their limitations. We conclude in Section~\ref{sec:conclusions}.

This work supersedes an early version (hereafter the \emph{Sirius} version), described in \citet{Mandelbaum:2017}, of the bright-star masks that were used in the S15B and S16A internal data releases, as well as in the first public data release DR1 \citep{Aihara:2017}. This early version contained about 9\% of bright galaxies (making it problematic for a number of science cases), and was over conservative in the choice of mask radius for stars brighter than the 5th magnitude, causing an unnecessary loss in survey area. The version of the bright-star masks described in this work is called the \emph{Arcturus} version, and will be implemented in subsequent HSC pipeline versions and data releases. The full masks and tools to flag objects in the HSC-SSP footprint can be downloaded at \url{ftp://obsftp.unige.ch/pub/coupon/brightStarMasks/HSC-SSP/}.


\section{Bright stars in the HSC-SSP survey}
\label{sec:pb}

\subsection{The HSC-SSP survey}

Here we briefly describe the HSC-SSP survey. For more details we refer the reader to the camera \citep{Miyazaki:2012,Miyazaki:2017}, the survey overview\footnote{\url{http://hsc.mtk.nao.ac.jp/ssp/}} \citep{HSC:2017}, the HSC software pipeline \citep{Bosch:2017}, the DR1\footnote{\url{https://hsc-release.mtk.nao.ac.jp/doc/}} \citep{Aihara:2017}, and the photometric redshifts \citep{Tanaka:2017} papers. { The HSC-SSP data are calibrated with the PanSTARRS1 data. We refer the reader to \citet{Tonry:2012}, \citet{Schlafly:2012} and \citet{Magnier:2013} for additional details on the PanSTARRS1 data.}

The HSC-SSP project is a 300-night deep imaging survey which started in March 2014, carried out at the Subaru telescope with the 1.5~deg-diameter HSC camera. The goal of the survey is to explore the properties of the dark universe (dark matter and dark energy) and the formation and evolution of the galaxies from the local Universe until the epoch of reionisation, using a number of different probes, including weak and strong gravitational lensing, galaxy clustering and galaxy abundances. 

The survey is split into three layers:
\begin{itemize}
\item a Wide layer ($r < 26$\footnote{$5\sigma$~point-source limiting magnitudes. Note that all magnitudes quoted in this work are in the AB system.}) over $1\,400$~deg$^2$, composed of three main fields (\emph{Spring}, \emph{Fall}, and \emph{Northern}) and one pointing for calibration (\emph{Aegis}),
\item a Deep layer ($r < 27$) over 28~deg$^2$ composed of four fields (\emph{XMM-LSS}, \emph{Elais-N1}, \emph{E-COSMOS} and \emph{DEEP2-3}), 
\item and an UltraDeep layer ($r < 28$) over 4~deg$^2$, composed of two fields (\emph{SXDS/UDS} and \emph{COSMOS}). 
\end{itemize}
The three layers are observed in the five optical and near infrared (NIR) broad-band filters $g$, $r$, $i$, $z$ and $Y$. In addition, the Deep layer is observed in the three narrow-band filters NB387, NB816, and 
NB921, and the UltraDeep layer is observed in the three narrow-band filters NB816, N921, and NB101.

\subsection{Why masking the bright stars?}

Our goal is to limit the impact of saturated stars on nearby sources, to prevent the use of spurious sources in the science analyses, and to record the loss in area due to the occultation of the bright stars for extragalactic studies. The adopted strategy is to flag all sources near a bright star and propagate the information all the way from the data processing to the database (i.e. we keep all the sources but we flag those affected), so that one can filter a sample of interest during data retrieval from the database.

It is also important to record the fraction of the survey that is occulted by the bright-star masks, so in parallel we produce a catalogue of points randomly placed in the survey area, flagged in the same way as the sources. { In short, the random points are drawn with a density of 100 points per square arcmin, for each patch and each filter. Then, the survey information (field geometry, number of co-added exposures, PSF, pixel variance, etc.) is recorded at each random-point position using a \texttt{python} module running within the HSC pipeline environment\footnote{The source code can be accessed at \url{https://github.com/jcoupon/maskUtils} and the random-point sample is described at \url{https://hsc-release.mtk.nao.ac.jp/doc/index.php/random-points-for-dr1/}.}. Finally, we add a flag indicating whether a point is inside or outside the bright-star masks (more details on the flagging procedure are given in Appendix~\ref{sec:flagging}
). The masked fraction is computed from the ratio between the random points inside the masks and the total number of points. With a density of 100 points per square arcmin and assuming a mask coverage of $\sim10\%$, the precision on the masked area is $0.5\%$ on the scale of a degree.}

\subsection{The impact of saturated stars on HSC images}
\label{sec:impact}

\begin{figure}
 \begin{center}
  \includegraphics[width=0.4\textwidth]{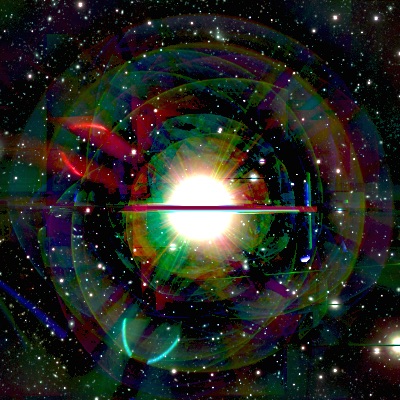} \\
  \includegraphics[width=0.4\textwidth]{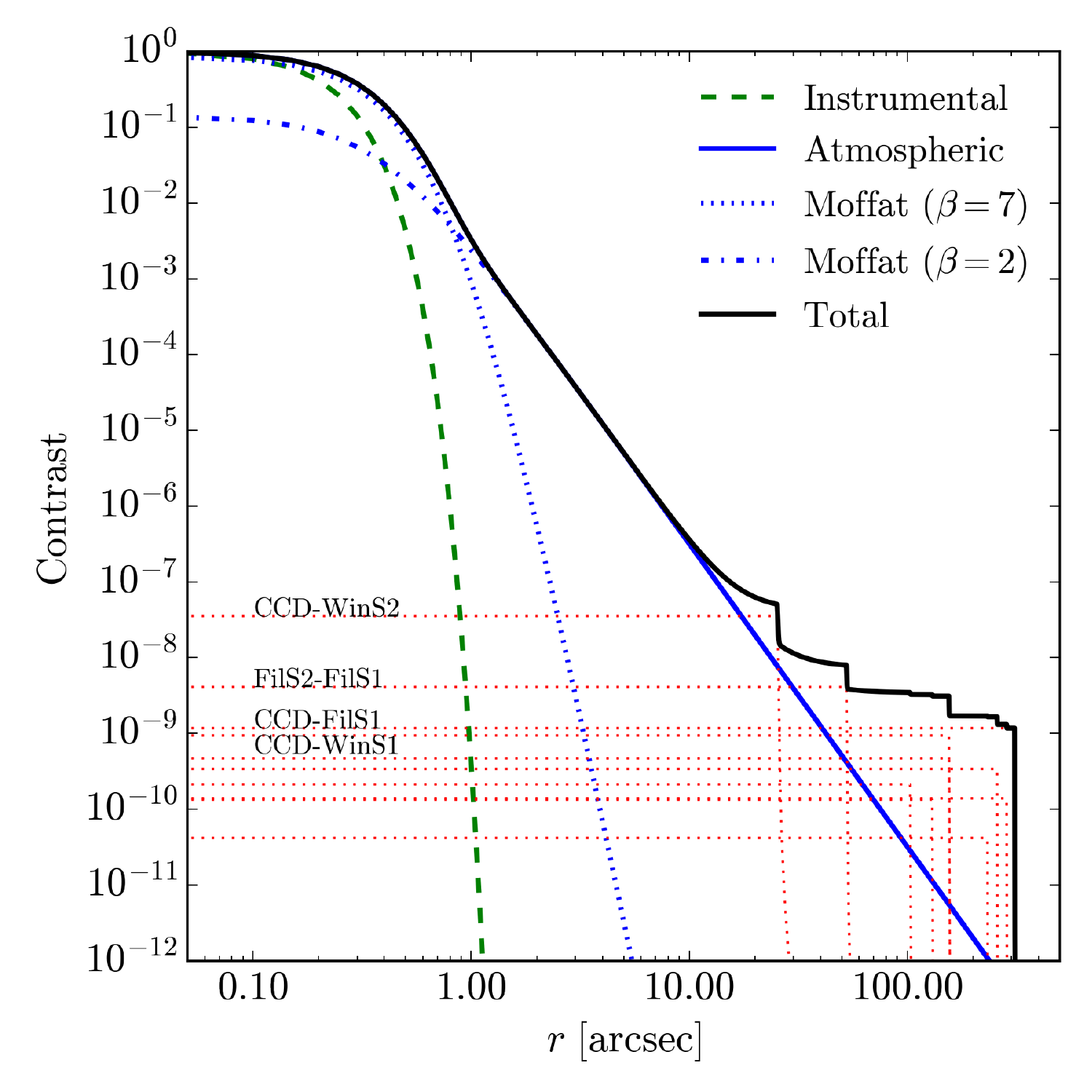}
 \end{center}
    \caption{Top: an example of a saturated star ($G_{\rm Gaia} = 3.9$) in the HSC-SSP survey. The image is a $gri$ composite image. One can see the PSF-shaped diffraction patterns, the extended luminous halos, and the horizontal bleed trails. Bottom: { a theoretical model of the radial effects caused by a bright star on an HSC image. The PSF pattern is the sum of the instrumental effects (green dashed line) and the atmospheric blurring effects (blue solid line), modelled as the sum of two Moffat functions with $\beta=7$ (blue dotted line) and $\beta=7$ (blue dash-dotted line). The various large-scale effects (red dotted lines) are caused by optical reflections.}}
    \label{fig:saturated_star}
\end{figure}

The main artefacts caused by the saturated bright stars can be divided into optical effects (diffraction pattern, optical reflections and ghosts) and electronics effects (electrons leaking to nearby pixels, NIR photons scattering, brighter-fatter effect). On the image, these effects result in (1) Point Spread Function (PSF)-shaped luminous patterns, (2) extended luminous haloes and ghosts, (3) linear ``bleed trails'', and { (4) spikes in the $Y$-band images in the row direction in addition to bleed trails in the column direction.  This is the result of scattering off the channel stops at the back of the detector.  The sensors are too opaque to bluer photons for this effect to appear in other bands}. An example of a typical saturated bright star on a HSC image is shown in the top panel of Figure~\ref{fig:saturated_star}. 

The size of the PSF diffraction pattern and the bleed trails depend on the seeing and the position on the focal plane. The radial extent of the luminous PSF pattern depends on the exposure time and the brightness of the star. The brighter-fatter effect \citep{Antilogus:2014} also mildly increases the extent of the diffraction pattern as a function of star brightness. Optical reflections in the instrument are responsible for bright images (ghosts) that spread in various places on the image, also depending on the position on the focal plane. In particular, the ghosts include a circular extended halo generated by the incident light which is reflected at the CCD surface and again reflected by the filter surface. The size of this extended halo is determined solely by the distance between the two reflecting surfaces, being in theory independent on the brightness of the ghost source star, but appear only for very bright stars.

In the bottom panel of Figure~\ref{fig:saturated_star}, we show an example of the radial luminosity profile created by the instrumental and the atmospheric PSF, as well as a number of optical reflections.
The blue line is the Kolmogorov profile which describes the atmospheric turbulence with a full width half maximum (FWHM) seeing size of \timeform{0.5''}. The combination of $\beta=7$ and $2$ Moffat functions (plotted as blue dotted and dash-dotted lines, respectively) is used to describe the Kolmogorov profile (see \cite{Racine:1996}). The instrumental PSF, which is simplified as a Gaussian with FWHM=\timeform{0.357''} \citep{Miyazaki:2017} and plotted as the green line, is smaller compared to atmospheric seeing profile and can be neglected (it only affects the image on scales on the order of \timeform{0.1''}). 

When the contrast reaches $10^{-7}$, the circular ghosts emerge. The size and contrast are calculated for different combinations of reflecting plane surfaces (CCD, and incident surface, S1) and exit surface (S2) of the window and the filter, based on the optics model assuming that the reflectance of each surface is known. The profiles are represented as red dotted lines and the significant ones are labeled. The sum of all components is the black solid line. Note that not all of the low-contrast components -- in particular those with no  theoretical models (including the aureole, \cite{Racine:1996}) -- are included in this figure (it must be obtained from the real images). Also, the extended halo is not exactly circular except in the central part of the camera, due to the vignetting of the wide-field corrector cutting the pupil image of the primary mirror.

\subsection{Saturation limits in the HSC-SSP survey}
\label{sec:satlimits}

In this work we limit ourselves to the sample of bright stars that saturate on HSC-SSP images (the information on non-saturated stars -- such as brightness and PSF-size -- can be easily gathered directly from the data through the HSC pipeline estimates). Here, we evaluate the typical saturation limits in the HSC-SSP survey.


The saturation limit on a given co-added image mainly depends on the single-visit exposure time, which is different for each survey layer as shown in Table~\ref{tab:exp_time},
\begin{table}
  \tbl{Single-visit exposure times (in seconds) for each filter and survey layer.}{%
  \begin{tabular}{lccccc}
\hline
 Layer & $g$ & $r$ & $i$ & $z$ & $Y$ \\
\hline
Wide & 150 & 150 & 200 & 200 & 200 \\
Deep & 180 & 180 & 270 & 270 & 270 \\
Ultra Deep & 300 & 300 & 300 & 300 & 300\\
\hline
\end{tabular}}\label{tab:exp_time}
\end{table}
and the PSF size (the sharper the PSF, the fainter the saturation limit). The PSF at a given position is computed by the HSC pipeline from fitting the (non-saturated) stars with a smooth spatially varying PSF model, per single exposure, and co-added using a sigma-weighted mean (see \cite{Bosch:2017}). 

\begin{figure*}
 \begin{center}
  \includegraphics[width=0.32\textwidth]{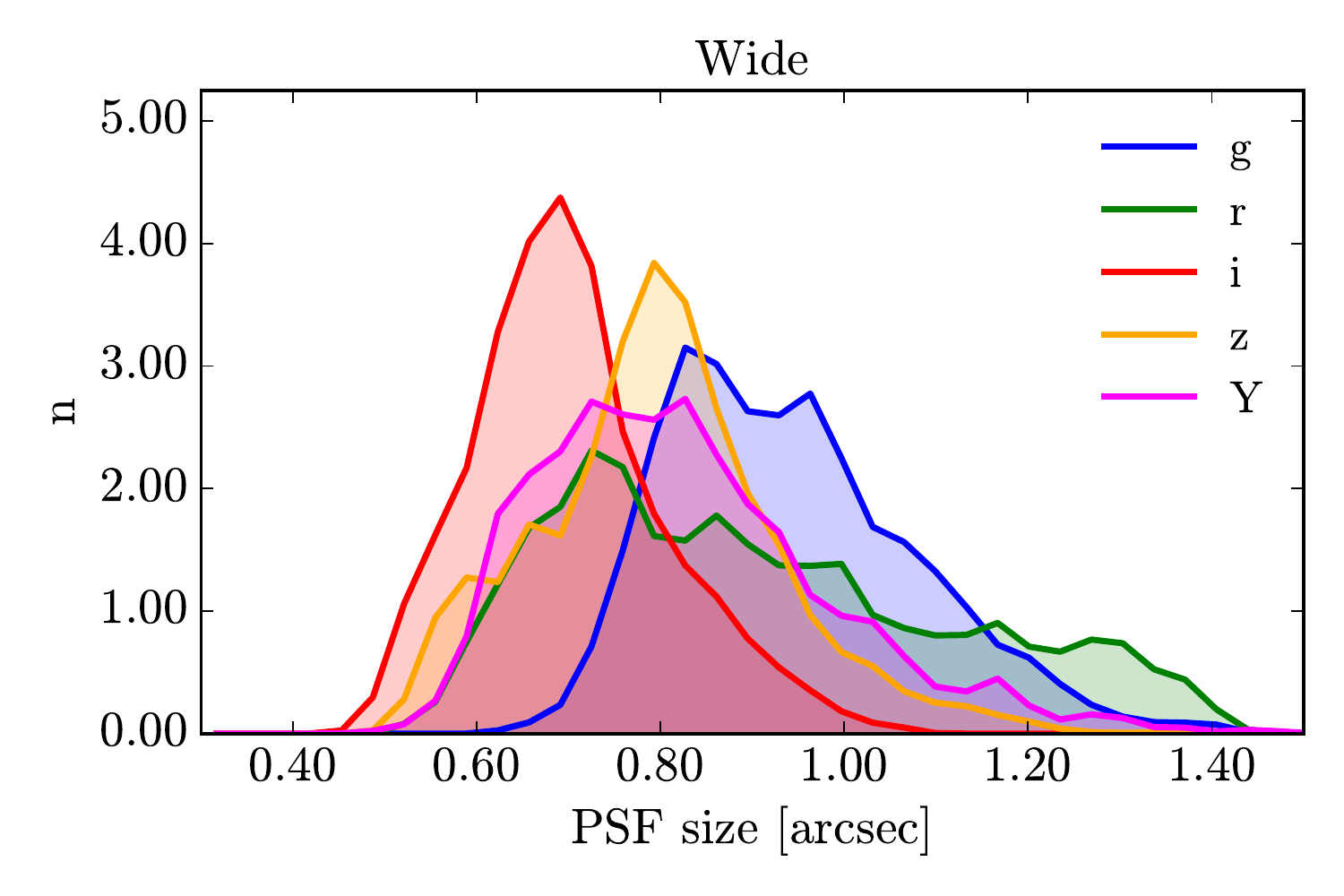}
  \includegraphics[width=0.32\textwidth]{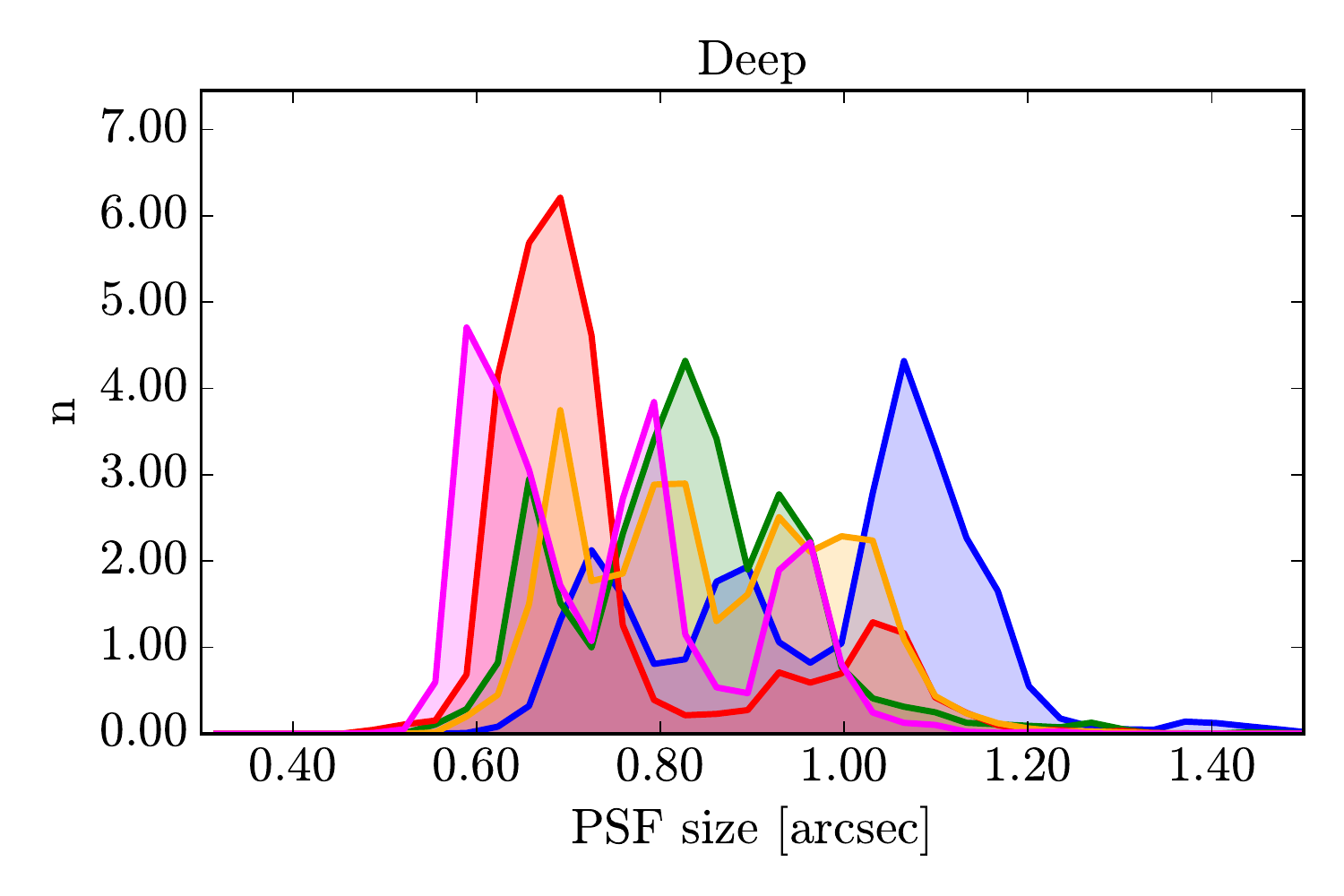}
  \includegraphics[width=0.32\textwidth]{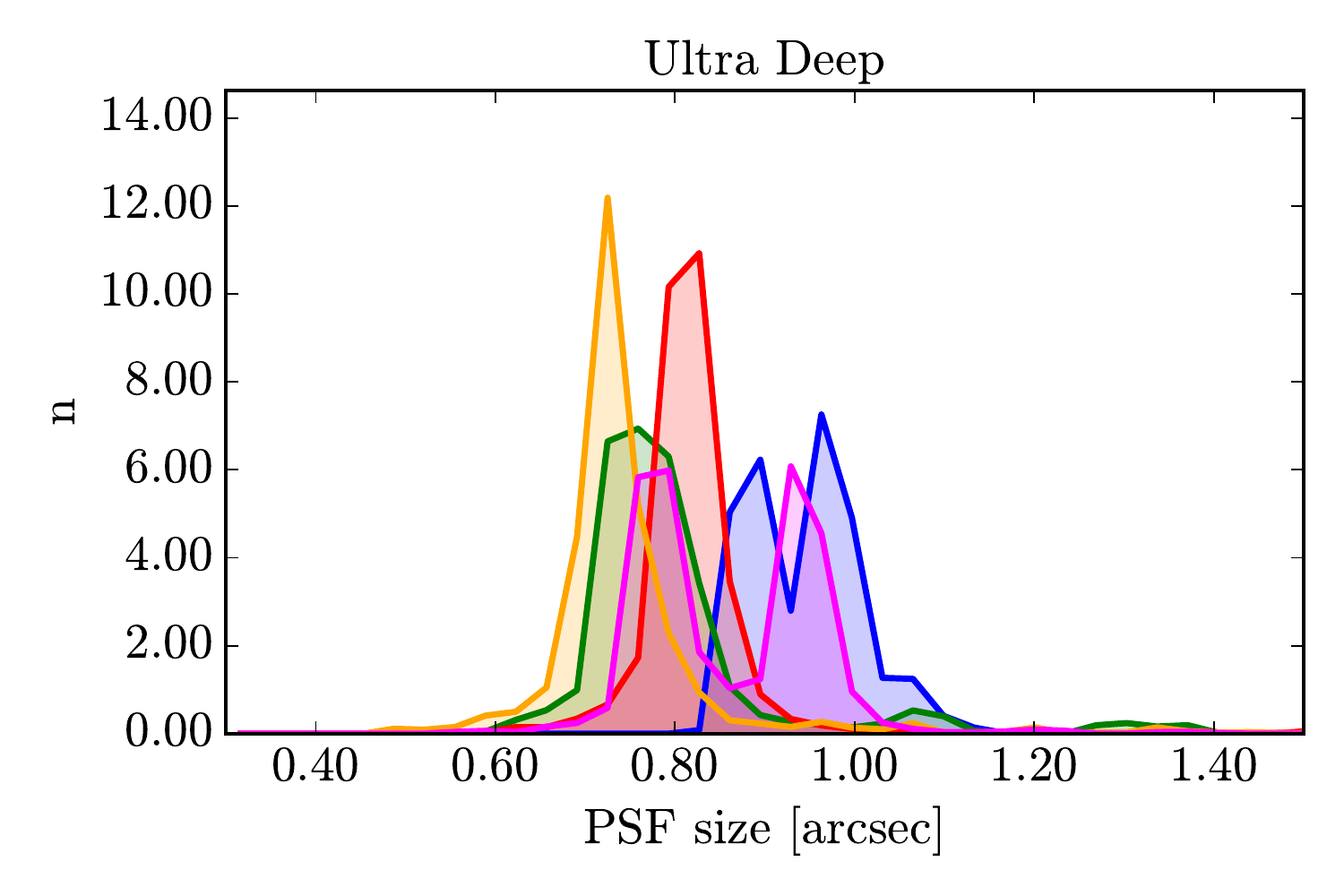}
  \includegraphics[width=0.32\textwidth]{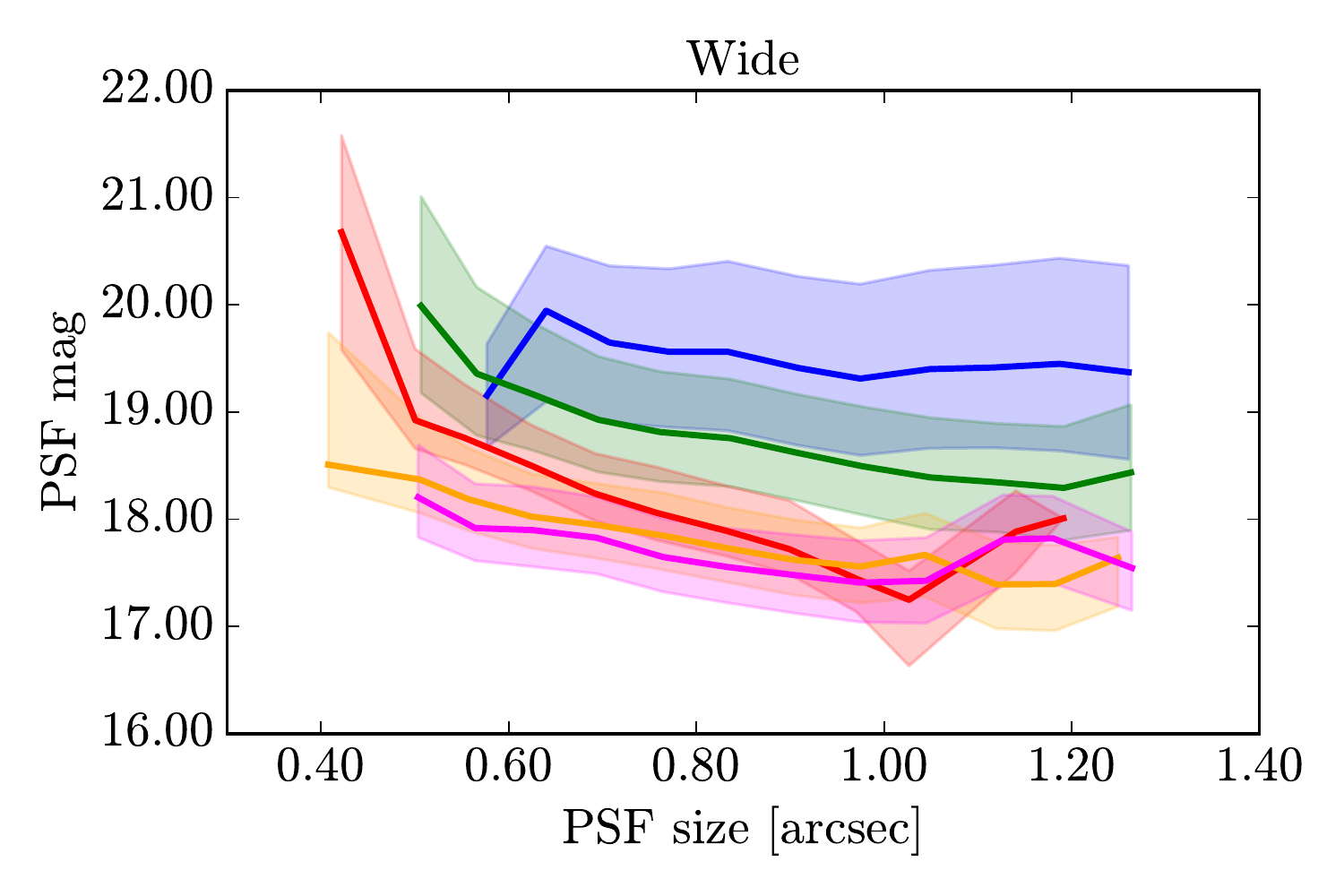}
  \includegraphics[width=0.32\textwidth]{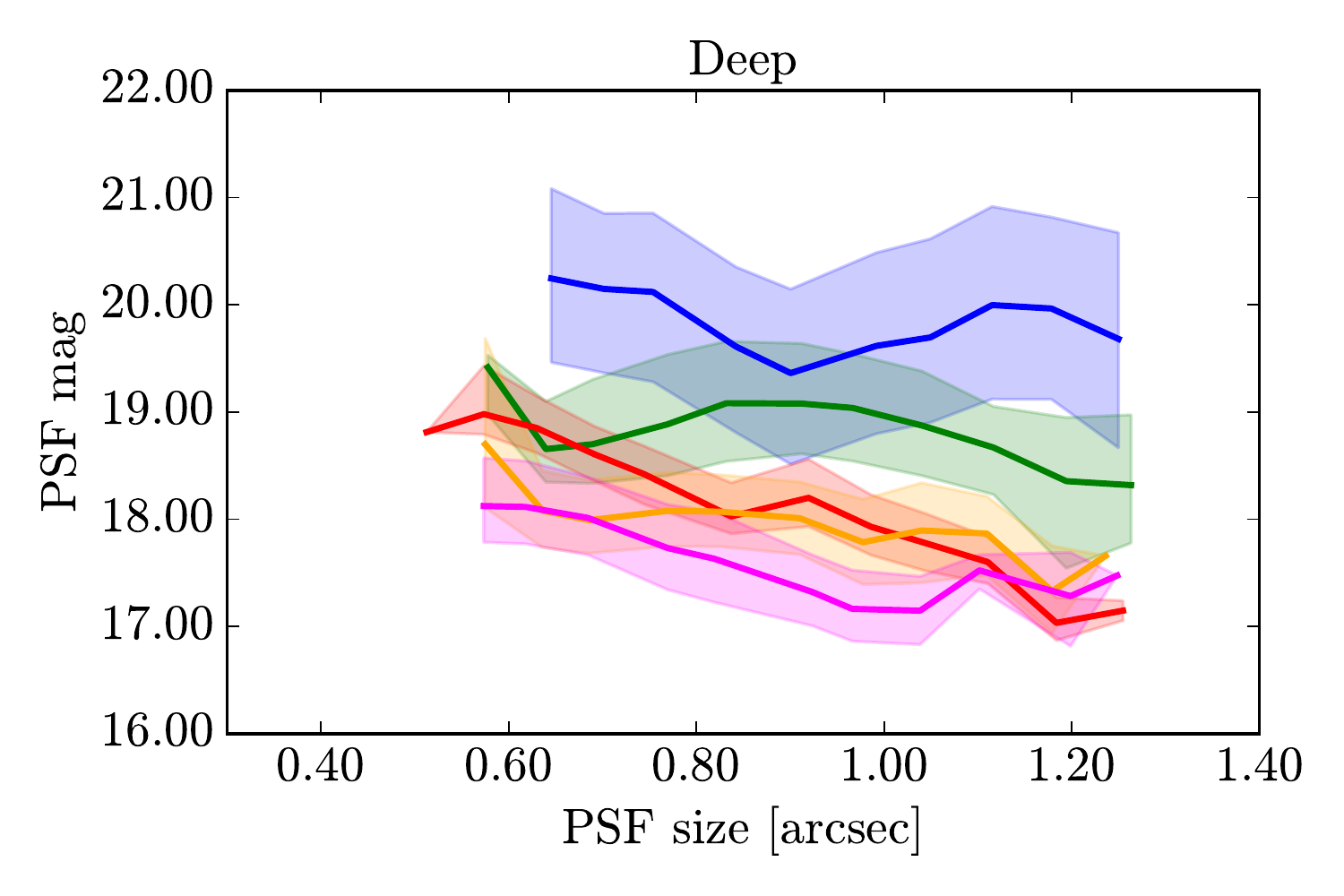}
  \includegraphics[width=0.32\textwidth]{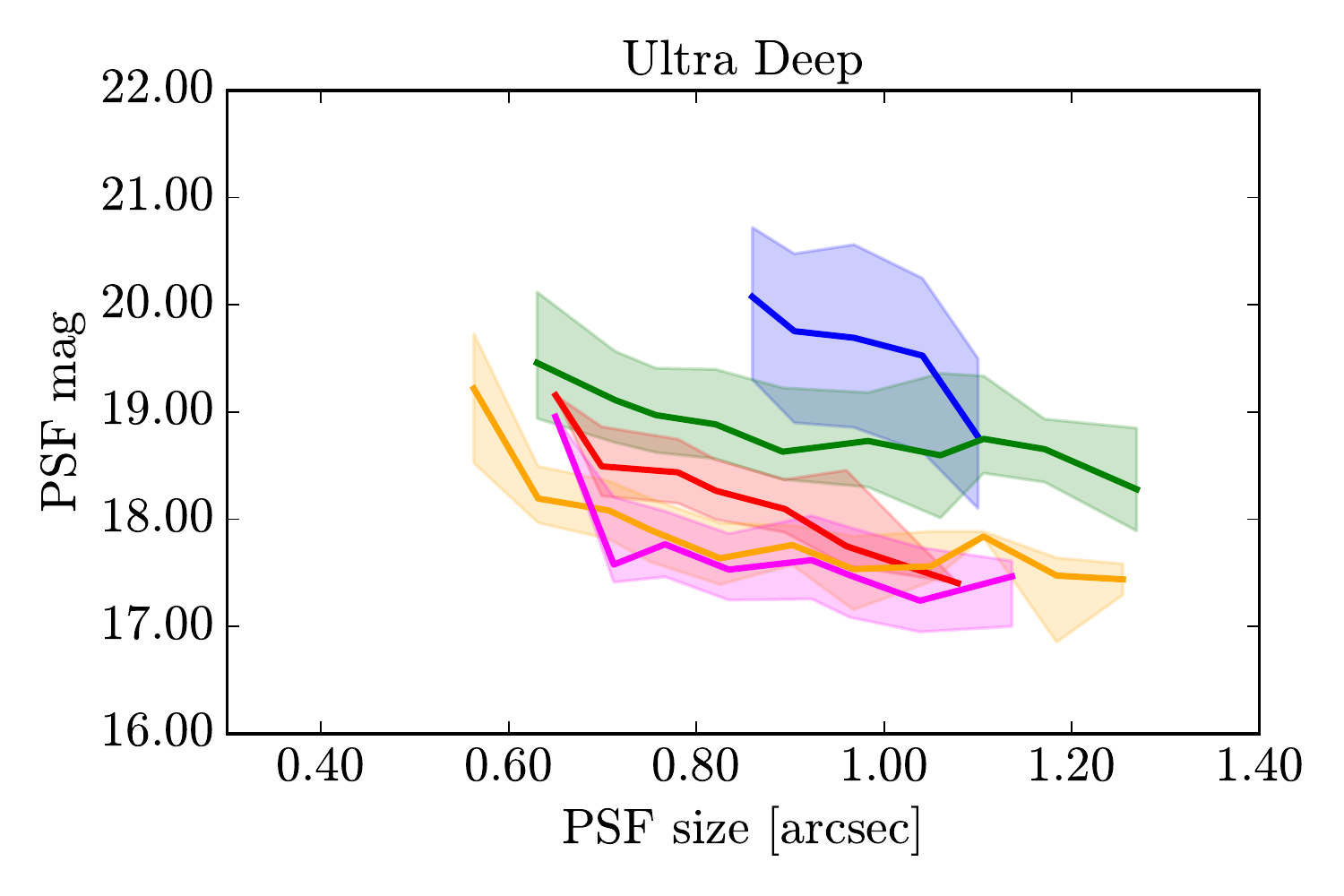}
 \end{center}
    \caption{Top: normalised PSF size (FWHM) distributions in the three HSC-SSP layers { (co-added images)} for the S16A internal release (left:Wide, middle:Deep, and right:Ultra Deep), and for each broad-band filter. Bottom: saturation magnitudes (PSF estimate) for bright stars as a function of PSF size in the three HSC-SSP layers { (co-added images)} for the S16A internal release (left:Wide, middle:Deep, and right:Ultra Deep), and for each filter. The solid line shows the median, whereas the shaded area represents the 25\% and 75\% percentiles.}
    \label{fig:seeingDist}
\end{figure*}

To evaluate the typical PSF sizes { on co-added images} in HSC-SSP, we record, for many random positions in the S16A release footprint, the PSF sizes from the second moment of the best-fit PSF model, multiplied by $2\sqrt{2}$ to convert into the commonly used FWHM estimate, under the assumption that the PSF luminosity profile is gaussian\footnote{Although it is not really gaussian, this assumption is good enough for the relative comparison between two positions in the survey.}.
The top row of Figure~\ref{fig:seeingDist} shows the PSF size (FWHM) distribution in each survey layer and each of the $grizY$ filters. 
The bottom row of Figure~\ref{fig:seeingDist} shows the median saturation magnitudes in each survey layer and each filter. Saturation magnitude limits are computed from selecting all bright point sources detected by the pipeline, with a successfully measured magnitude (PSF estimate), imposing that the central pixel is flagged as saturated and that the source is not deblended. The typical magnitude limits range from $17$ to $20$. The figure shows the expected trend that the saturation limit becomes fainter as the PSF gets smaller. We also confirm that the longer exposure times in the Ultra Deep layer result in fainter saturation limits as compared to the Wide layer, although the difference is not very large. The changing limit observed in the different filters is linked to the varying filter transparency and chromatic effects of the PSF.

{ The saturation limits in best-PSF data are summarised in Table~\ref{tab:saturLimits}. These represent the faintest possible magnitudes at which stars saturate in the HSC-SSP data. We observe that, depending on the filter, the faintest magnitude occurs in a different survey layer. Several effects are competing with each other: on the one hand, we expect that the longer single-exposure times in the Deep and Ultra-Deep layers will lead to fainter saturation limits (this is the case for the $z$ and $Y$ bands). On the other hand, the shorter exposure times in the Wide layer are compensated by usually better PSFs (by construction, e.g. for the $i$ band, or by chance), so that the saturation limits are equivalent or fainter than in the deeper layers (note for example the $i$-band magnitude of $20.7$ that occurs in very good PSF data, $<\timeform{0.5''}$)}.
\begin{table}
  \tbl{Point-source saturation magnitudes on co-added images for the S16A release, given for the best (smallest) PSFs.}{%
\begin{tabular}{lccccc}
\hline
 Layer & $g$ & $r$ & $i$ & $z$ & $Y$ \\
\hline
Wide & 19.2 & 20.0 & 20.7 & 18.5 & 18.2 \\
Deep & 20.2 & 19.4 & 18.8 & 18.7 & 18.1 \\
Ultra Deep & 20.1 & 19.5 & 19.2 & 19.2 & 19.0 \\
\hline
\end{tabular}}\label{tab:saturLimits}
\end{table}

Eventually, the goal is to build a bright-star sample which goes as faint as the best-PSF saturation limits, so that all saturated stars are masked in the HSC-SSP footprint. We will see that the Gaia sample is well suited for this purpose.

\section{The bright-star sample}
\label{sec:catalogue}

The reference bright-star\footnote{Here, other bright point sources, such as Quasi Stellar Objects (QSOs), are included in the term \emph{bright stars}.} sample should meet several requirements to be used for masking:
\begin{itemize}
\item it { should} be pure and complete over the magnitude range where stars saturate on HSC images,
\item it must cover the entire area of the HSC-SSP survey,
\item and it must not rely on HSC-SSP data (since the bright star masks are made before data reduction and because we do not trust the brightness estimates where pixels saturate).
\end{itemize}

We build the star sample from the Gaia Data Release 1 (DR1), completed with the Tycho-2 star sample to address a number of issues associated with the first Gaia release. We use the $G_{\rm Gaia}$ magnitude -- a broad-band filter covering the range 4000--10000~\AA{} -- as the primary characterisation of star brightness, and we use the SDSS to remove the extended sources wrongly identified as stars by Gaia in the DR1. We detail below the steps we follow to build the sample. A quick summary is given in Section~\ref{sec:summary}.

\subsection{Gaia}

Gaia \citep{Gaia-Collaboration:2016ab} is an ESA space mission launched in 2013 whose goal is to observe one billion stars (1\% of the total Milky Way stars), and to provide accurate positions, radial velocities, and spectrophotometry in optical\footnote{The spectrophotometry will be limited to relatively bright stars, $G_{\rm Gaia} < 15$.}. The first Gaia data release \citep{Gaia-Collaboration:2016aa} was publicly delivered in September 2016. It includes position measurements and $G_{\rm Gaia}$-band photometry over the entire sky. 

The Gaia star sample is ideally suited for our purpose. With a survey covering the entire sky at a limiting magnitude of $G_{\rm Gaia} < 20.7$ and with optical spectrophotometry, it perfectly meets our requirements. However, the Gaia DR1 comes with a number of limitations:
\begin{itemize}
\item only the $G_{\rm Gaia}$ photometry is provided,
\item the sample is incomplete at bright magnitude,
\item the limiting magnitude is ill-defined and varies across the sky,
\item some large areas are missing on the ecliptic due to quality filtering and not-yet fully scanned regions,
\item and some galaxies are included in the sample towards faint magnitudes.
\end{itemize}

\subsection{Tycho-2}

To complete the Gaia sample at bright magnitude and in under-dense (or empty) areas, we use the Tycho-2 star sample. 

The Tycho-2 star sample \citep{Hog:2000aa} is a complete all-sky catalogue of the 2.5 million brightest stars to a magnitude $V<11.5$, with accurate positions and proper motions, observed by the ESA Hypparcos satellite. It provides two broad-band photometry in $B_{T}$ and $V_{T}$ filters. Here we use the catalogue released by \citet{Pickles:2010aa}, which provides emulated SDSS-like  $ugriz$ photometry for all Tycho-2 stars. To be consistent with the Gaia stars, we transform the emulated SDSS magnitudes for each Tycho-2 star into a $G_{\rm Gaia}$ magnitude, according to the formula given in \citet{Jordi:2010aa}:
\begin{eqnarray}
\label{eq:emulGaia}
& G_{\rm Gaia} = & g_{\rm sdss} - 0.0940 - 0.5310\times (g_{\rm sdss}-i_{\rm sdss}) \\ 
& & -0.0974\times (g_{\rm sdss}-i_{\rm sdss})^2 \nonumber \\ 
& & + 0.0052\times(g_{\rm sdss}-i_{\rm sdss})^3 \, . \nonumber
\end{eqnarray}
We check the accuracy of this expression in Appendix~\ref{sec:magComp}.

Since a star's brightness varies across the spectrum, it would be, in principle, required to set the size of the masks in each band, separately, using the color information when available. However, we take the simpler (and limited by the Gaia DR1 release) approach to characterise the star brightness in all HSC bands using the unique $G_{\rm Gaia}$ magnitude. We come back to this issue in Section~\ref{sec:flterDepence}.


\subsection{Gathering a complete star sample}

\begin{figure*}
 \begin{center}
  \includegraphics[width=0.9\textwidth]{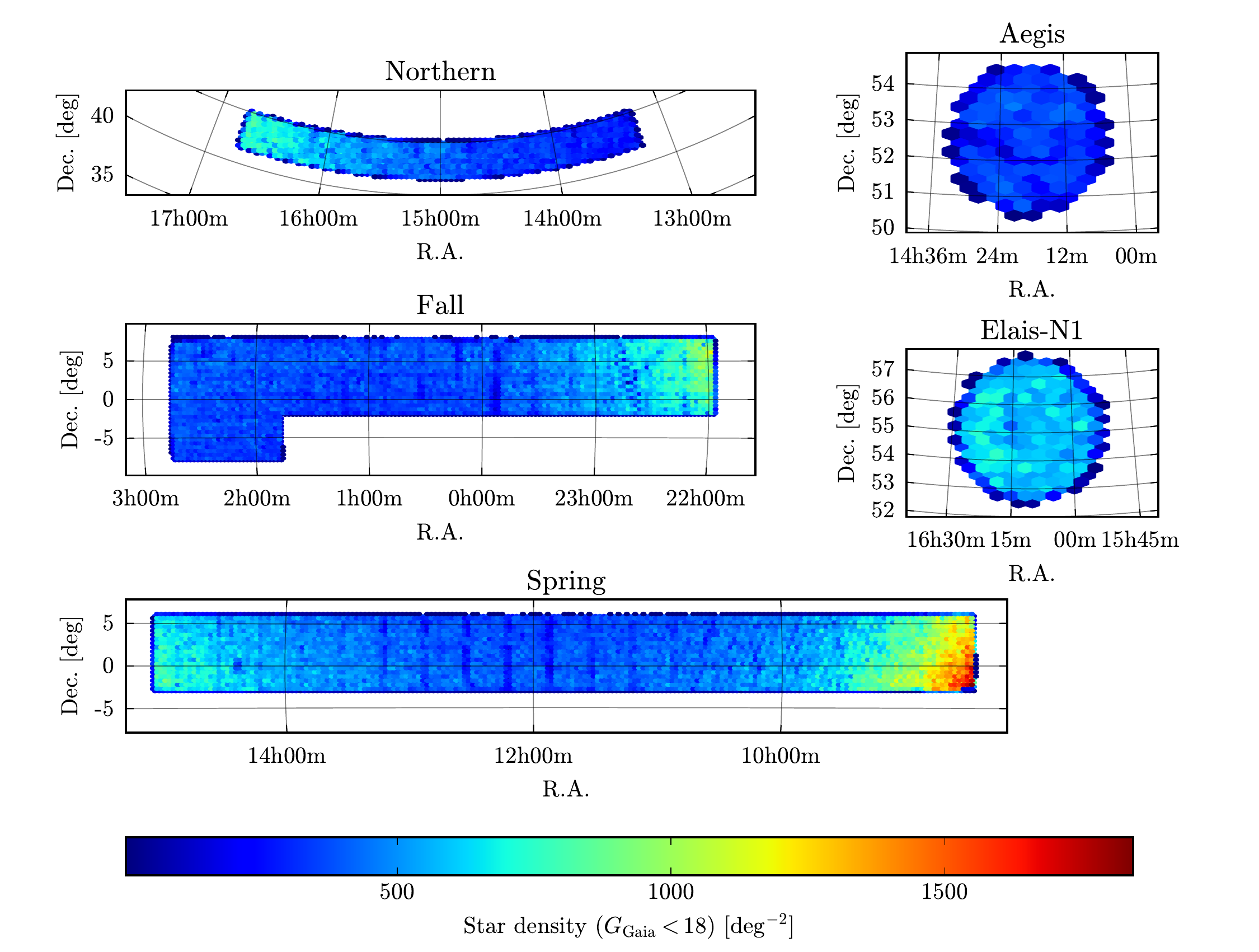}
  \end{center}
 \caption{Sky density in equatorial coordinates of stars brighter than $G_{\rm Gaia}=18$, within the HSC-SSP footprint (when completed). We show the Elais-N1 Deep field, the Aegis, the Northern, Fall, and Spring Wide fields. The other Deep and Ultra-deep fields lie within the Wide-field footprints. In the Spring and Fall fields, one can see the inhomogeneity of the DR1 sky distribution of Gaia sources (the dark patterns show the low-density areas).}
    \label{fig:skyDensity}
\end{figure*}

In practice, we first obtain the DR1 Gaia and Tycho-2 catalogues in an area enclosing the full planned HSC-SSP footprint, extended to one degree around the borders to account for bright stars outside the survey footprint that can potentially affect the sources inside. The full HSC-SSP footprint will be eventually composed of three Wide fields: Spring, Fall, and Northern, as well as one pointing in the AEGIS field used for photometric redshift calibration. All the Deep and Ultra Deep fields are within the Wide footprint, except the Elais-N1 Deep field, in which we add to our star sample the corresponding Gaia and Tycho-2 stars. We show in Figure~\ref{fig:skyDensity} the sky density of the bright-star catalogue in the planned HSC-SSP footprint. Note the higher star density near the Galactic plane at R.A.~$\sim22$h and $\sim9$h in the Fall and Spring fields, respectively.

Then, we combine the Tycho-2 and Gaia catalogues. We select all sources from the Gaia DR1 release, filtering out those flagged as \texttt{duplicated\_source}.  When a source is found in both Gaia and Tycho-2 catalogues, we keep the Gaia position and magnitude. For the Tycho-2 stars that are not in the Gaia sample, we emulate the $G_{\rm gaia}$ magnitude following Equation~\ref{eq:emulGaia}.

\begin{figure}
 \begin{center}
  \includegraphics[width=0.49\textwidth]{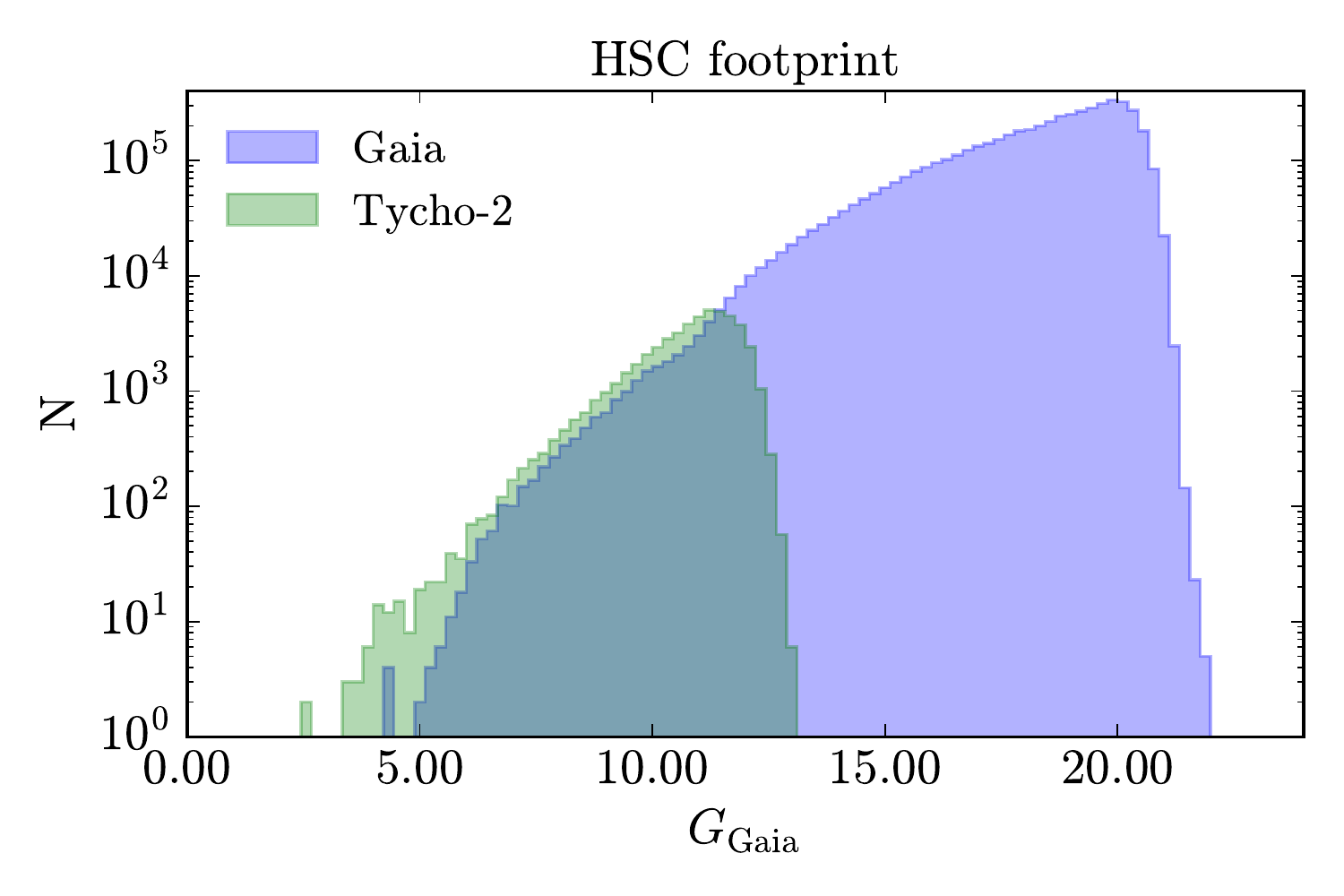}
 \end{center}
 \caption{Magnitude distributions for Gaia sources (measured magnitude) and Tycho-2 stars (emulated magnitude) in the planned HSC-SSP footprint. As the DR1 Gaia sample is incomplete at bright magnitude, we use the Tycho-2 stars to complete it.}
    \label{fig:magDist}
\end{figure}

In Figure~\ref{fig:magDist} we show the magnitude distribution of Gaia and Tycho-2 sources. The Gaia completeness is especially affected at bright magnitude due to the selection imposed in DR1, but also at all magnitudes due to some spatial incompleteness, as illustrated in Figure~\ref{fig:skyDensity}. The addition of the Tycho-2 sample ensure at least a complete star sample at $G_{\rm gaia}\lesssim12$.

\subsection{Ensuring a pure star sample}
\label{sec:pureSample}

To remove the galaxies incorrectly identified as stars by Gaia, we make use of the photometric information provided by the SDSS survey \citep{York:2000aa,Albareti:2016}. To do so, we match the Gaia+Tycho-2 sample previously assembled to the SDSS photometric sources downloaded from the DR13 release database\footnote{\url{http://www.sdss.org/dr13/}}. We define a source as extended in the SDSS using the photometric information provided in the three $g,r$ and $i$ bands with the highest signal-to-noise ratio as folows:
\begin{verbatim}
clean == 1
AND
(obj.flags_g & dbo.fPhotoFlags('SATUR_CENTER'))==0
AND
(obj.flags_r & dbo.fPhotoFlags('SATUR_CENTER'))==0
AND
(obj.flags_i & dbo.fPhotoFlags('SATUR_CENTER'))==0
AND (  
     (0.1 < psfMag_g-cModelMag_g 
      AND psfMag_g-cModelMag_g < 20.0)
     OR 
     (0.1 < psfMag_r-cModelMag_r 
      AND psfMag_r-cModelMag_r < 20.0)
     OR 
     (0.1 < psfMag_i-cModelMag_i 
      AND psfMag_i-cModelMag_i < 20.0) )
\end{verbatim}

The first three conditions impose that the center of the source is not flagged as saturated in any of the $gri$ bands. We do not expect galaxies -- even bright -- to saturate in the SDSS. The next three conditions are similar to the definition adopted by the SDSS collaboration to define extended sources, but are more conservative to ensure a 100\% complete sample of extended sources, the goal being to remove \emph{all} galaxies from the bight-star sample. In short, this condition means that all sources with a luminosity profile 10\% brighter than the PSF luminosity profile in any of the $gri$ bands are flagged as extended.

To validate our approach we also match the star catalogue to the S16A HSC-SSP data. The HSC images are much deeper and with a better resolution than SDSS, allowing to do a more accurate extended/point-source separation. As mentioned above, our objective is to build an HSC-SSP-agnostic bright-star sample, so the comparison with HSC-SSP serves only as validating our approach. 

We flag extended sources in HSC-SSP using exclusively the $i$-band photometry in the Wide, as the $i$-band observations are always done in best-seeing conditions. We define an extended source as:
\begin{verbatim}
parent_id == 0
AND NOT iflags_pixel_saturated_center 
AND 0.01 < imag_psf-icmodel_mag 
AND imag_psf-icmodel_mag  < 20.0
\end{verbatim}

This selection closely follows that of the SDSS, with the exception that it is only done in the $i$-band and that the luminosity profile should only change by 1\% to be flagged as extended. The first condition ensures that the source is not deblended. Overall such a definition is also extremely conservative to ensure that all galaxies are properly identified and removed from the star sample.

\begin{figure}
 \begin{center}
  \includegraphics[width=0.49\textwidth]{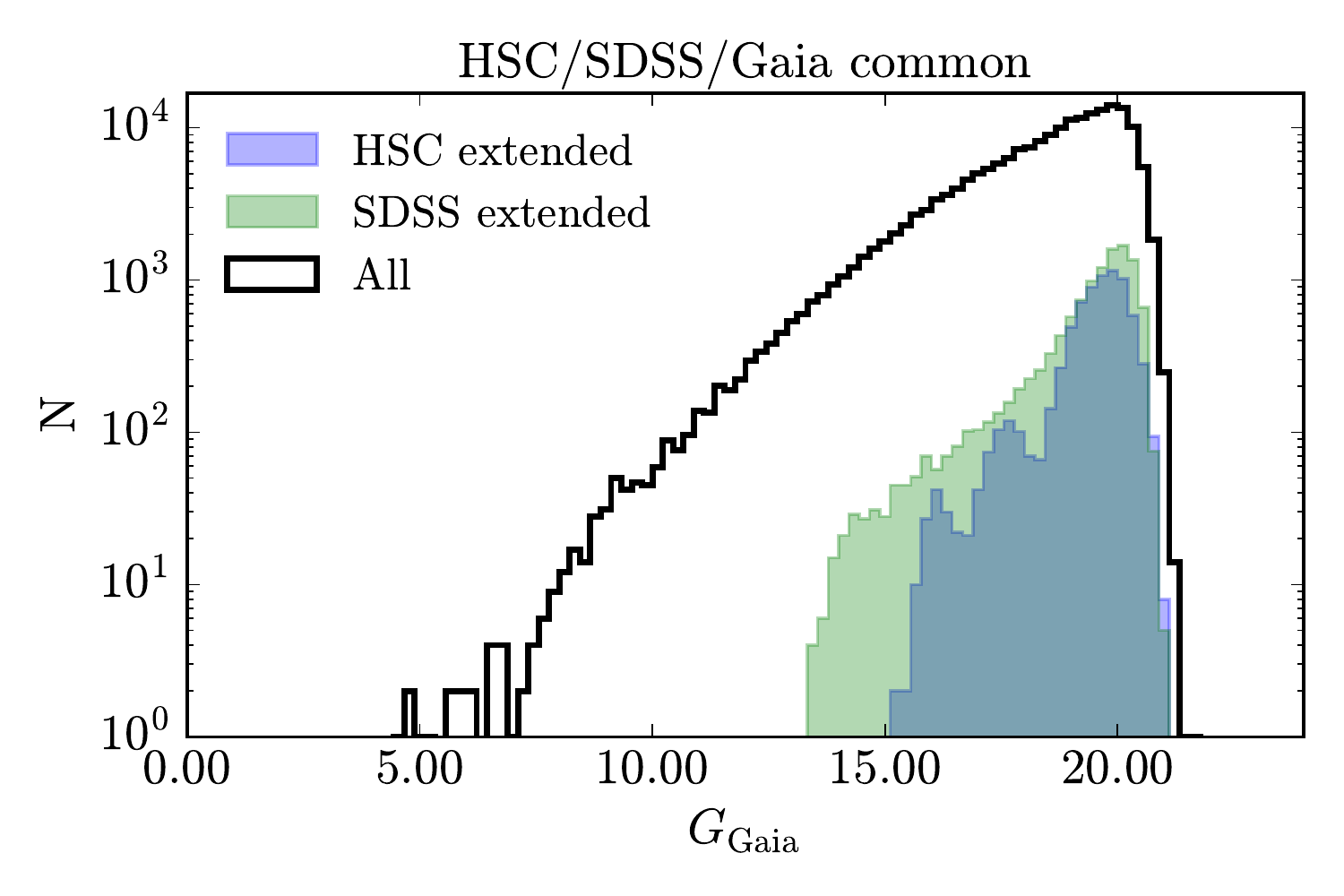}\\
  \includegraphics[width=0.49\textwidth]{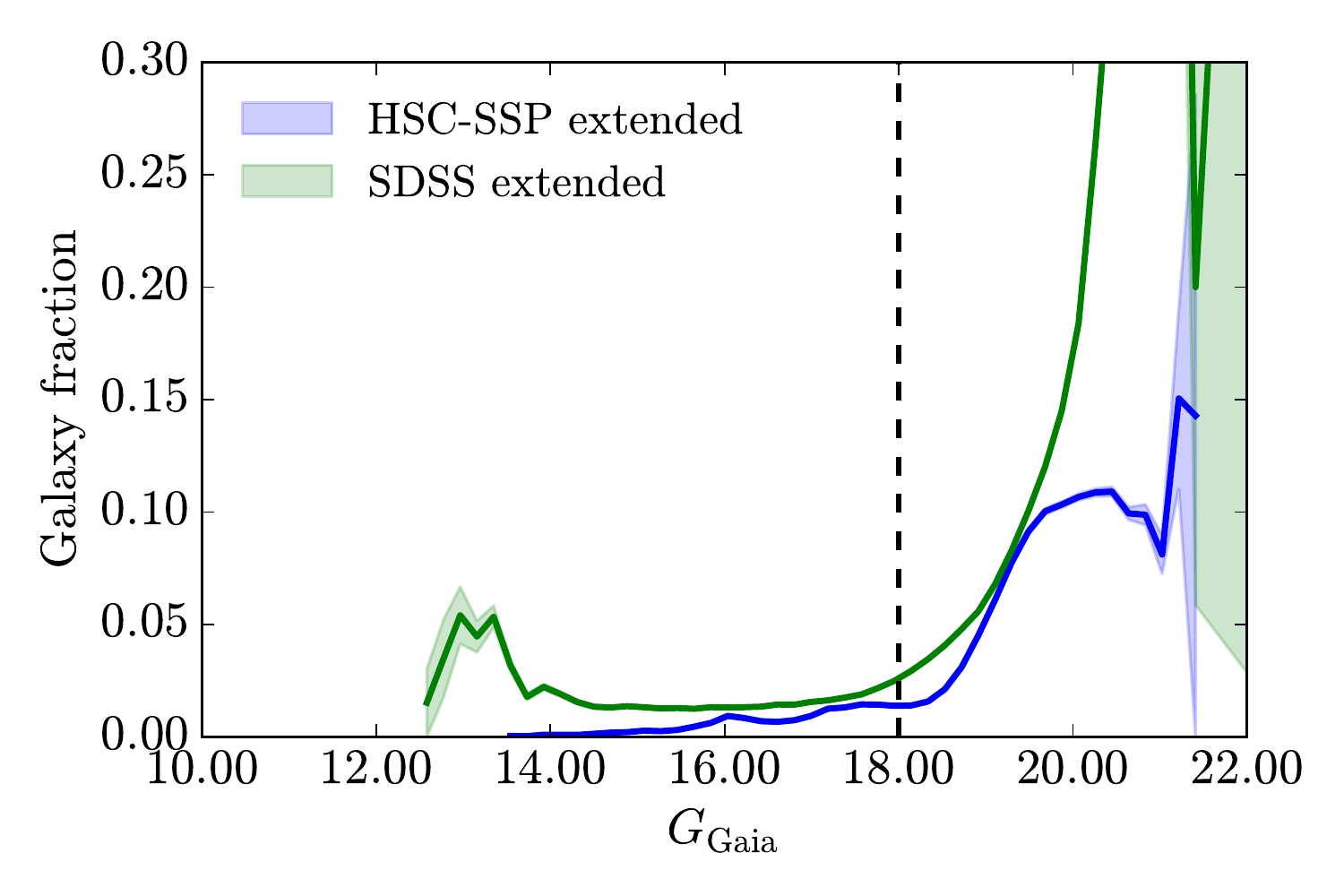}
	\end{center}
 \caption{Magnitude distributions for Gaia+Tycho-2 sources. The top panel shows the distributions of all sources in the Gaia+Tycho-2 combined catalogue (black solid line), the extended sources in HSC-SSP (blue), and the extended objects in SDSS (green), in a common area. The bottom panel shows the relative fraction of HSC (blue) and SDSS (green) extended sources in Gaia. { The vertical black dashed line shows the brightness limit ($G_{\rm gaia}<18$) adopted in this work.}}
    \label{fig:extended}
\end{figure}

The fraction and magnitude distribution of extended sources in Gaia, as a function of $G_{\rm gaia}$ magnitude, are shown in Figure~\ref{fig:extended}.
Extended sources start contaminating the star sample in the regime fainter than $G_{\rm gaia}\sim12$, but remains below 3--5\%. The fraction starts becoming significantly larger beyond $G_{\rm gaia}\sim18$, reaching at least 10\% at the 19th magnitude. The fraction increases significantly beyond 19 if using the SDSS, probably due to the highly conservative cut in extendedness. We conclude that building a pure star sample (at least with the Gaia DR1 data) requires that we limit the Gaia star sample at $G_{\rm gaia}<18$, which is slightly brighter than the ideal limit for HSC-SSP (as shown in Section~\ref{sec:satlimits}).

In the regime $14 < G_{\rm gaia} < 18$, we remove all sources ($1.5\%$) that are flagged as extended in the SDSS according to the criterium given above. At $G_{\rm gaia} < 14$, visual inspection confirms that the ``bump'' seen in the SDSS source distribution is caused by stars with a significantly extended diffraction pattern.

The purity of our star sample is further investigated in Section~\ref{sec:purity}.
 
\subsection{Summary}
\label{sec:summary}

We summarise here the construction of the bright-star reference sample to be used for masking:
\begin{itemize}
\item our starting catalogue is the Gaia DR1 all-sky survey sample,
\item we restrict the sample to the footprint of the planned HSC-SSP survey, increased by one degree around the borders,
\item we complete the missing bright stars with the Tycho-2 star sample ($V<11.5$) and emulate the $G_{\rm Gaia}$ magnitude from the multi-band information,
\item the sample is cut at $G_{\rm Gaia} < 18$ beyond which both SDSS and HSC-SSP images show increased extended source contamination ($>10\%$),
\item we remove all sources ($1.5\%$) flagged as extended in the SDSS between $14 < G_{\rm Gaia} < 18$.
\end{itemize}

The final sample contains $1\,812\,106$ stars.

\section{Building the masks}
\label{sec:masks}

We describe in this section the adopted metric to empirically evaluate the impact of bright stars on the sources, and how we build the masks. Our goal is to account for the PSF-shaped luminous patterns, the extended luminous haloes, the $Y$-band vertical spikes and the linear bleed trails. The removal of the large-scale reflection ghosts, however, will be implemented in a future version of the HSC pipeline.

\subsection{Measuring the impact on detected sources}

\begin{figure*}
 \begin{center}
  \includegraphics[width=0.195\textwidth]{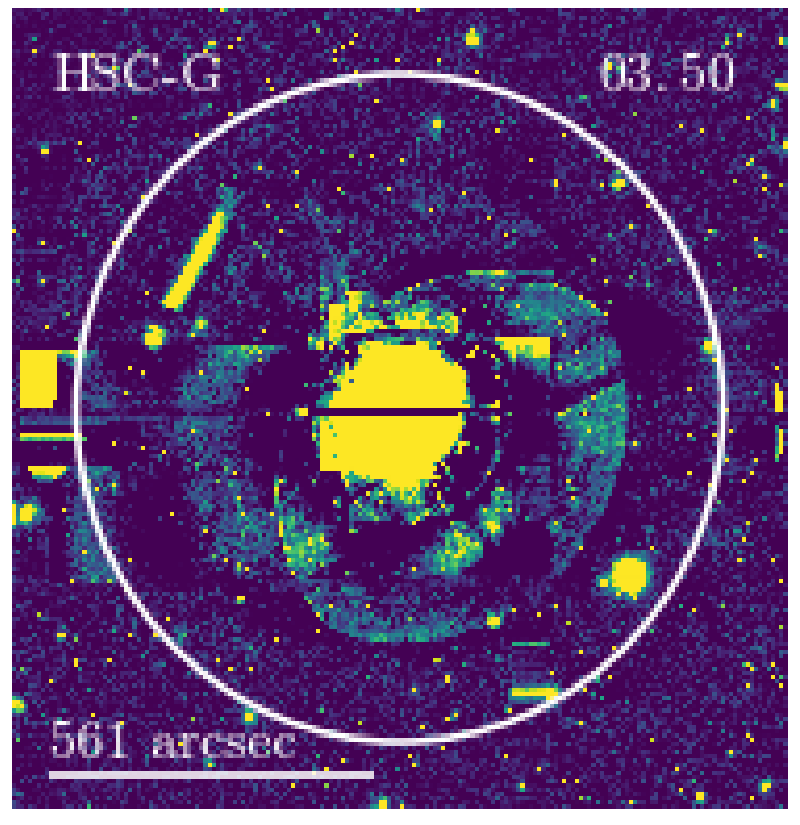}
  \includegraphics[width=0.195\textwidth]{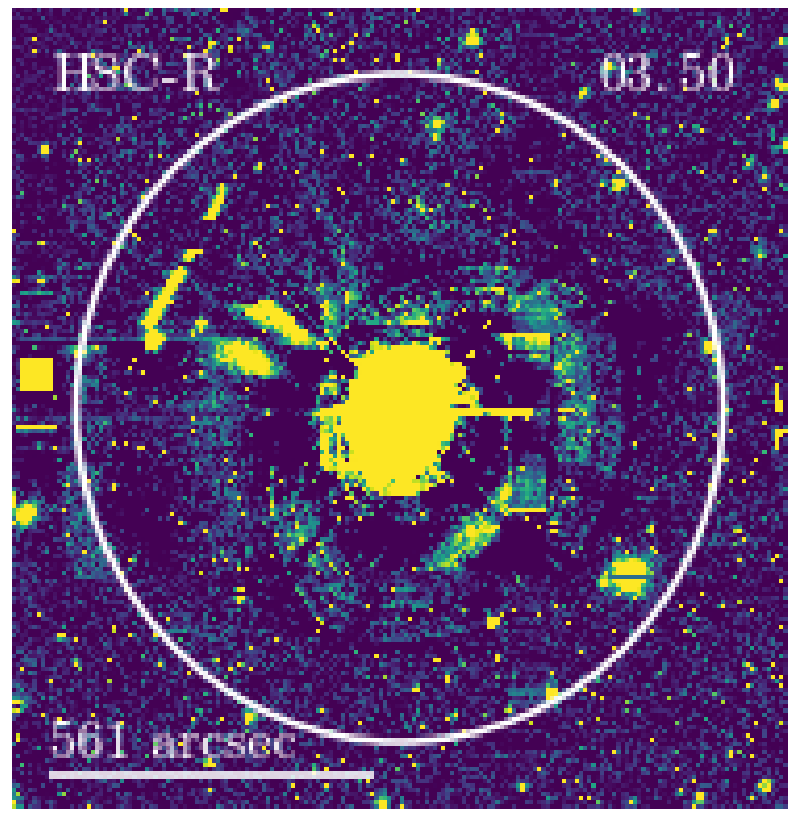}
  \includegraphics[width=0.195\textwidth]{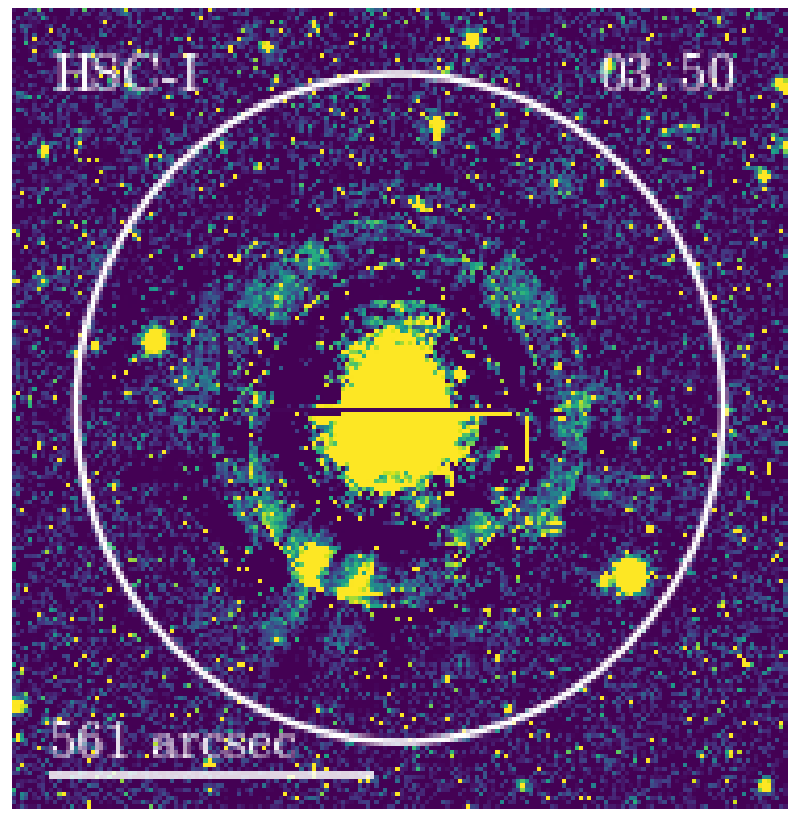}
  \includegraphics[width=0.195\textwidth]{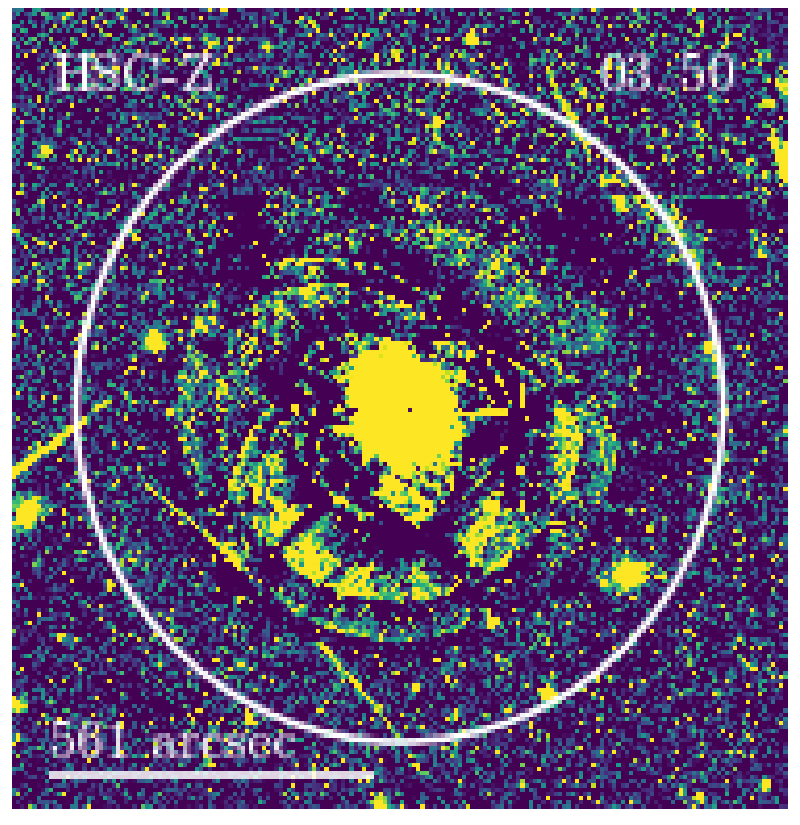}
  \includegraphics[width=0.195\textwidth]{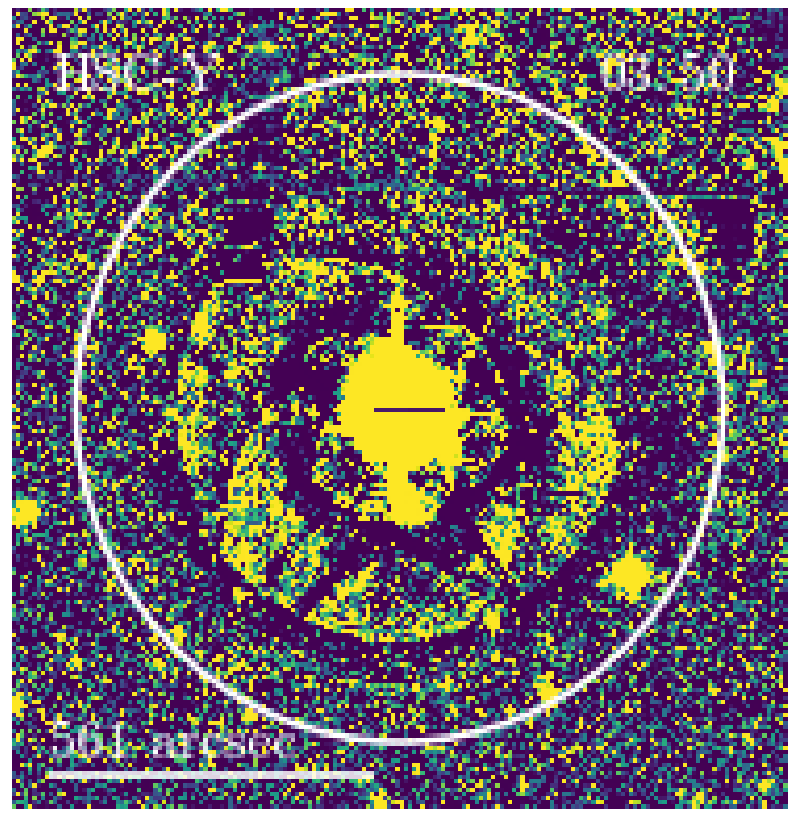}
  \includegraphics[width=0.195\textwidth]{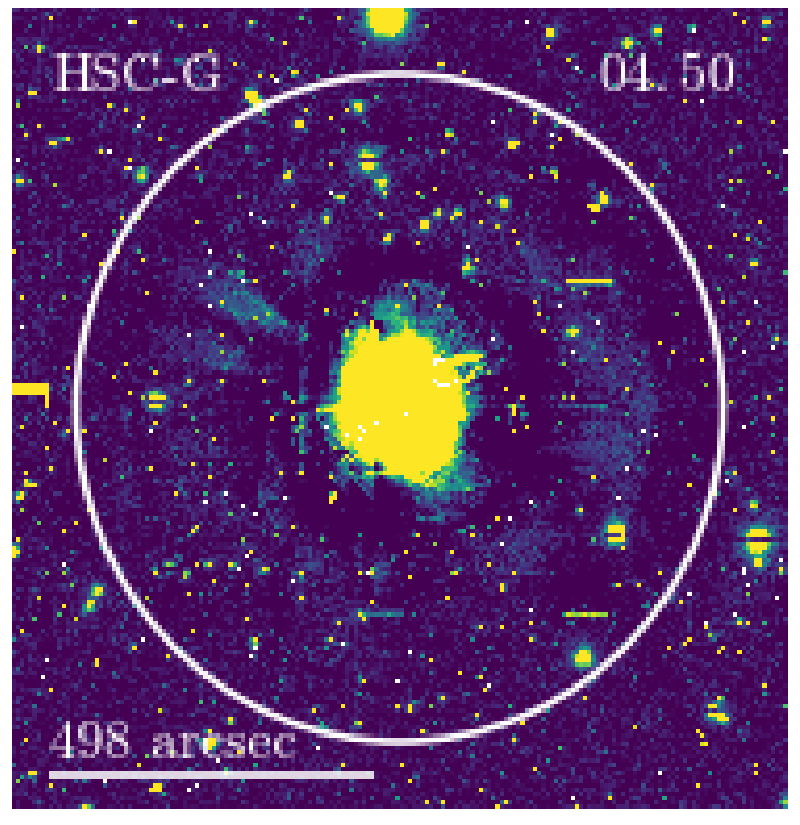}
  \includegraphics[width=0.195\textwidth]{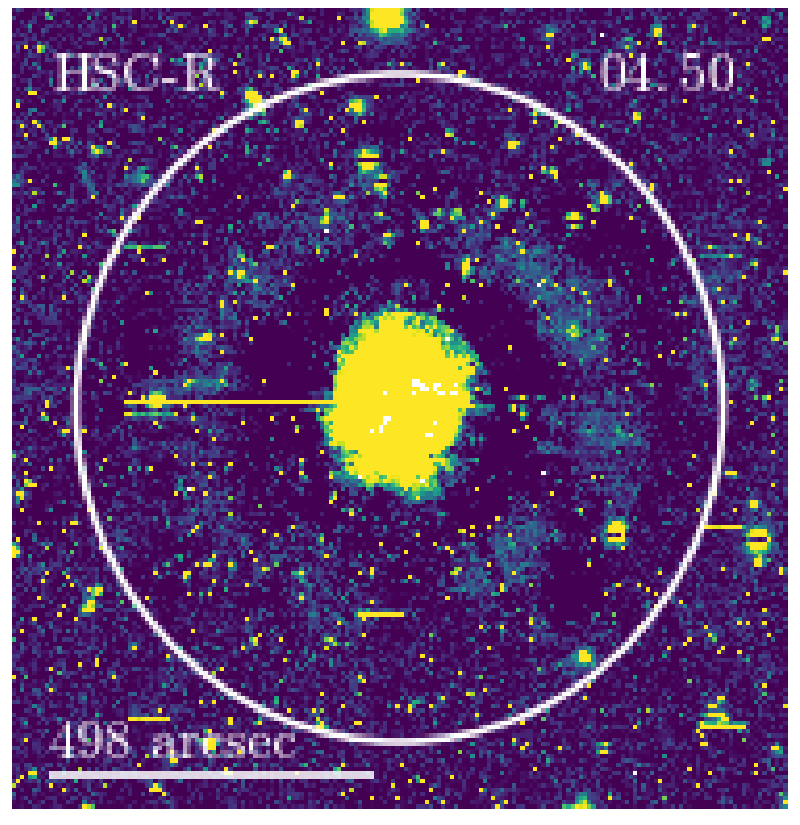}
  \includegraphics[width=0.195\textwidth]{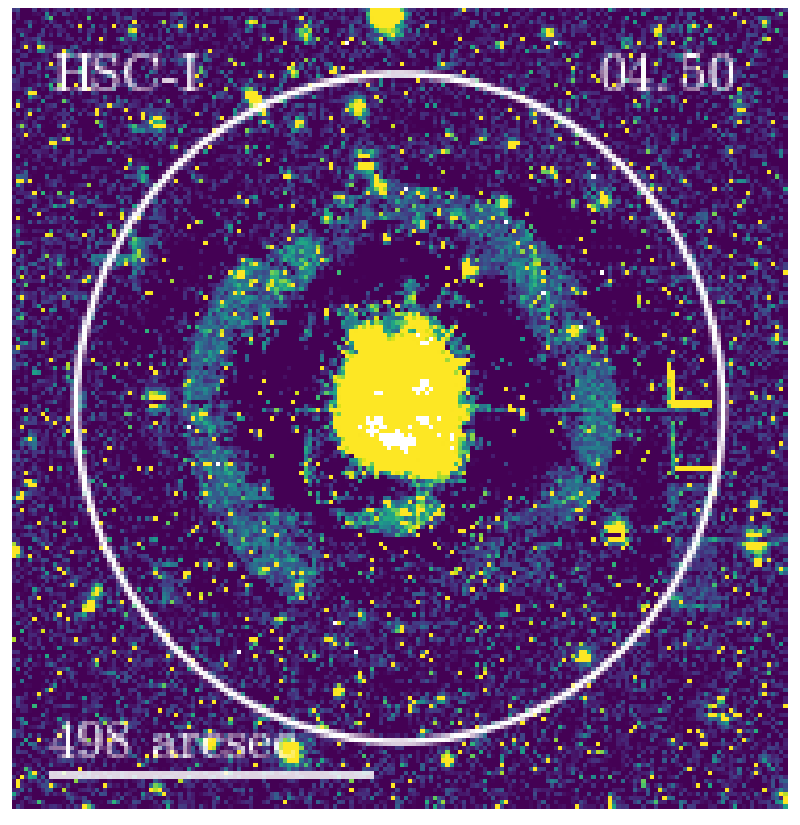}
  \includegraphics[width=0.195\textwidth]{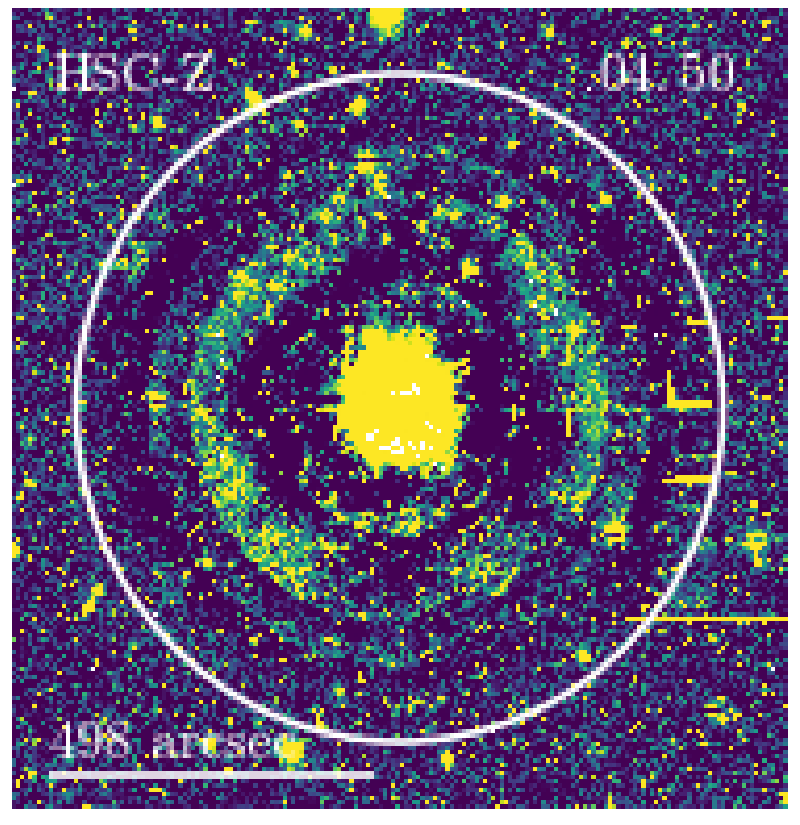}
  \includegraphics[width=0.195\textwidth]{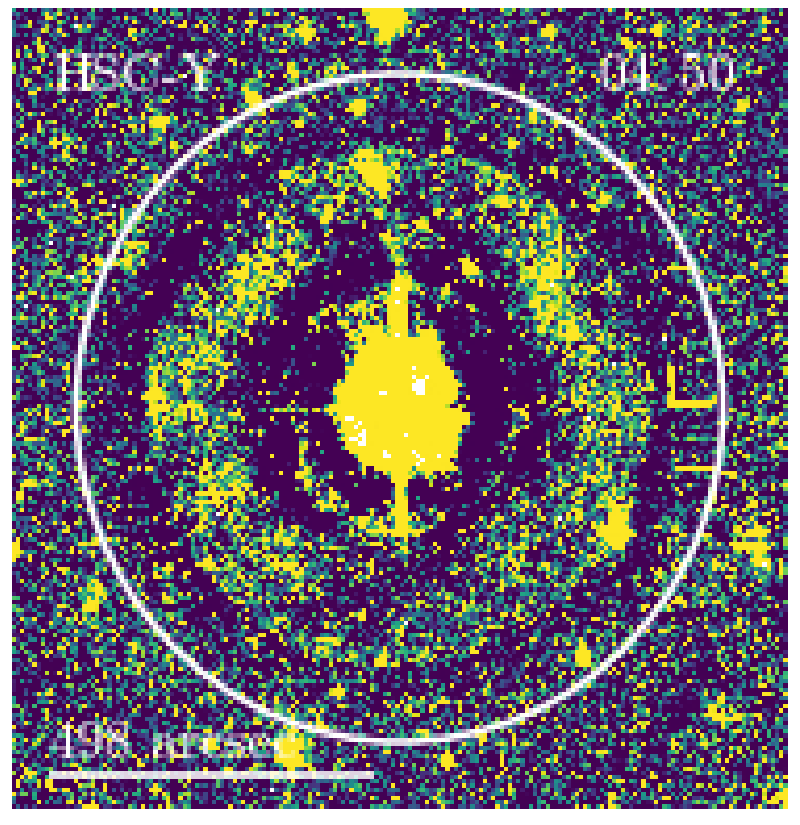}
  \includegraphics[width=0.195\textwidth]{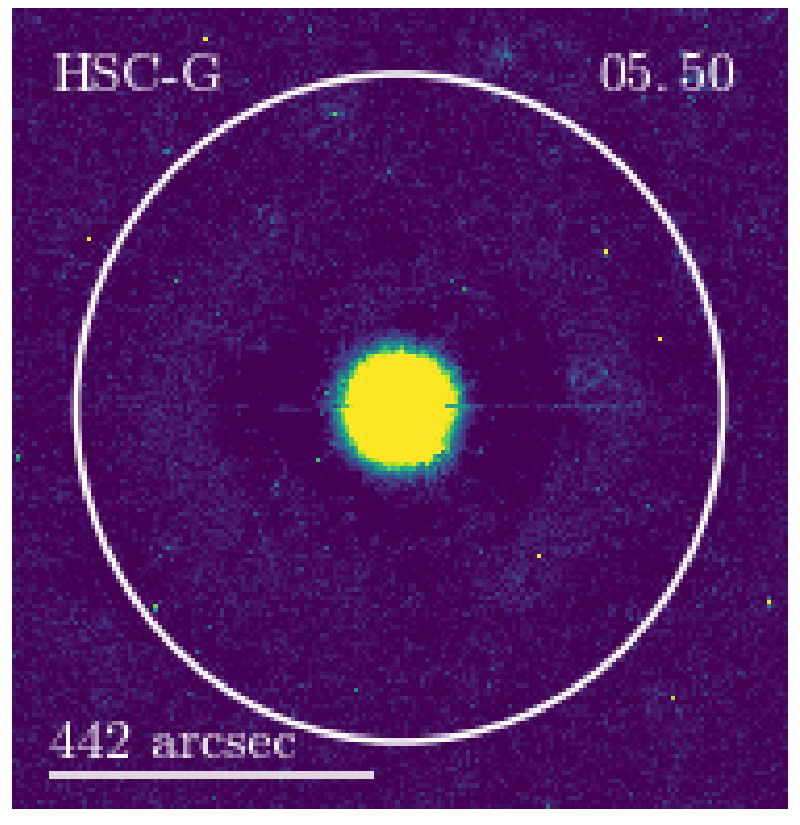}
  \includegraphics[width=0.195\textwidth]{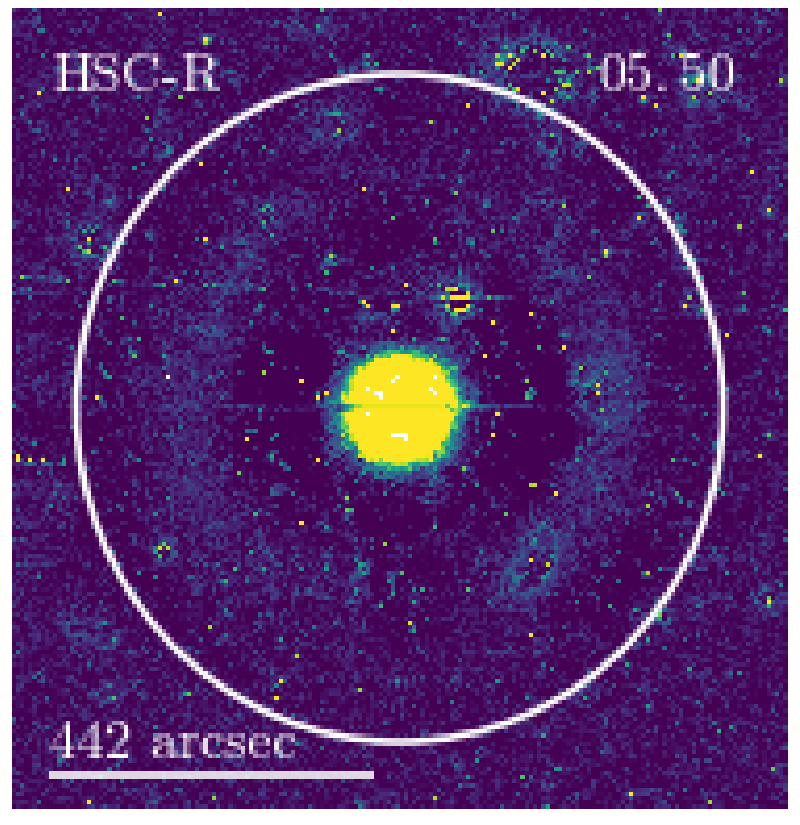}
  \includegraphics[width=0.195\textwidth]{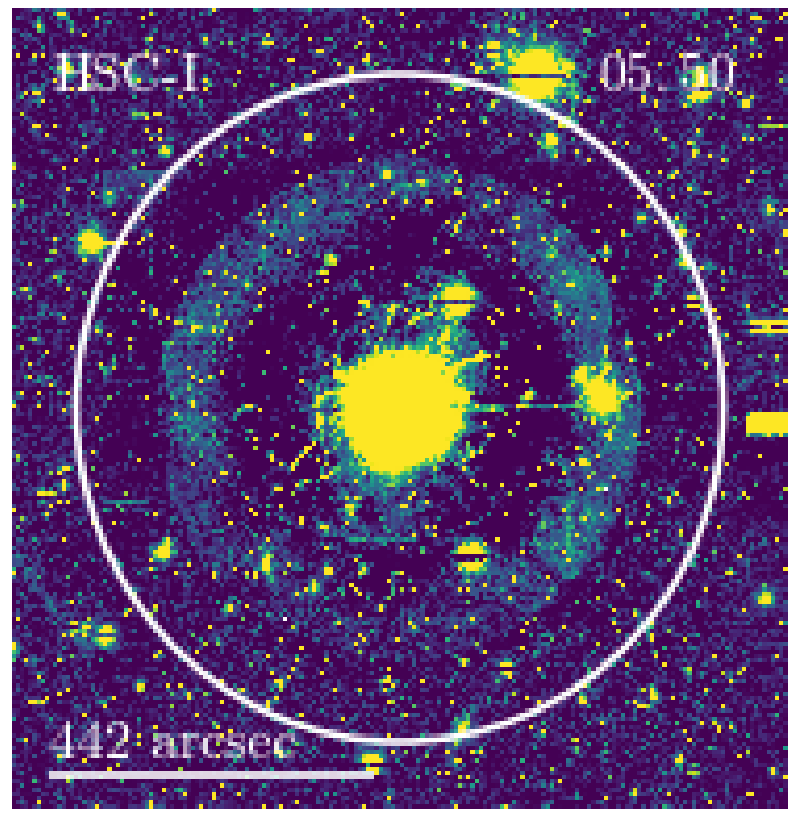}
  \includegraphics[width=0.195\textwidth]{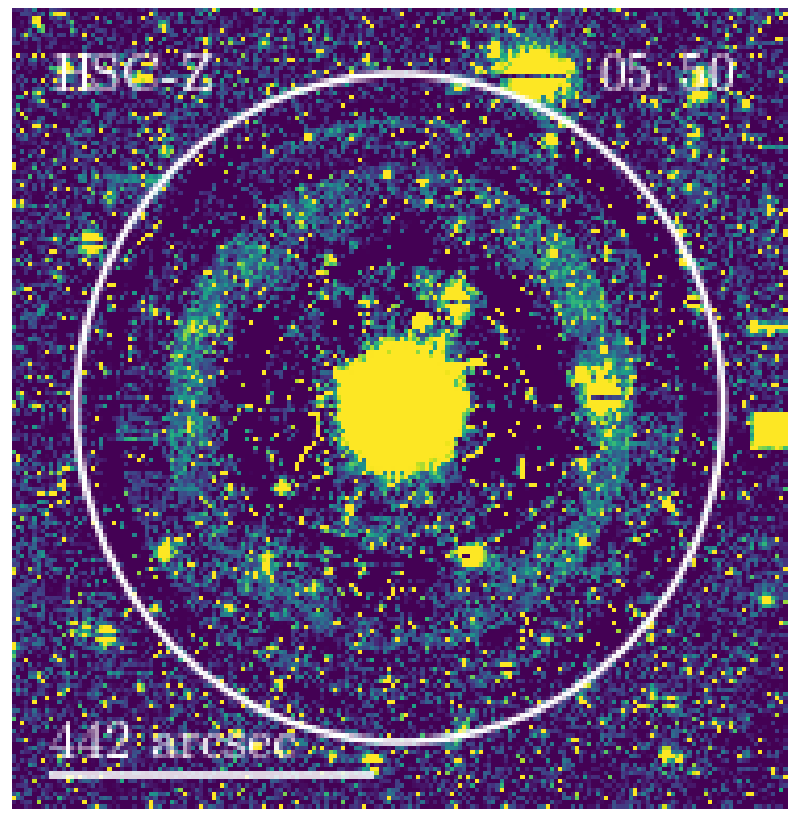}
  \includegraphics[width=0.195\textwidth]{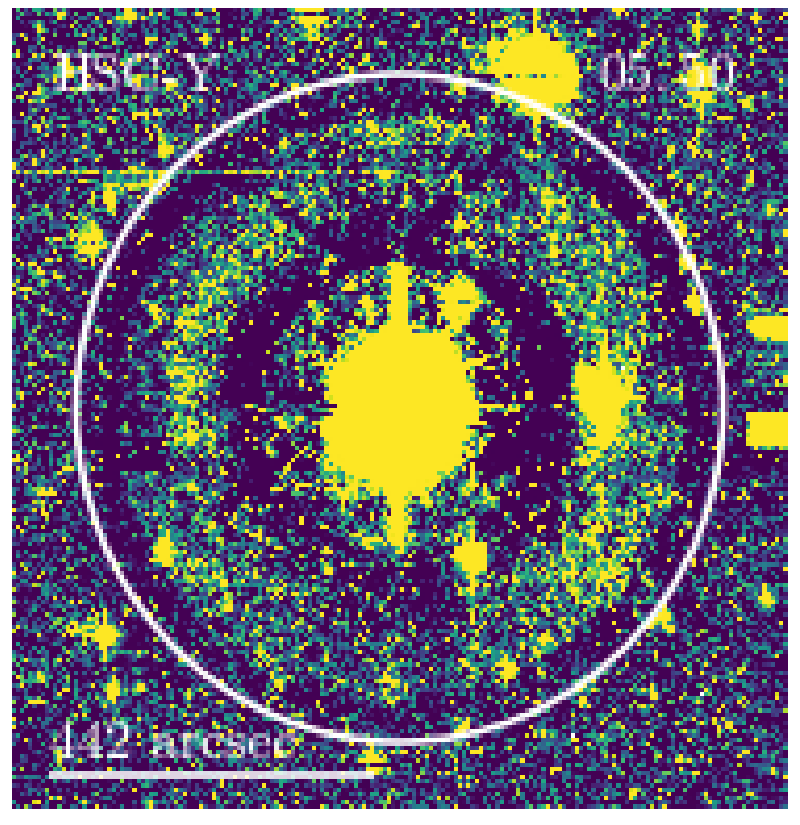}
  \includegraphics[width=0.195\textwidth]{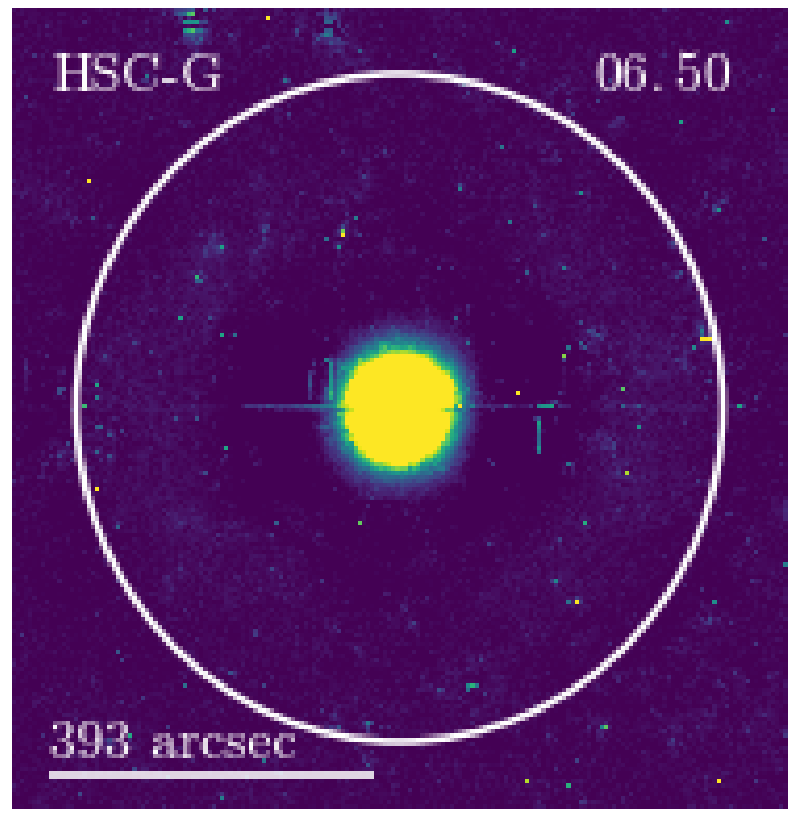}
  \includegraphics[width=0.195\textwidth]{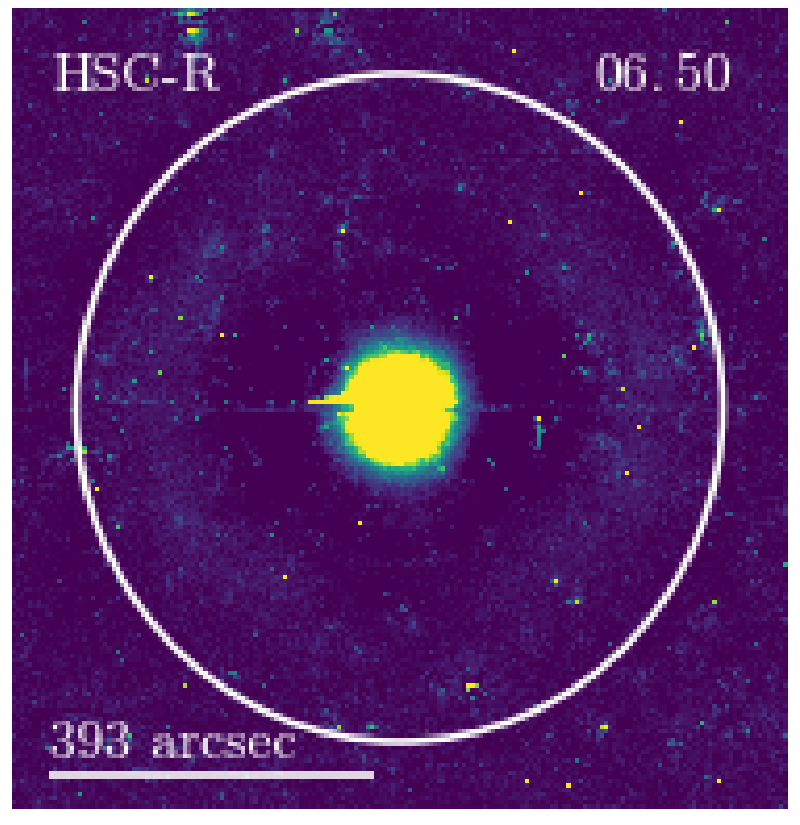}
  \includegraphics[width=0.195\textwidth]{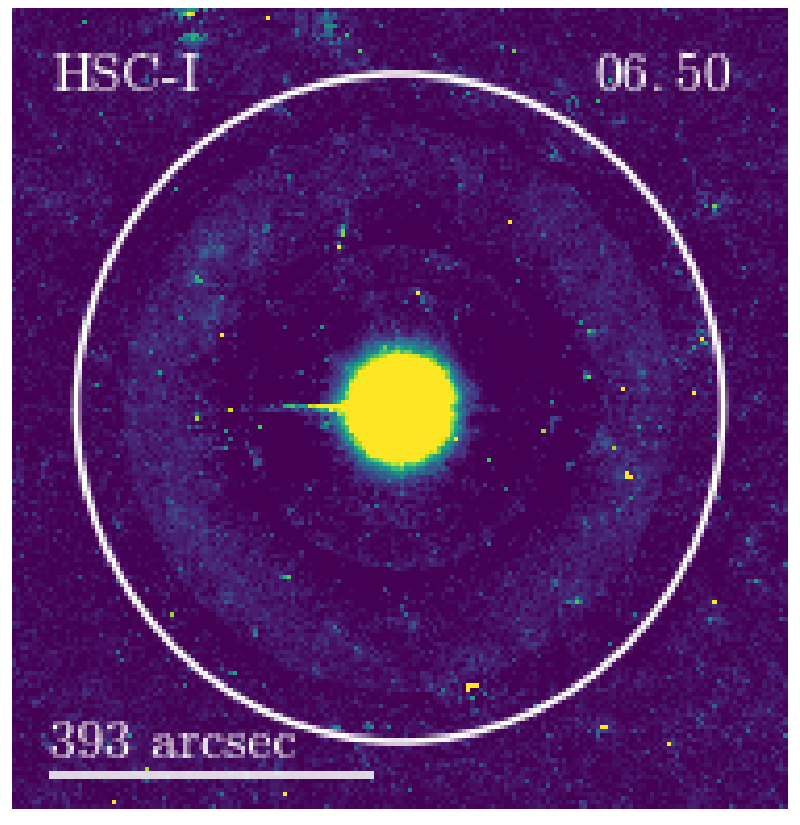}
  \includegraphics[width=0.195\textwidth]{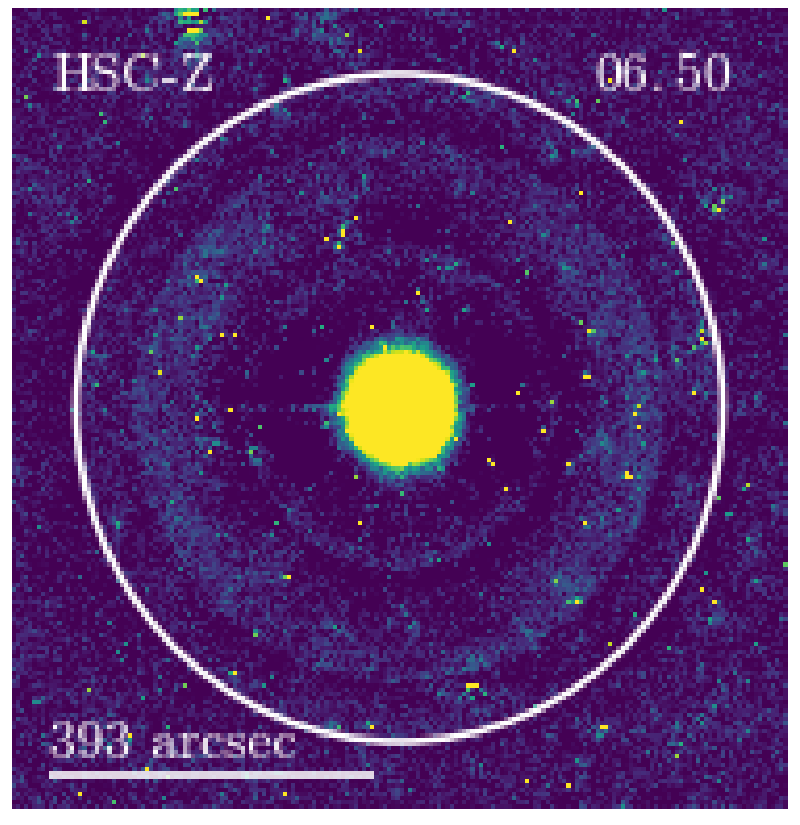}
  \includegraphics[width=0.195\textwidth]{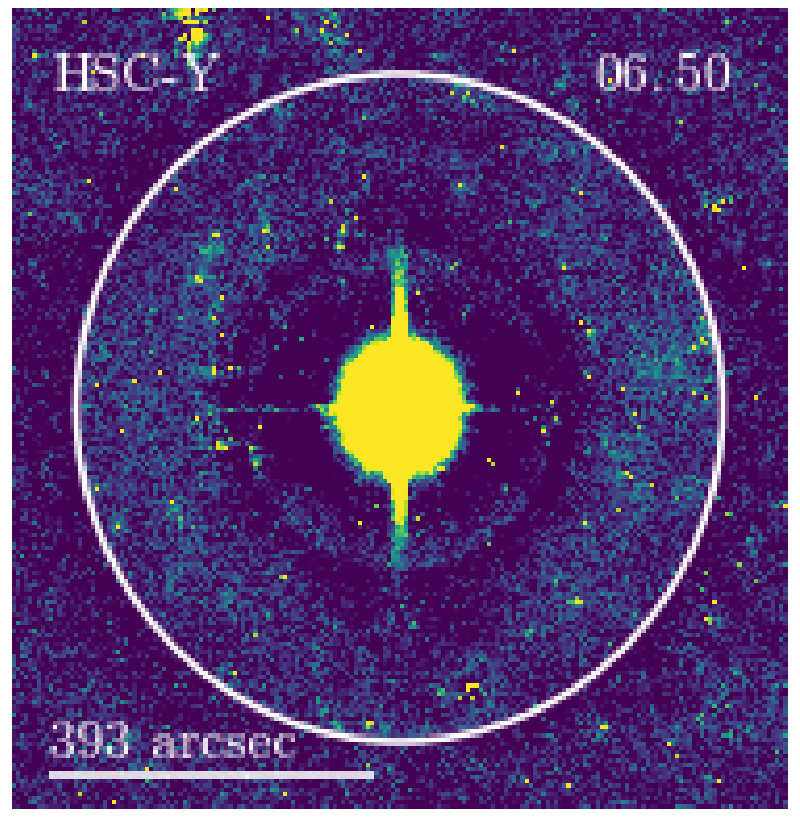}
  \includegraphics[width=0.195\textwidth]{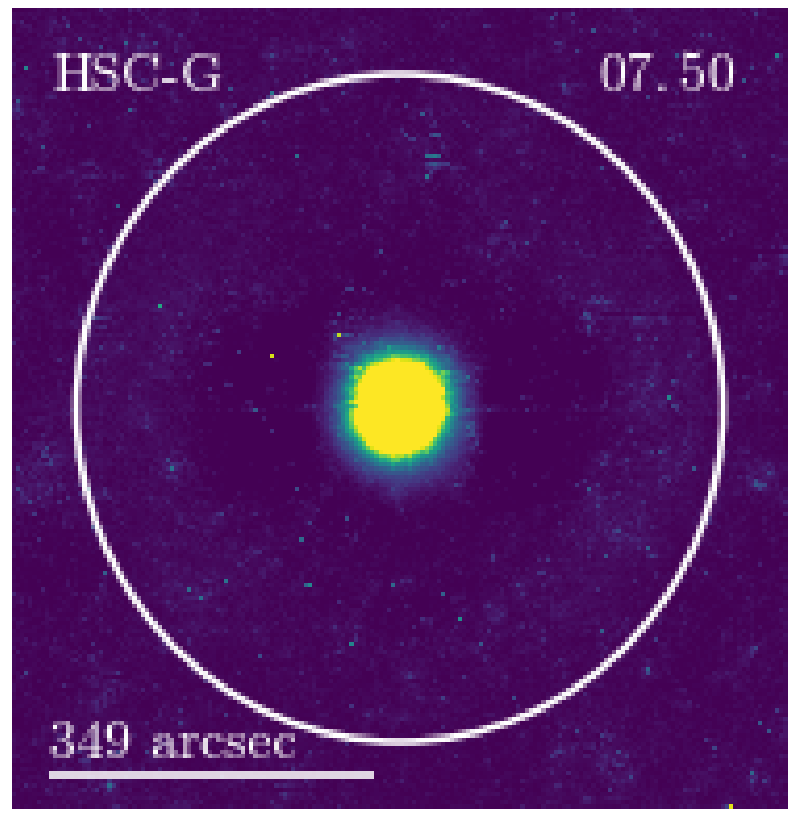}
  \includegraphics[width=0.195\textwidth]{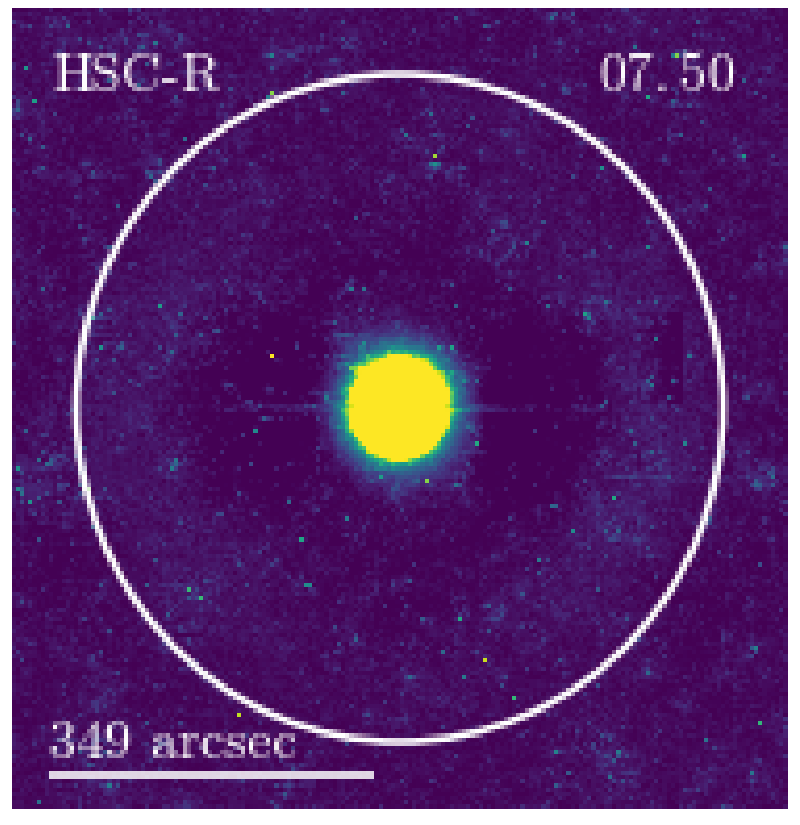}
  \includegraphics[width=0.195\textwidth]{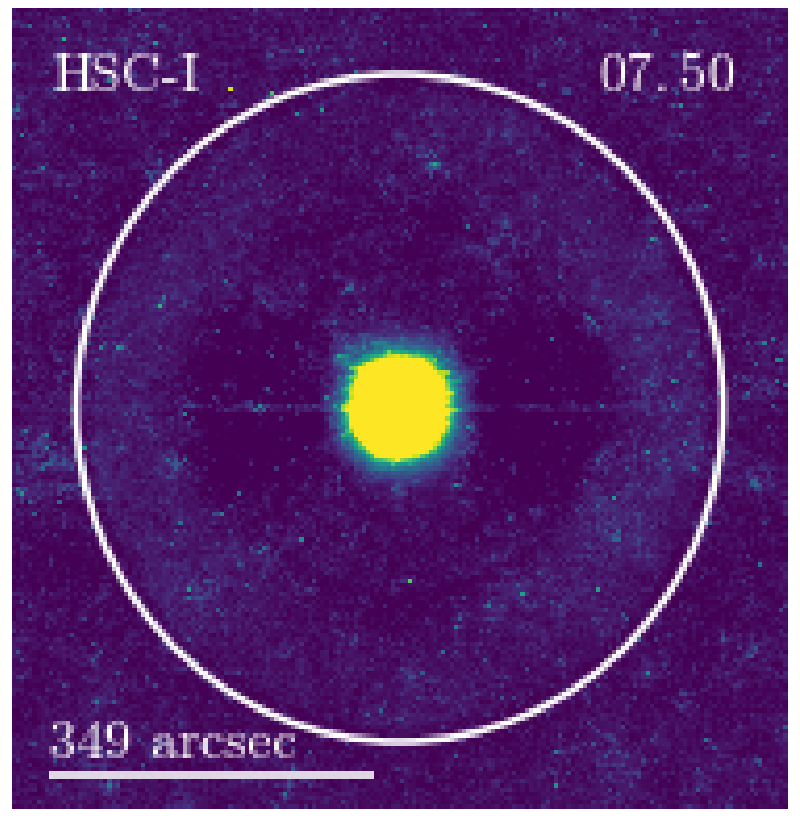}
  \includegraphics[width=0.195\textwidth]{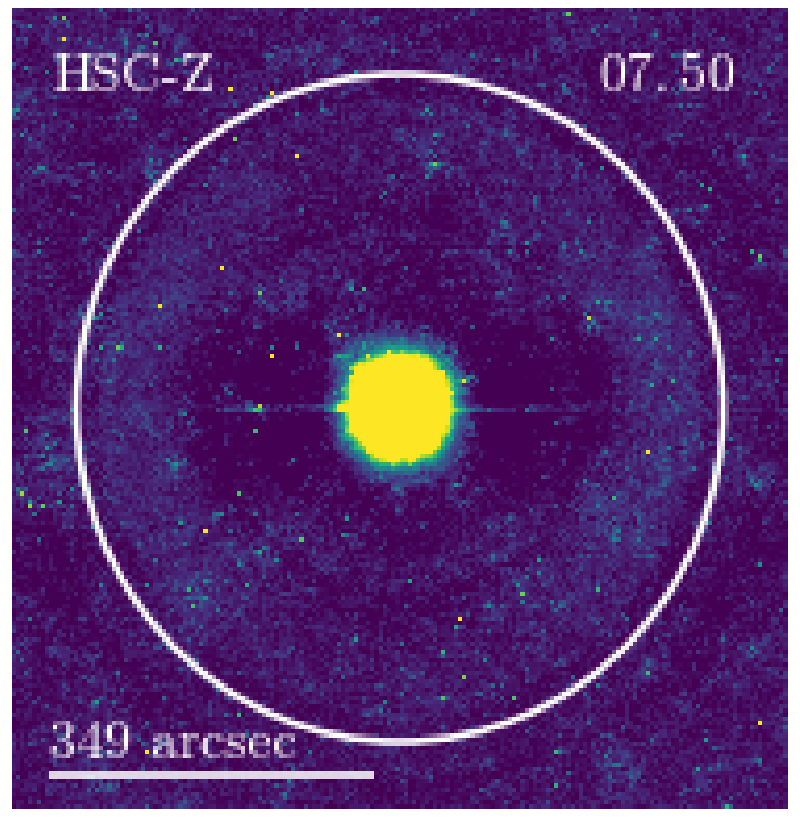}
  \includegraphics[width=0.195\textwidth]{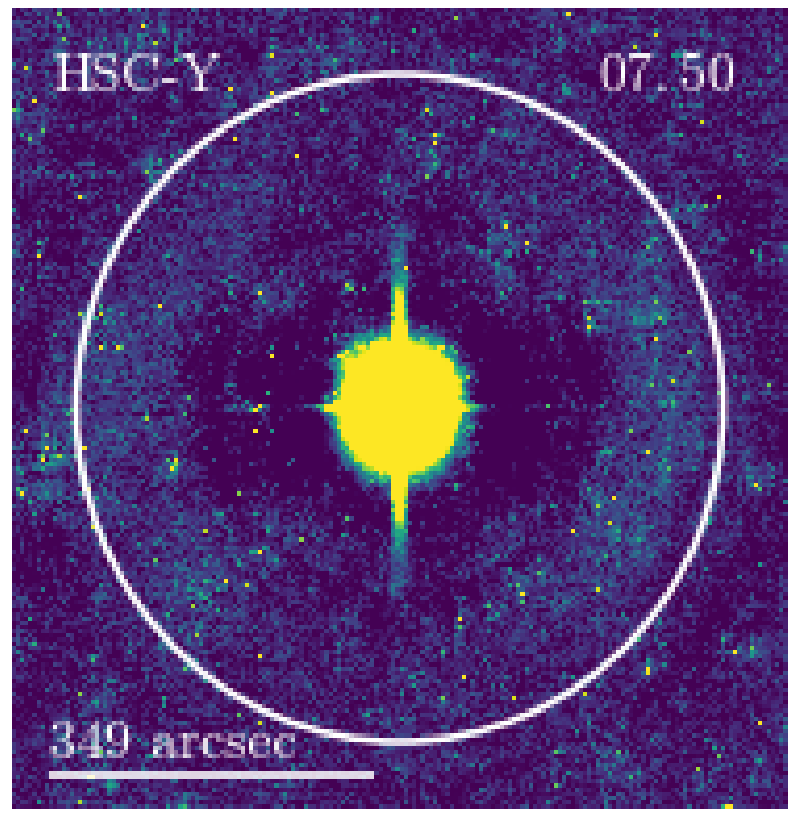}
\end{center}
       \caption{Stacked ($3\sigma$-clipped) images of stars in the range $3.5 \le G_{\rm Gaia} \le 7.5$ (the star brightness is decreasing from top to bottom) and for the $grizY$ filters (from left to right). The white circles show the circular masks to account for the isotropic effect on the nearby sources, whose radii are calculated from the variation in source density. The radius dependence on star magnitude is modelled by Equation~\ref{eq:rvsmag}.}
    \label{fig:stackedImages1}
\end{figure*}

\begin{figure*}
\begin{center}
  \includegraphics[width=0.195\textwidth]{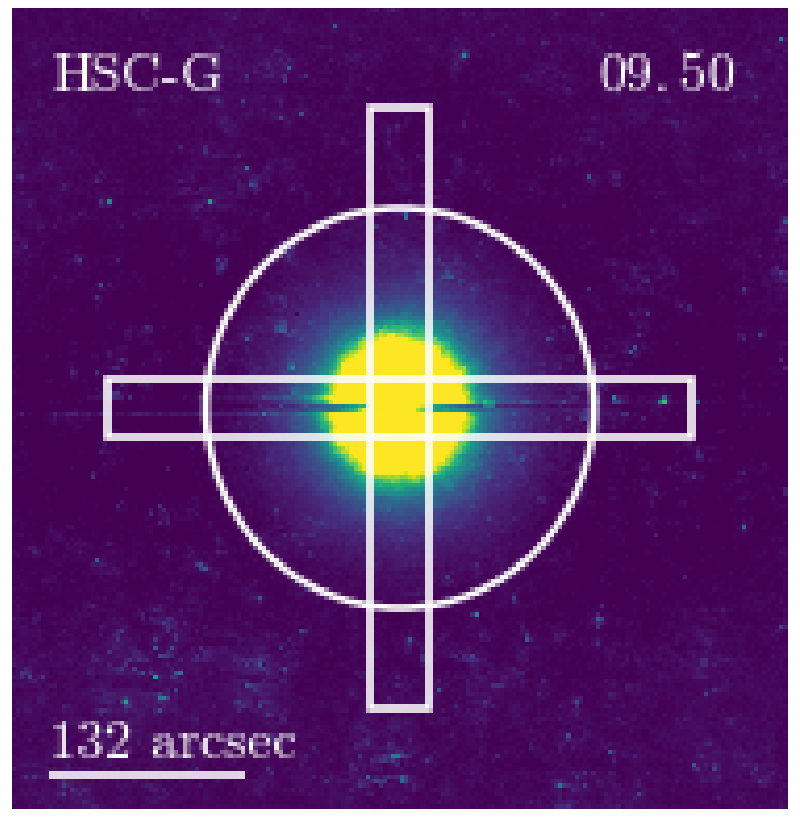}
  \includegraphics[width=0.195\textwidth]{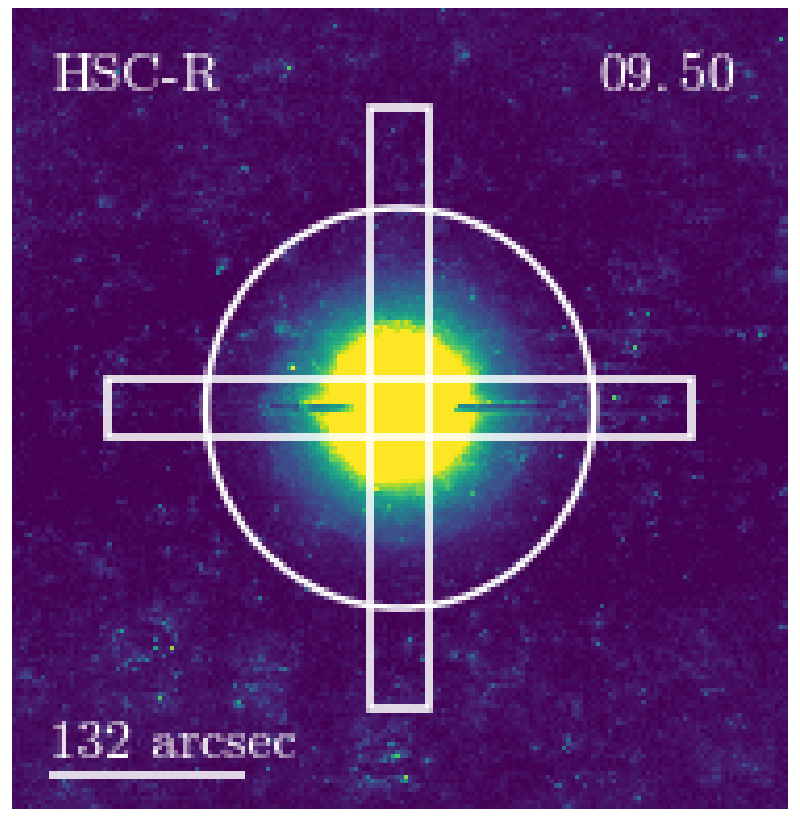}
  \includegraphics[width=0.195\textwidth]{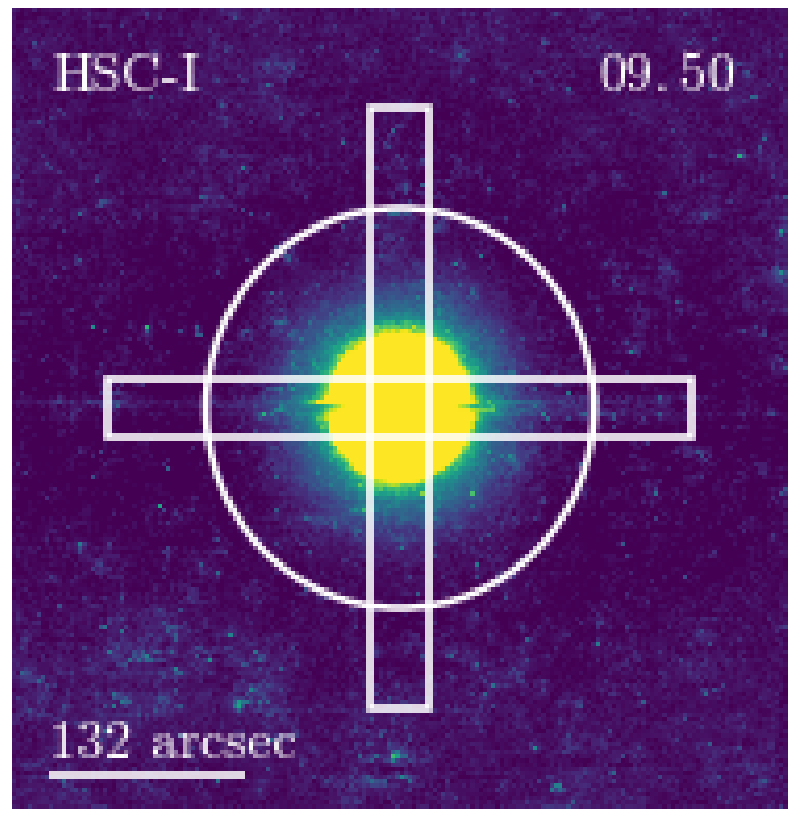}
  \includegraphics[width=0.195\textwidth]{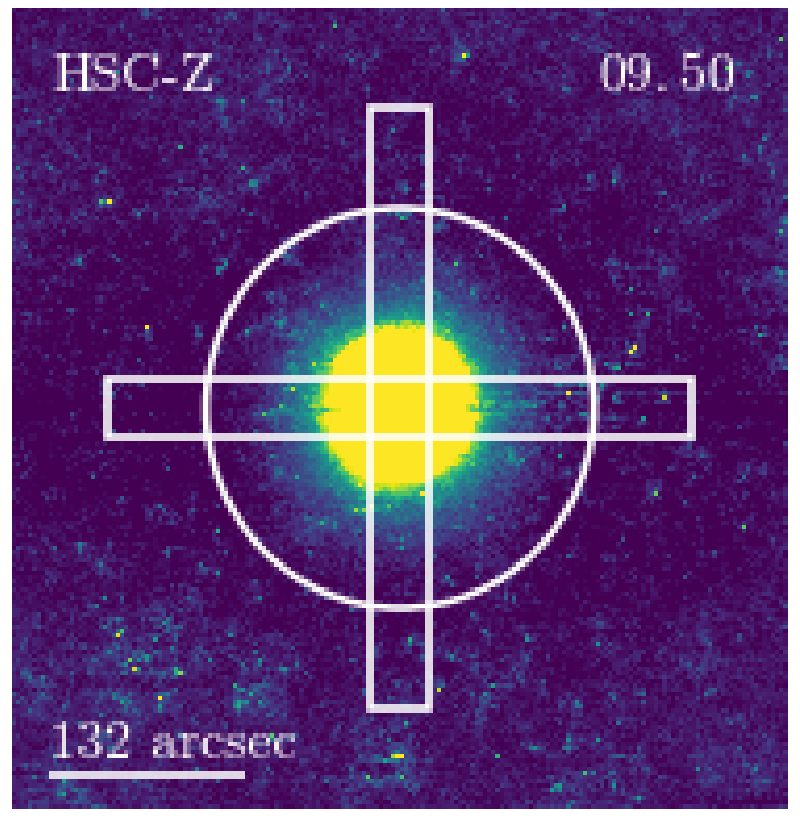}
  \includegraphics[width=0.195\textwidth]{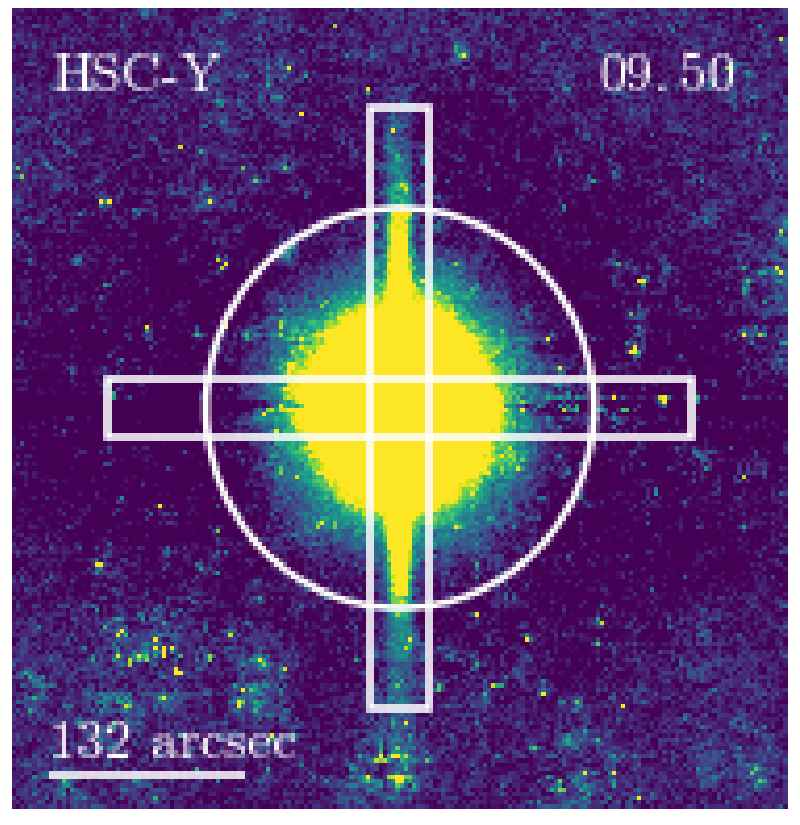}
  \includegraphics[width=0.195\textwidth]{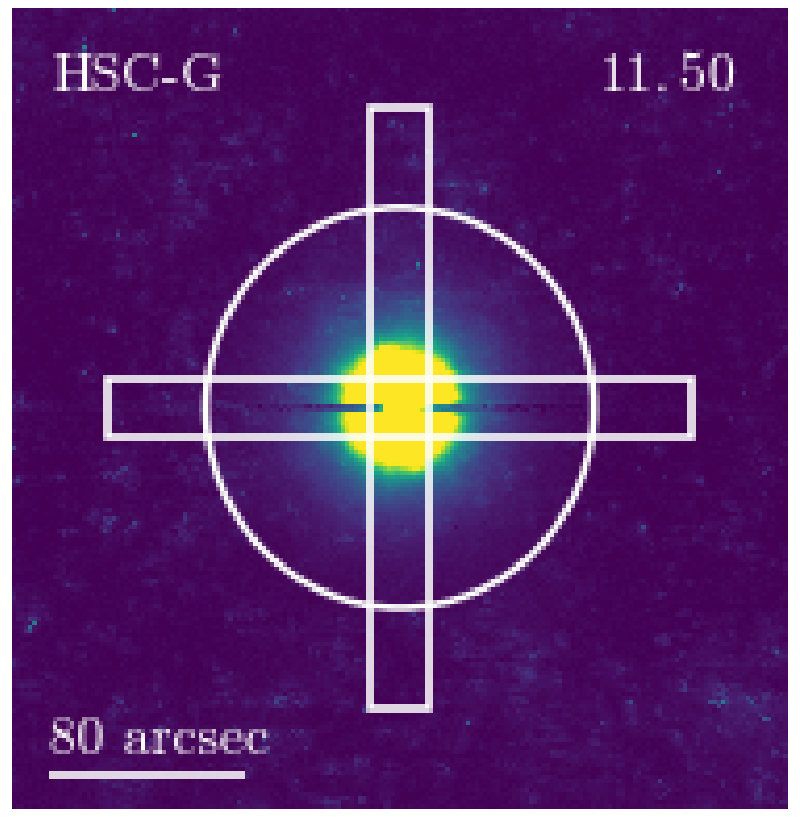}
  \includegraphics[width=0.195\textwidth]{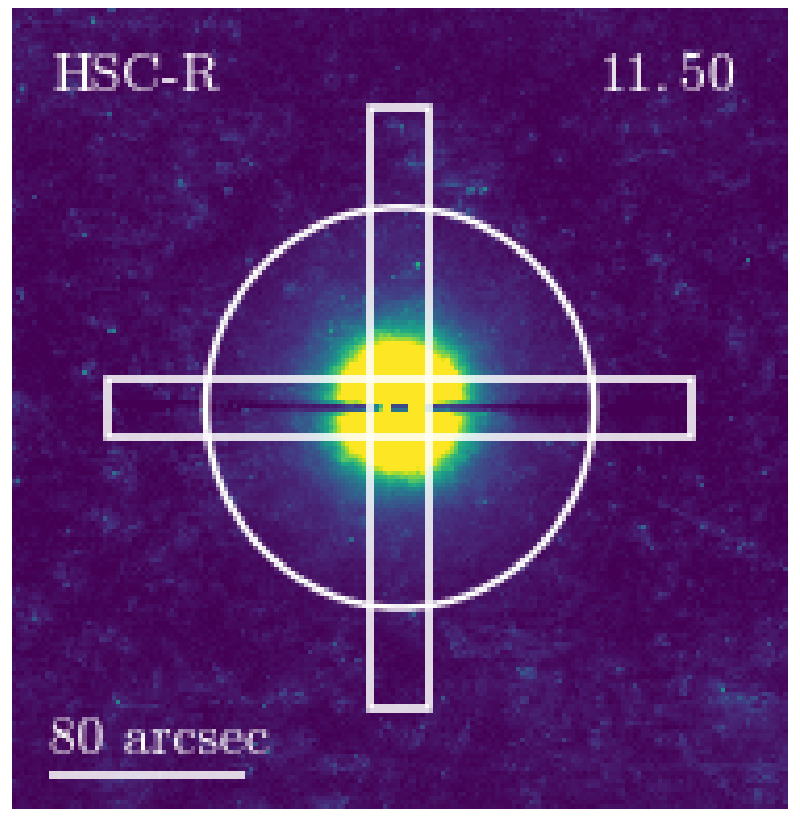}
  \includegraphics[width=0.195\textwidth]{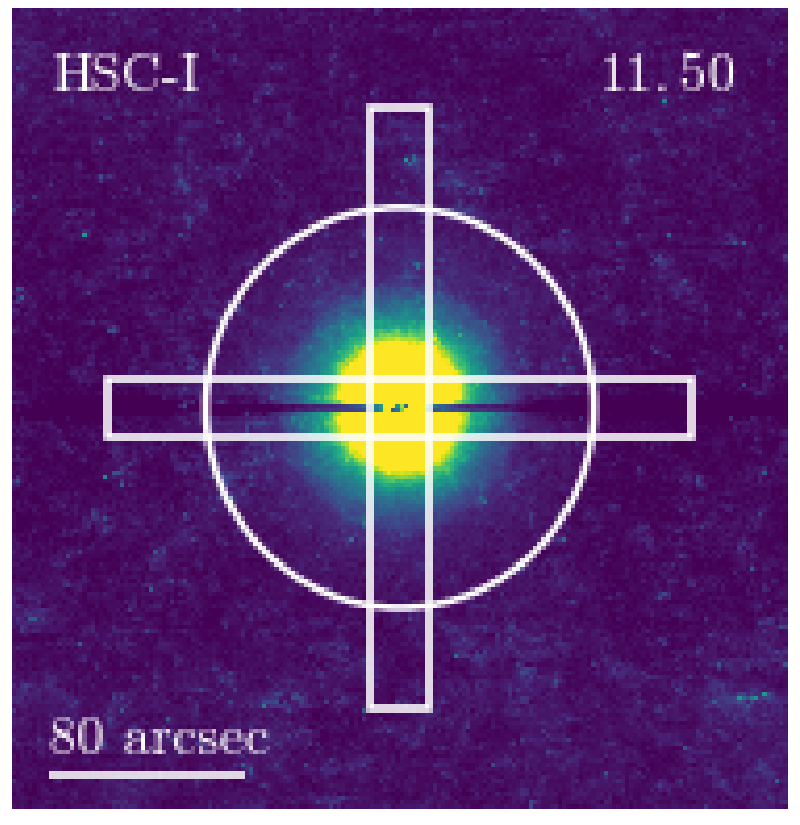}
  \includegraphics[width=0.195\textwidth]{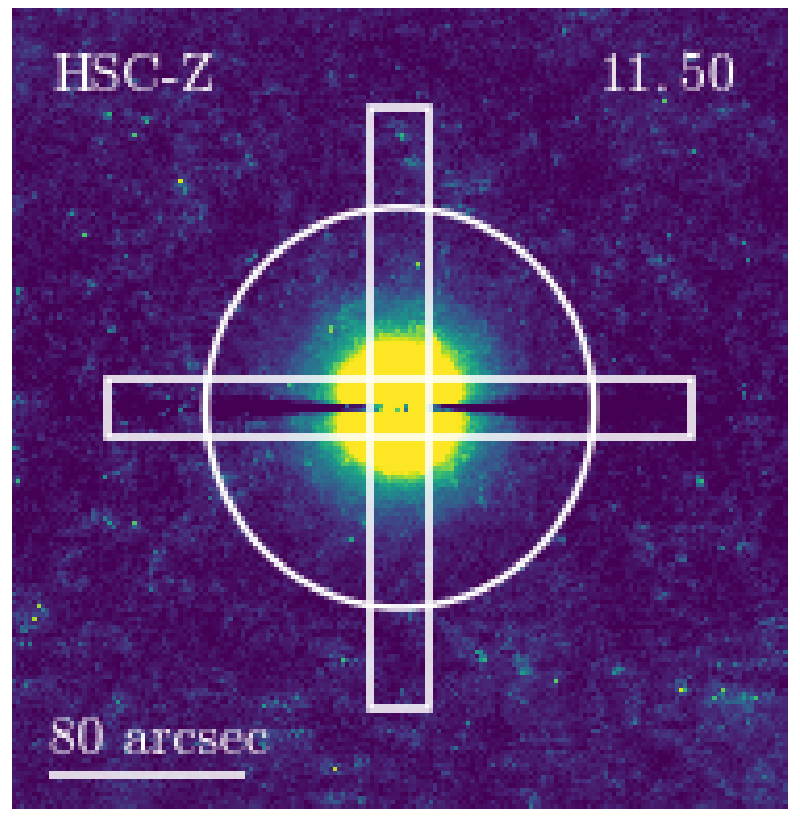}
  \includegraphics[width=0.195\textwidth]{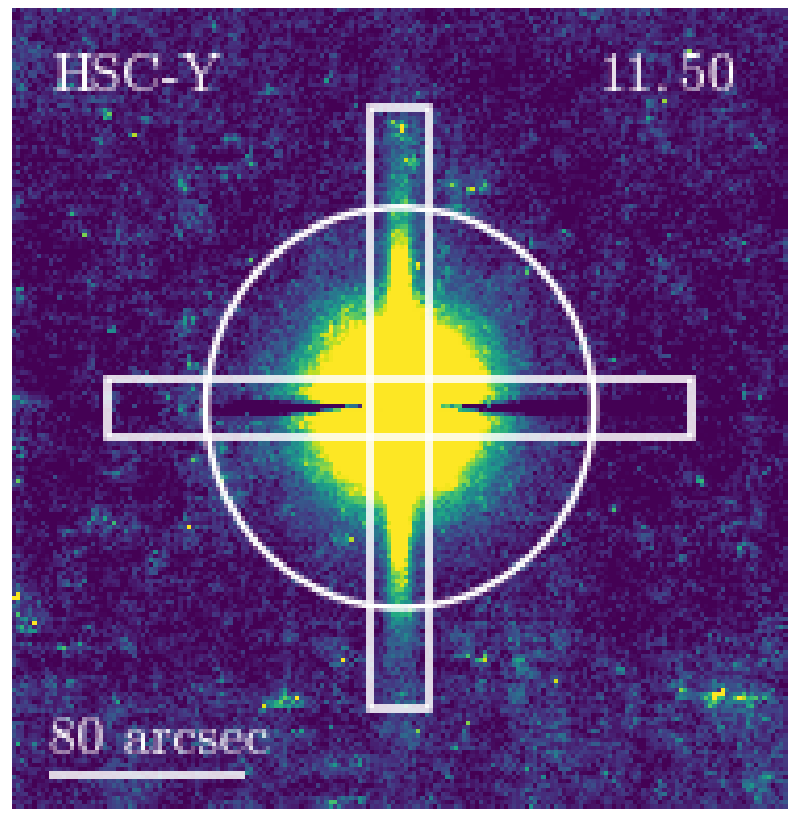}
  \includegraphics[width=0.195\textwidth]{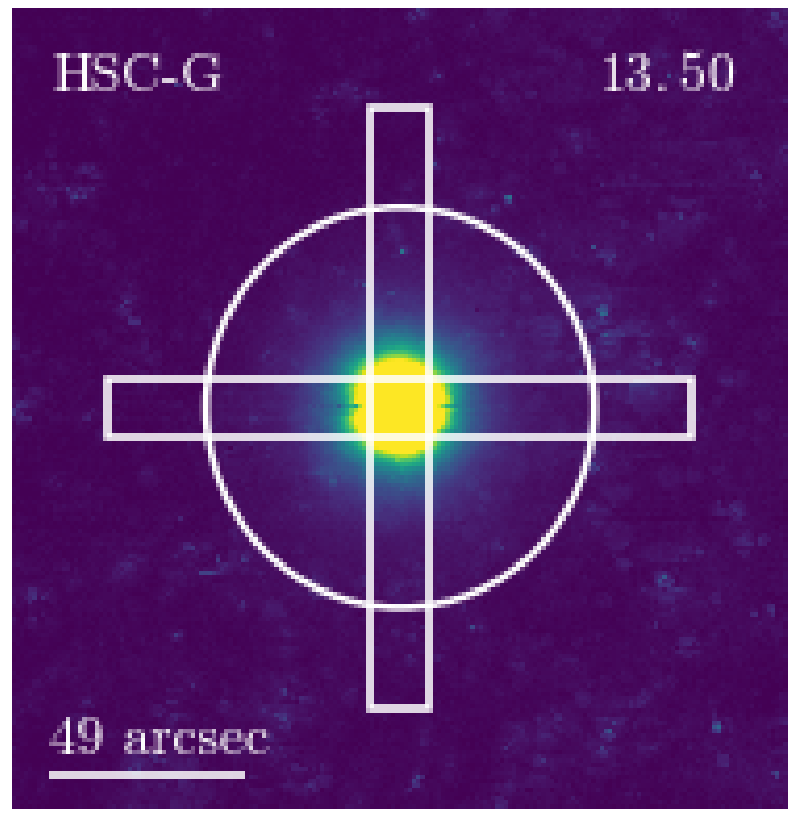}
  \includegraphics[width=0.195\textwidth]{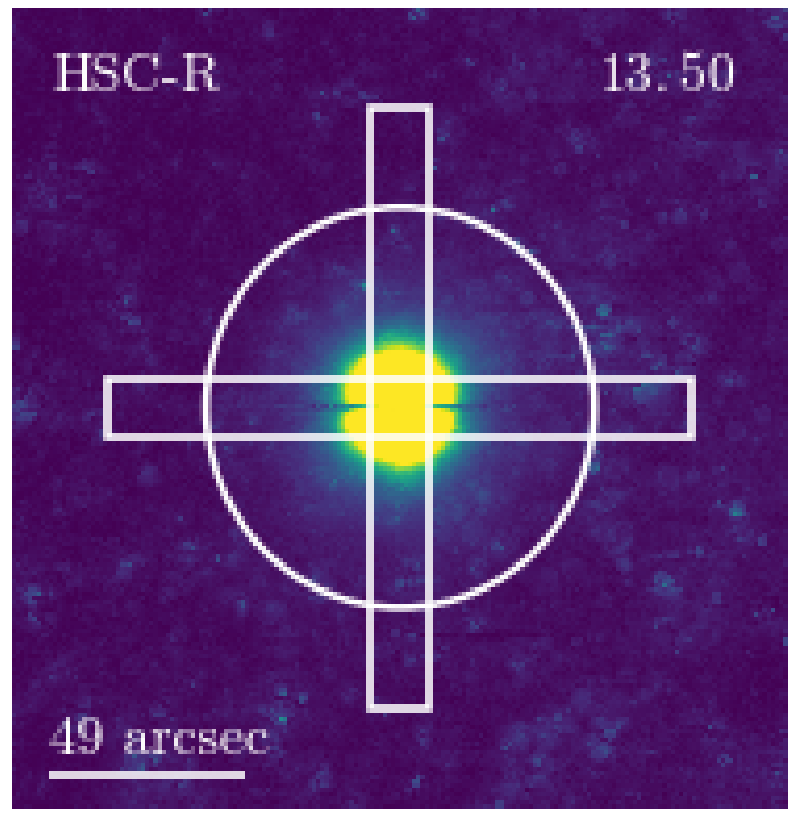}
  \includegraphics[width=0.195\textwidth]{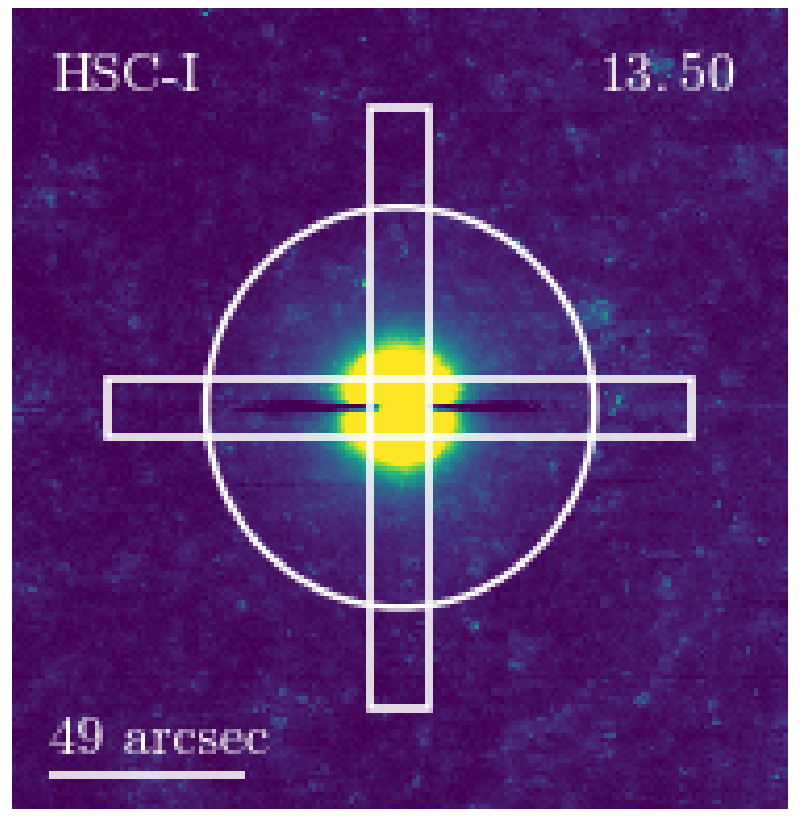}
  \includegraphics[width=0.195\textwidth]{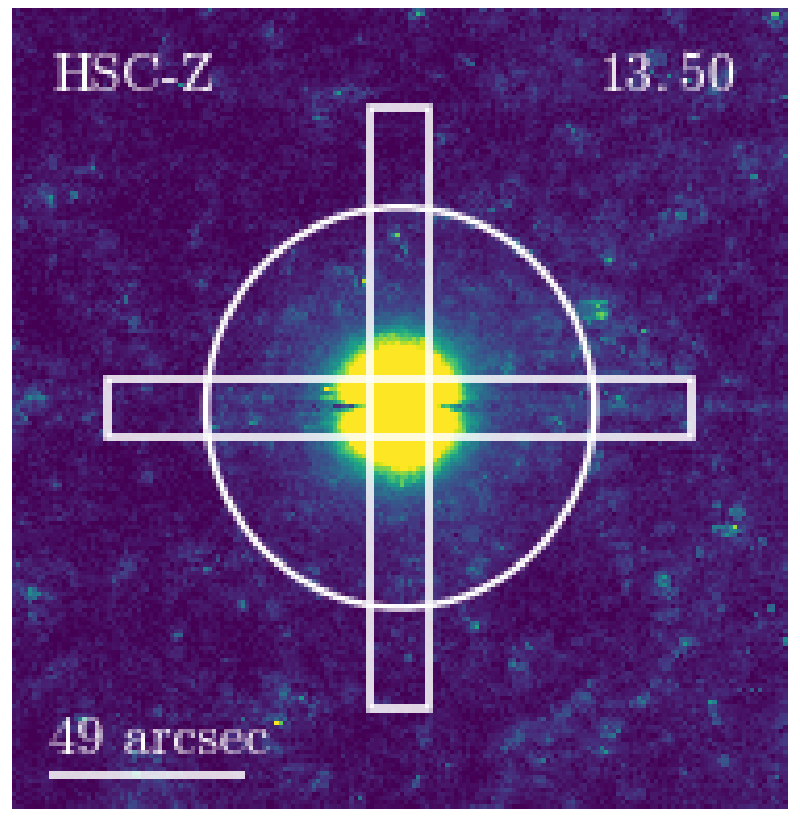}
  \includegraphics[width=0.195\textwidth]{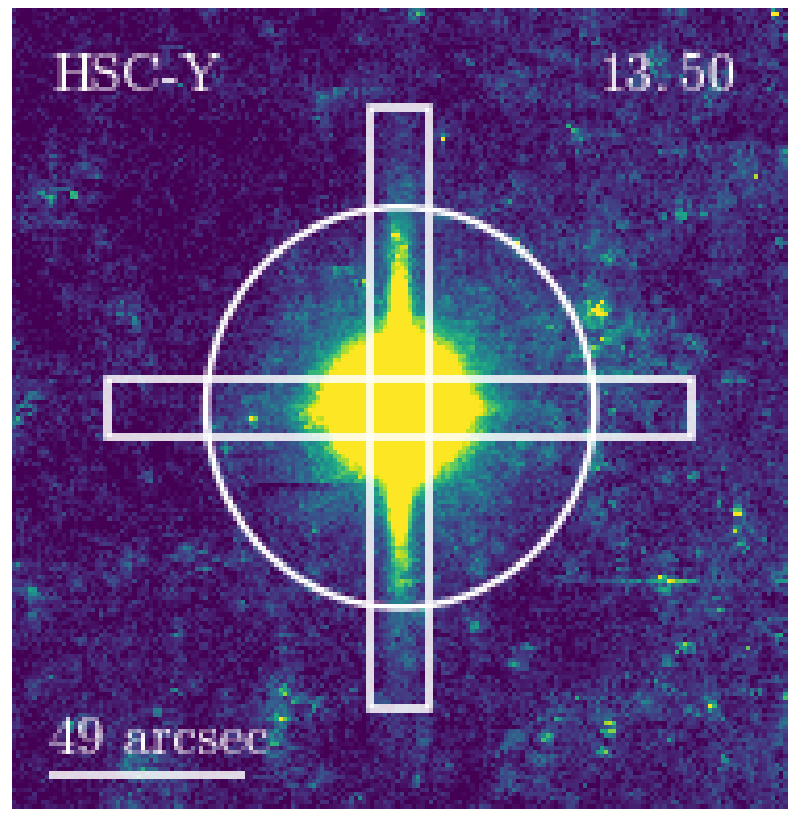}
  \includegraphics[width=0.195\textwidth]{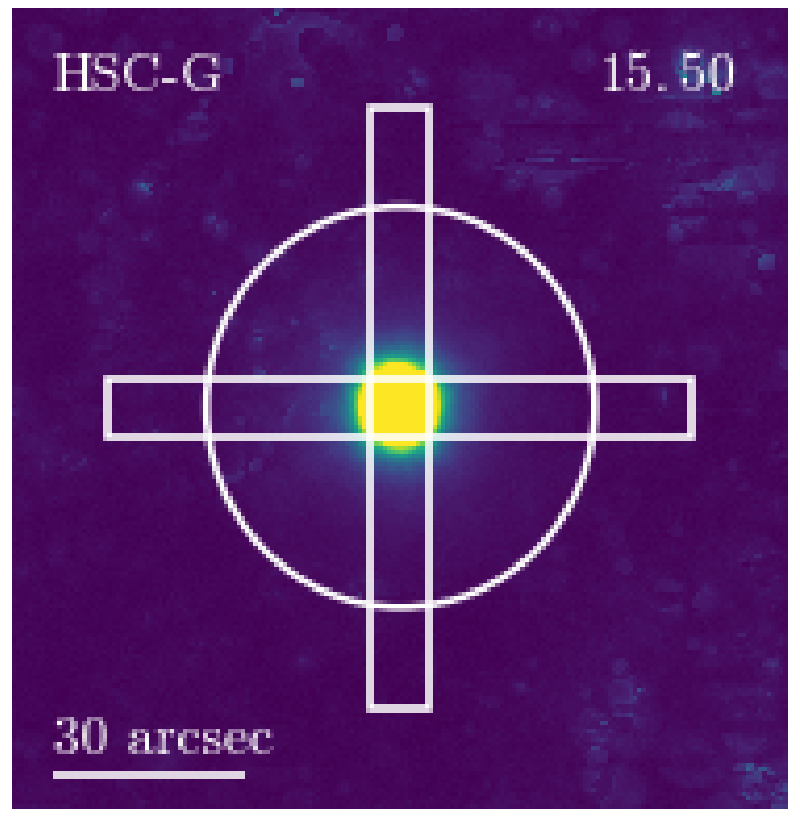}
  \includegraphics[width=0.195\textwidth]{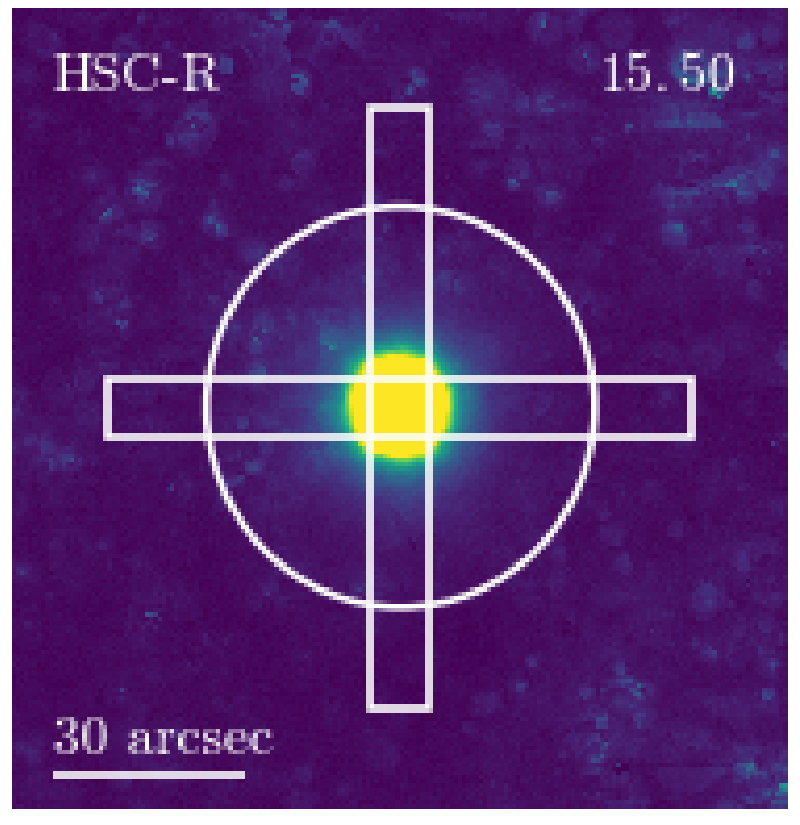}
  \includegraphics[width=0.195\textwidth]{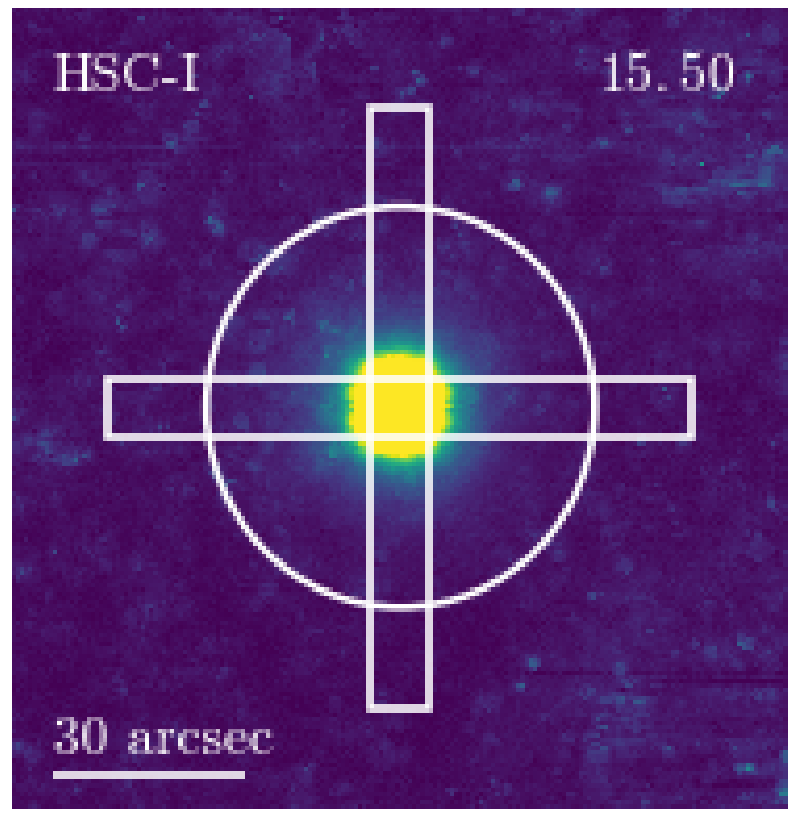}
  \includegraphics[width=0.195\textwidth]{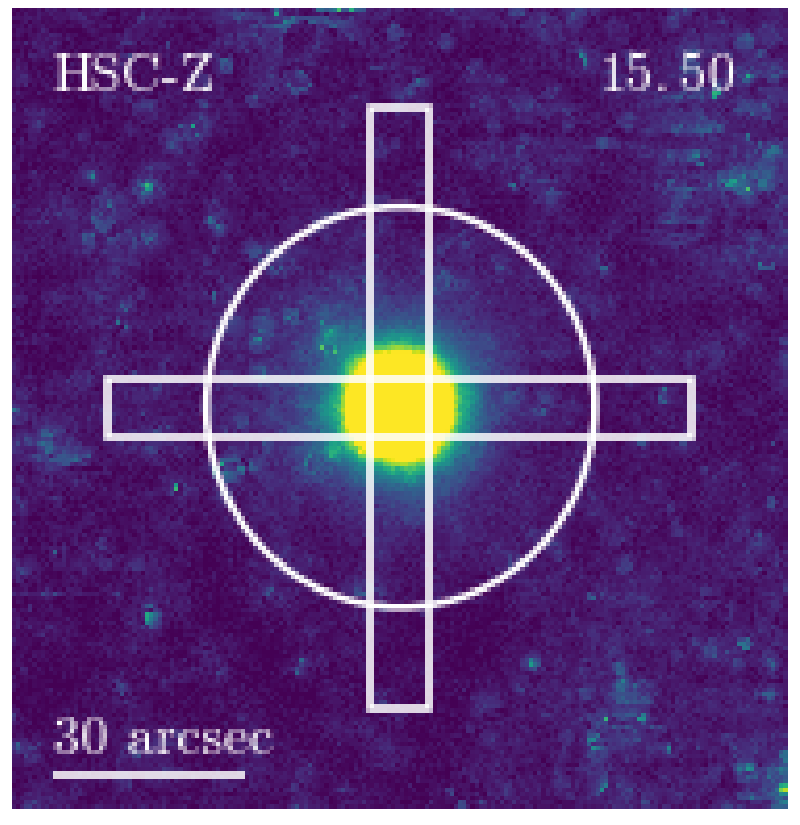}
  \includegraphics[width=0.195\textwidth]{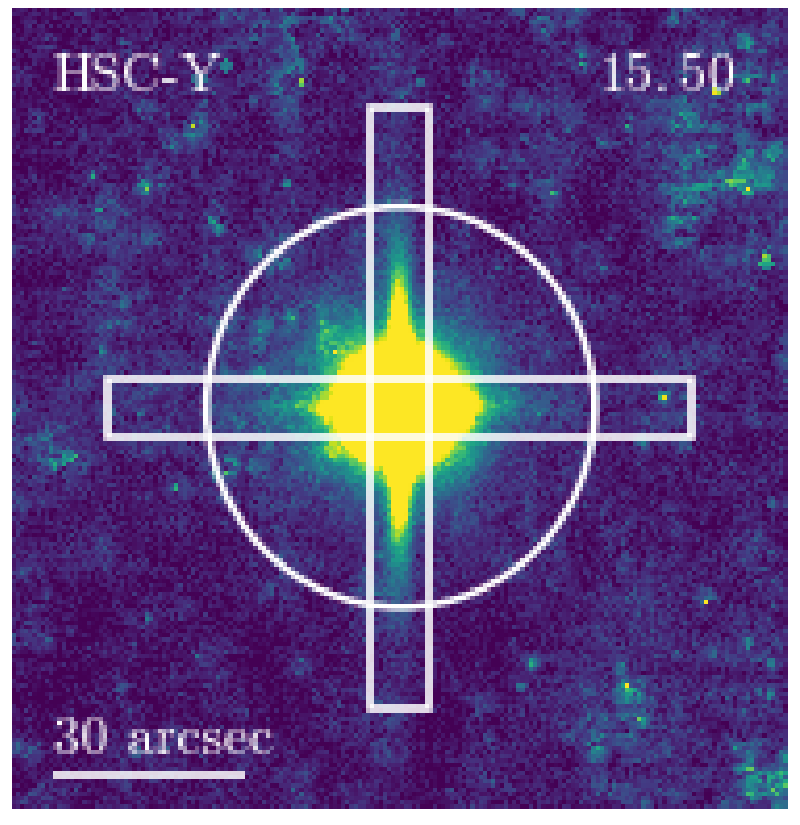}
  \includegraphics[width=0.195\textwidth]{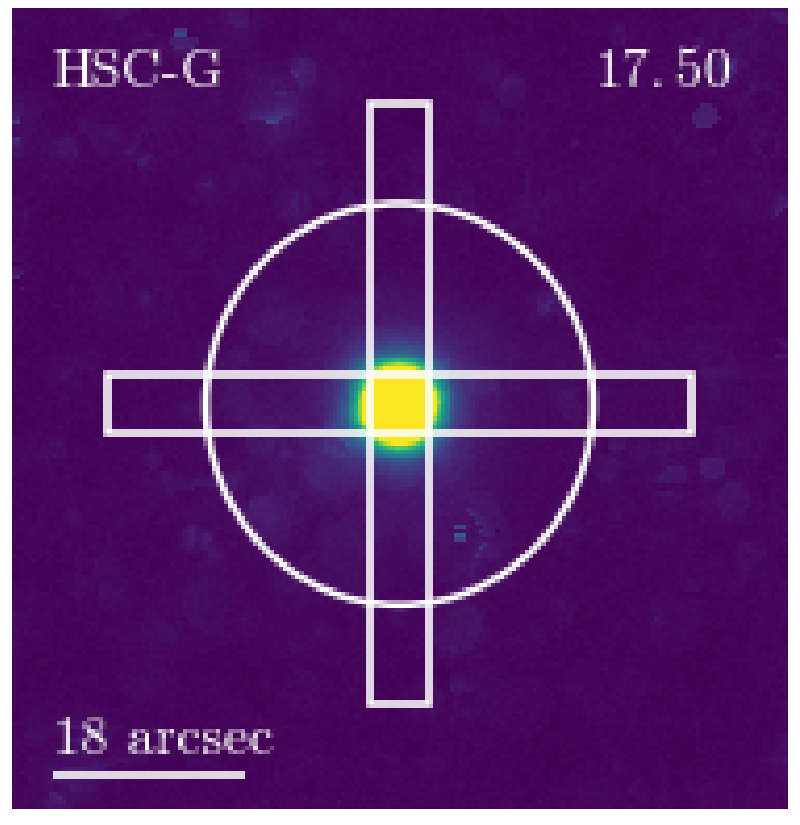}
  \includegraphics[width=0.195\textwidth]{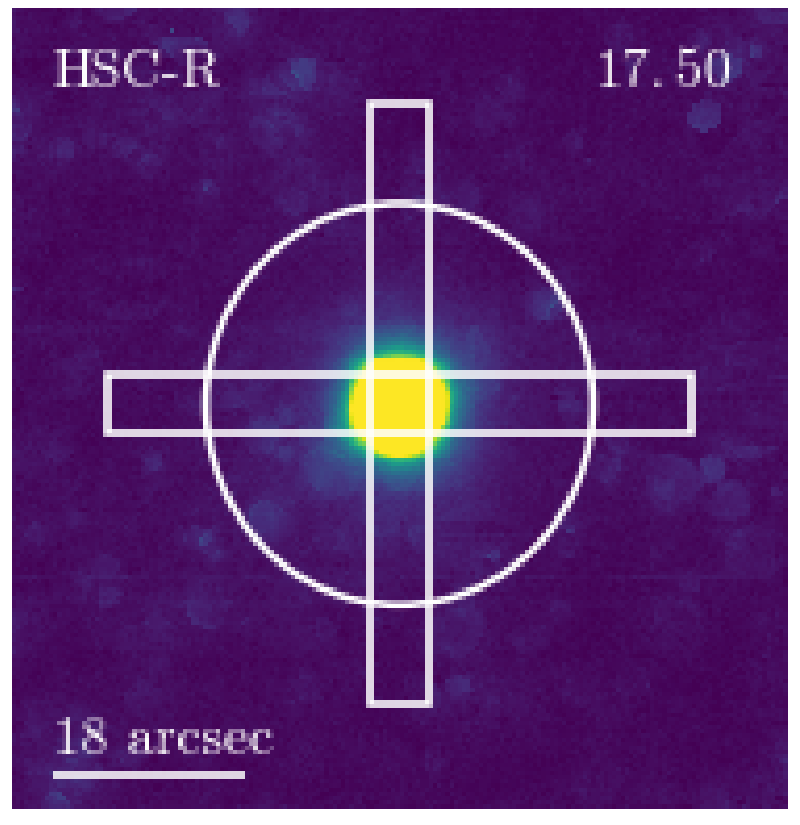}
  \includegraphics[width=0.195\textwidth]{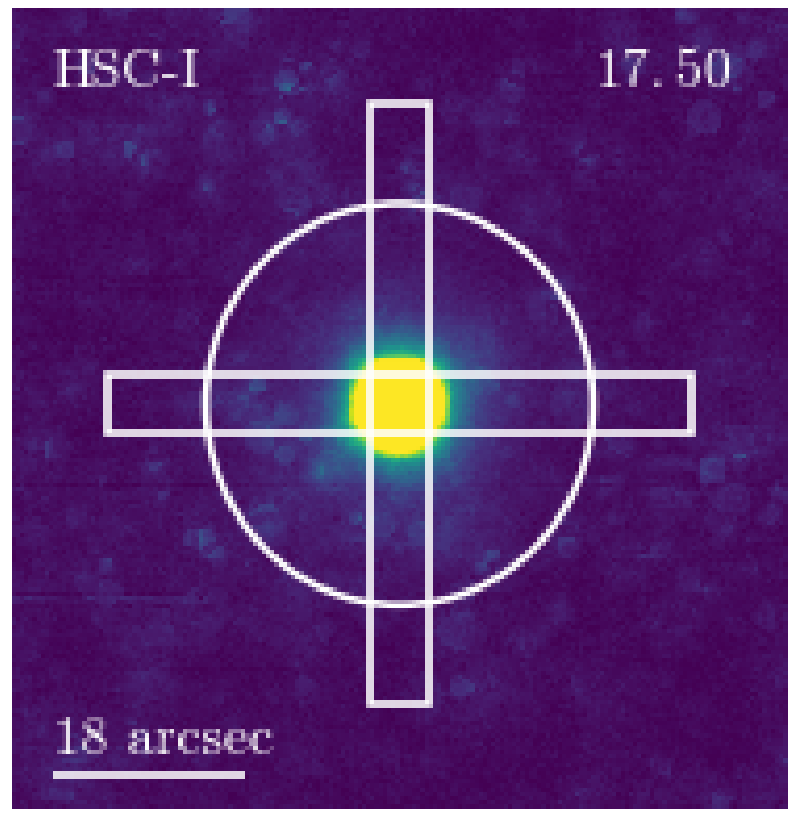}
  \includegraphics[width=0.195\textwidth]{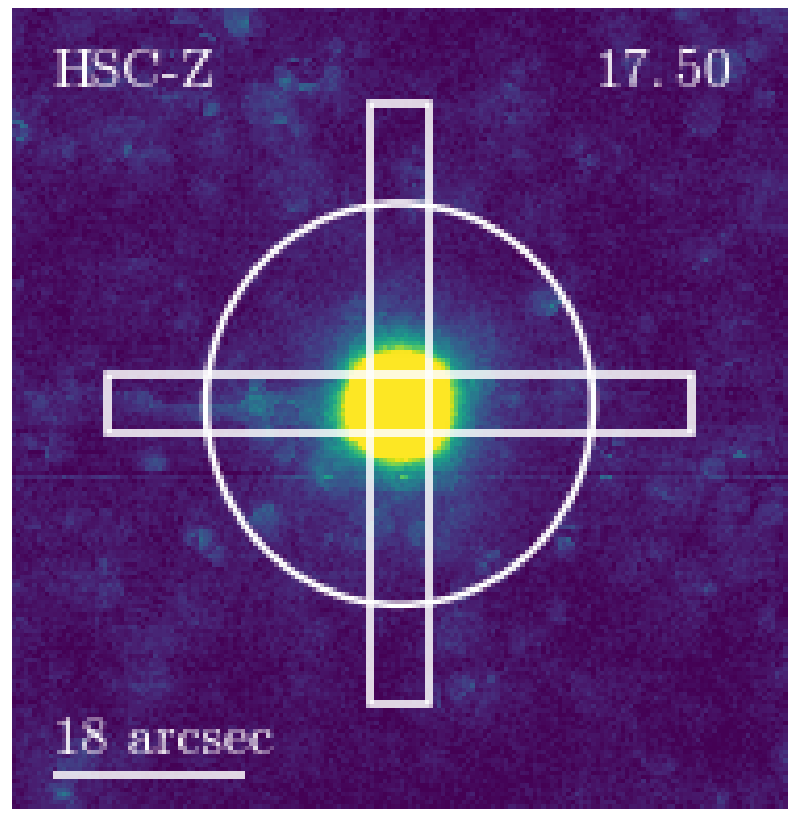}
  \includegraphics[width=0.195\textwidth]{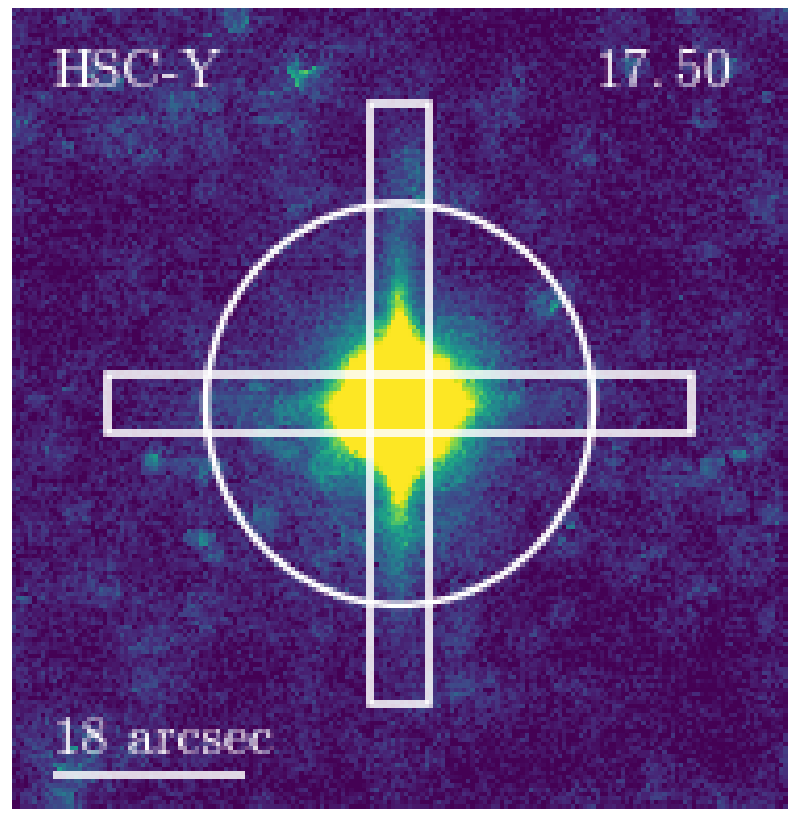}
\end{center}
  \caption{Same as Figure~\ref{fig:stackedImages1}, but for stars in the range $9.5 \le G_{\rm Gaia} \le 17.5$. The rectangles show the masks for the saturated bleed trails and the $Y$-band vertical spikes. Both rectangle lengths are set to 1.5 times the circular mask radius.}
    \label{fig:stackedImages2}
\end{figure*}

We show in Figures~\ref{fig:stackedImages1} and \ref{fig:stackedImages2} the 3$\sigma$-clipped stacked HSC images of the bright stars, stacked per bin of magnitude, and shown for each of the $g$, $r$, $i$, $z$ and $Y$ filters.
We use all bright stars at the brightest magnitudes ($G_{\rm Gaia}=3.5, 4.5$ and 5.5), whereas for fainter stars, to ease the computation, we only select stars in a narrow magnitude range around the magnitudes $G_{\rm Gaia}=6.5, 7.5, 9.5, 11.5, 13.5, 15.5$ and 17.5. Our binning scheme is summarised in Table~\ref{tab:binning}.
\begin{table}
  \tbl{Bright-star magnitude bins used for measuring the impact on nearby sources. The number of stars corresponds to the full HSC-SSP footprint. After matching to the S16A release data the number of stars amounts to roughly 25-30\%, and is given in parenthesis.}{%
\begin{tabular}{ lcccc }
\hline
 $G_{\rm Gaia}$  & mag range & $N_{\rm stars}$ (S16A) & $\langle G_{\rm Gaia} \rangle$   \\
\hline
03.50 & $\pm$0.5 & 14 (3) & 3.71 \\
04.50 & $\pm$0.5 & 61 (14) & 4.50 \\
05.50 & $\pm$0.5 & 151 (30) & 5.58 \\
06.50 & $\pm$0.25 & 217 (49) & 6.51 \\
07.50 & $\pm$0.25 & 589 (150) & 7.51 \\
09.50 & $\pm$0.05 & 570 (162) & 9.50 \\
11.50 & $\pm$0.02 & 1041 (290) & 11.50 \\
13.50 & $\pm$0.005 & 1108 (278) & 13.50 \\
15.50 & $\pm$0.002 & 1235 (343) & 15.50 \\
17.50 & $\pm$0.001 & 1253 (350) & 17.50 \\
\hline
\end{tabular}}
\label{tab:binning}
\end{table}
We see that for the brightest stars ($G_{\rm Gaia}<9$), the main effect is due to the circular bright halos. For stars fainter than $9$, the star halos are too faint to cause significant issues, and the dominant effect is the PSF-shaped pattern.

First, we assume that the impact of bright stars on nearby sources is isotropic and, to measure the extent of such impact, we match the bright-star sample, assembled in Section~\ref{sec:catalogue} and binned in $G_{\rm Gaia}$ magnitudes, with the HSC-SSP sources from the S16A release. We then measure the mean density of sources as a function of distance from the star position, normalised to the density measured in an outer annulus. 

The HSC pipeline proceeds in two steps to detect the sources (see \cite{Bosch:2017}). First it detects all aggregated pixels (footprints) above the background, called the ``parents''. { Second, the footprints featuring several significant peaks are deblended into ``children'', whereas the sources that are not deblended are called ``single''. We measure the mean density of sources around the bright stars for two separate samples: (1) the parent sources, and (2) the single and children sources, referred to as ``primary'' sources.} The drop in density of the parent sample is an indicator of the typical radius below which sources are being hidden by the PSF-shaped luminous pattern of the nearby bright star. The variation in density of the primary sample indicates where the HSC pipeline deblends artefacts (instead of real sources) caused by the bright star, and is an indicator of potential problems affecting the detection of real sources and their photometric measurements.

\begin{figure*}
 \begin{center}
  \includegraphics[width=0.32\textwidth]{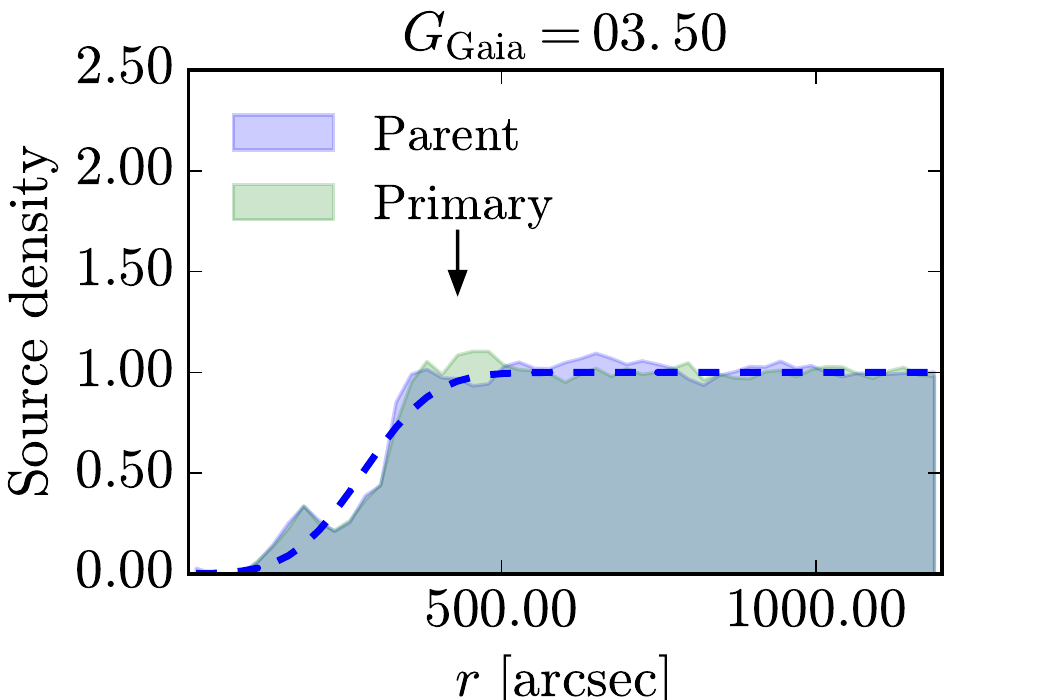}
  \includegraphics[width=0.32\textwidth]{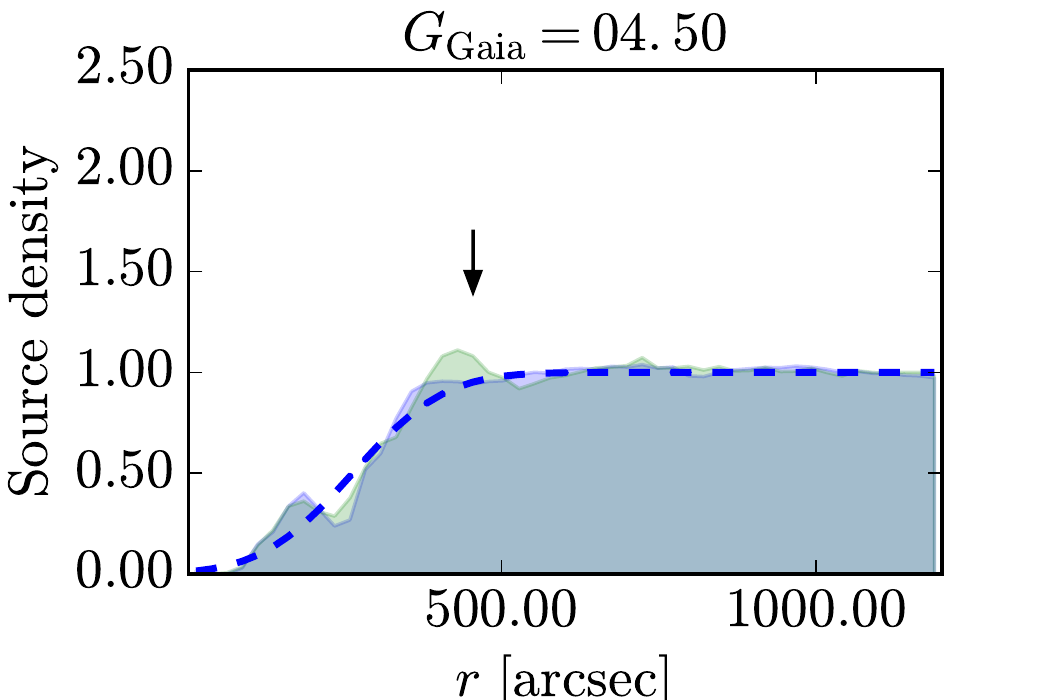}
  \includegraphics[width=0.32\textwidth]{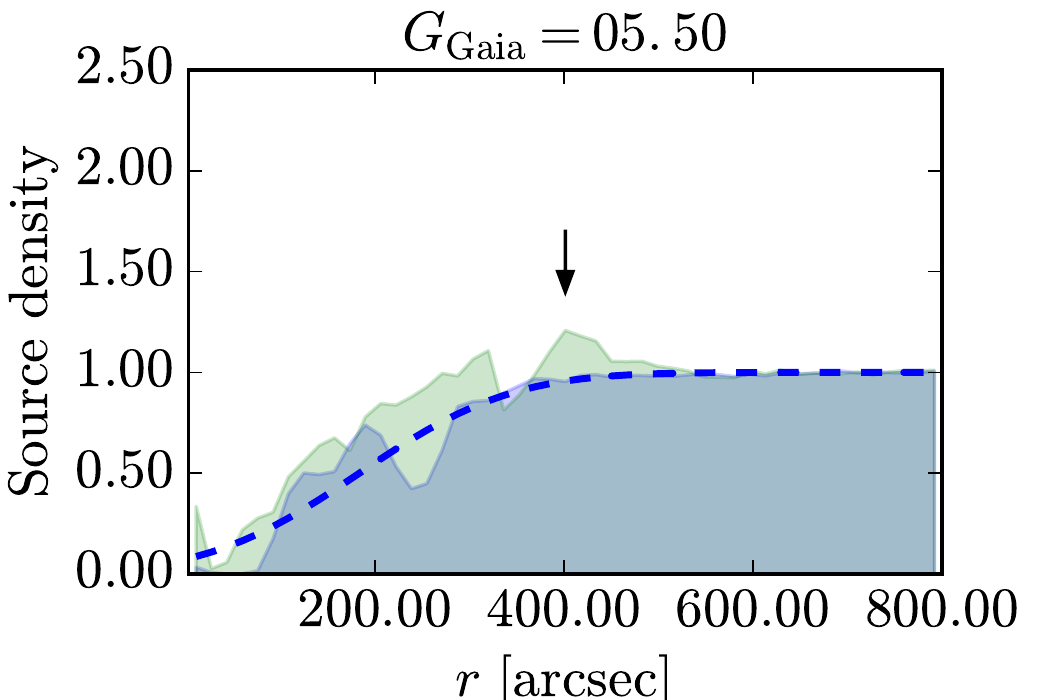}
  \includegraphics[width=0.32\textwidth]{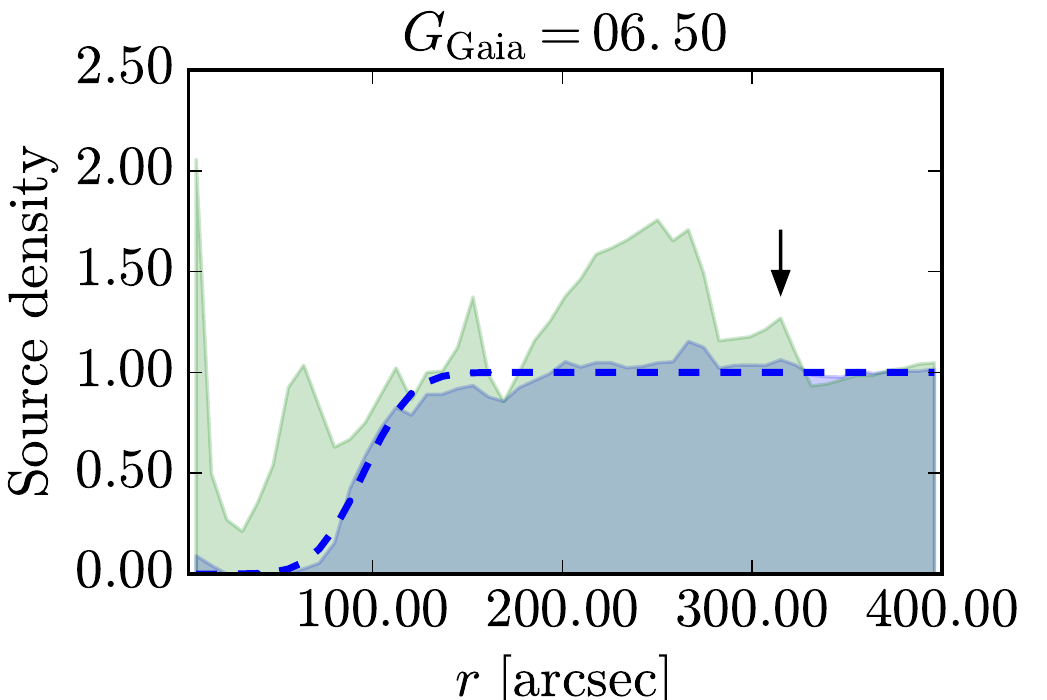}
  \includegraphics[width=0.32\textwidth]{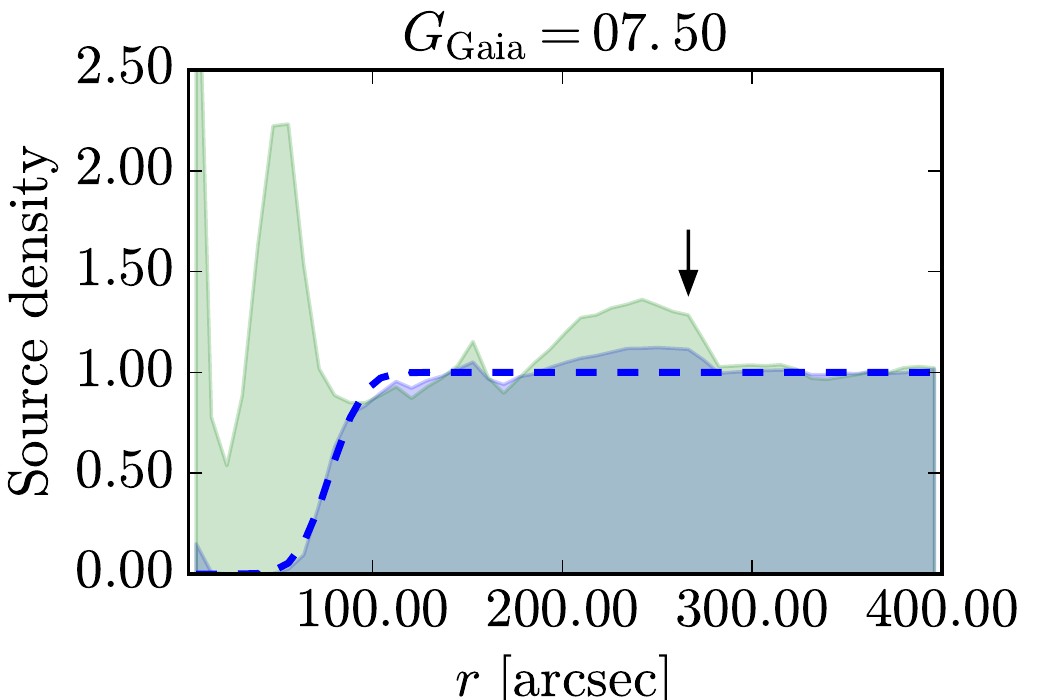}
  \includegraphics[width=0.32\textwidth]{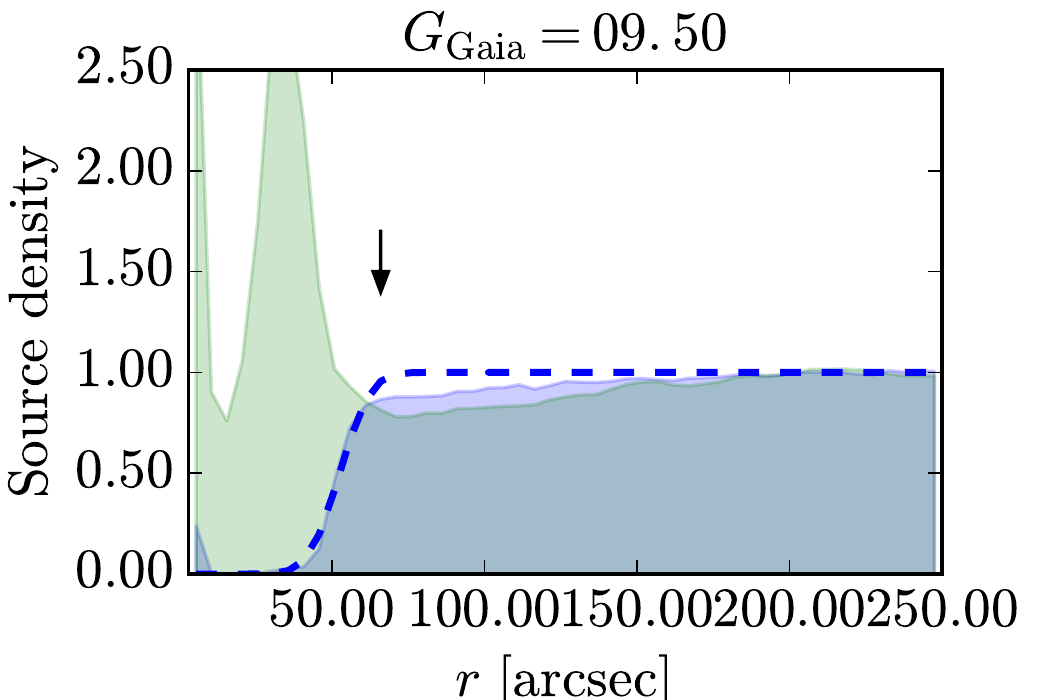}
  \includegraphics[width=0.32\textwidth]{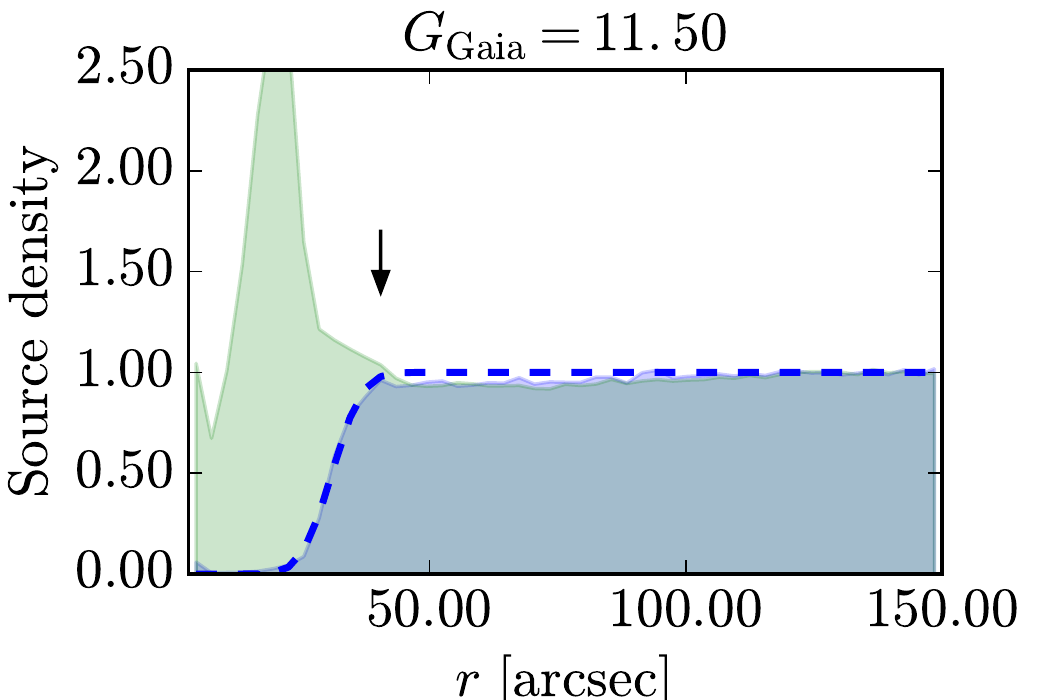}
  \includegraphics[width=0.32\textwidth]{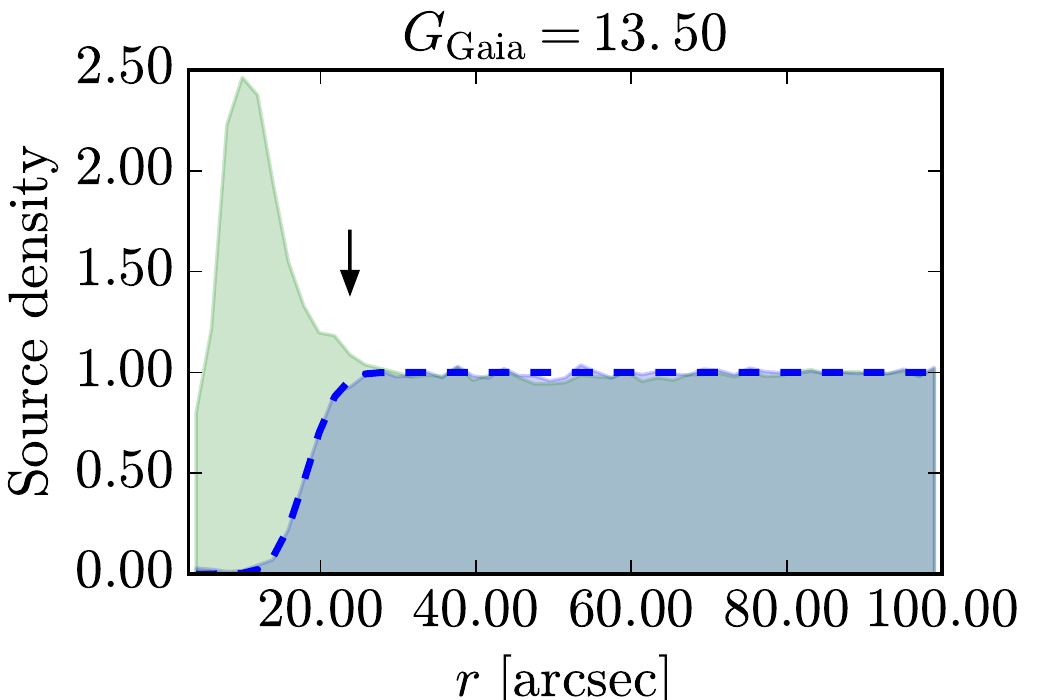}
  \includegraphics[width=0.32\textwidth]{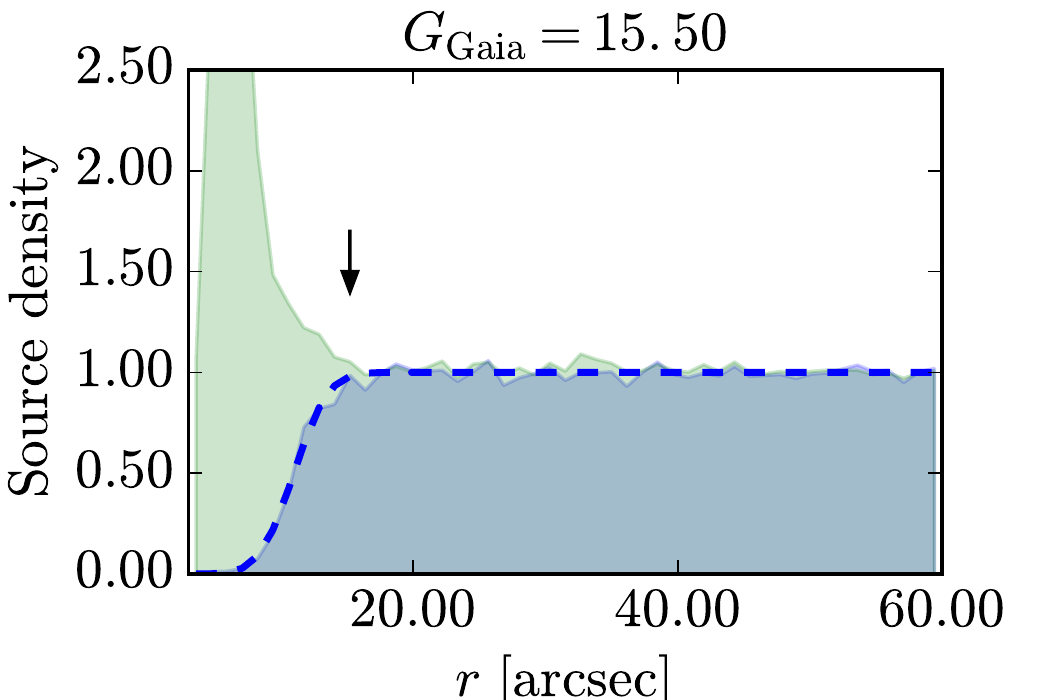}
  \includegraphics[width=0.32\textwidth]{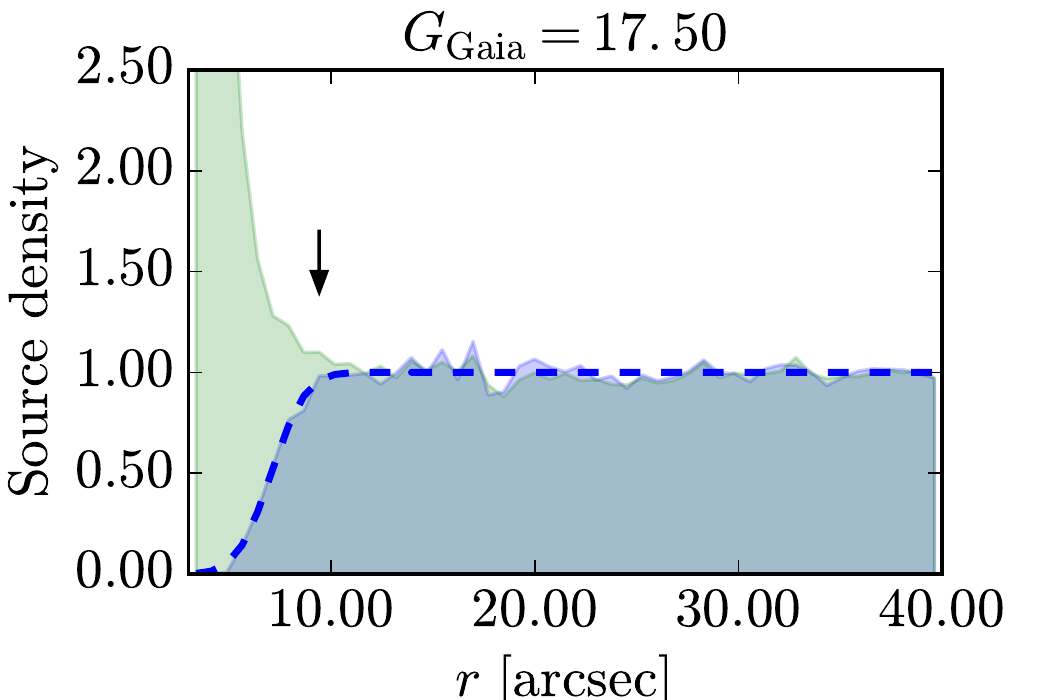}
\end{center}
   \caption{Normalised density of sources detected by the HSC pipeline averaged around samples of bright stars for decreasing star brightness (from left to right and top to bottom). The blue curve shows all the parent sources (unique or before deblending), whereas the green curve shows all the primary sources (unique or after deblending, i.e. the children). The blue dashed line is the best-fit error function to the blue curve. The black arrow shows the final mask radius.}
    \label{fig:sourceDensity}
\end{figure*}

We show in Figure~\ref{fig:sourceDensity} the normalised detected sources density for the parent (in blue) and primary sample (in green), averaged over the bright stars, binned per $G_\mathrm{Gaia}$ magnitude.
For the brightest stars ($G_\mathrm{Gaia} < 9$), the impact on sources occurs at larger radius, due to the concentric luminous halos around the star which artificially increase the density of primary sources compared to the mean density. For fainter stars ($G_\mathrm{Gaia} > 9$), the impact occurs at lower radius, and is characterised by a rapid drop of parent sources towards the star position.

\subsection{Setting up the mask sizes}
\label{ref:maskSize}

To quantify the isotropic extent of the masks, we use both the parent-source and the primary-source indicators. First, we fit an error function to the parent-source density distribution, where the free parameters are the position and width of the slope. The best-fit error function is shown as a blue dashed line on Figure~\ref{fig:sourceDensity}. We then measure the radius where the best-fit completeness reaches 95\%, and the radius where the primary-source density is 20\% above the mean source density. We finally set the mask radius to the most extended indicator, to guarantee that both the primary and parent source samples are unaffected by the bright star. The final mask radius is marked by a black arrow on Figure~\ref{fig:sourceDensity}.

\begin{figure}
\begin{center}
  \includegraphics[width=0.49\textwidth]{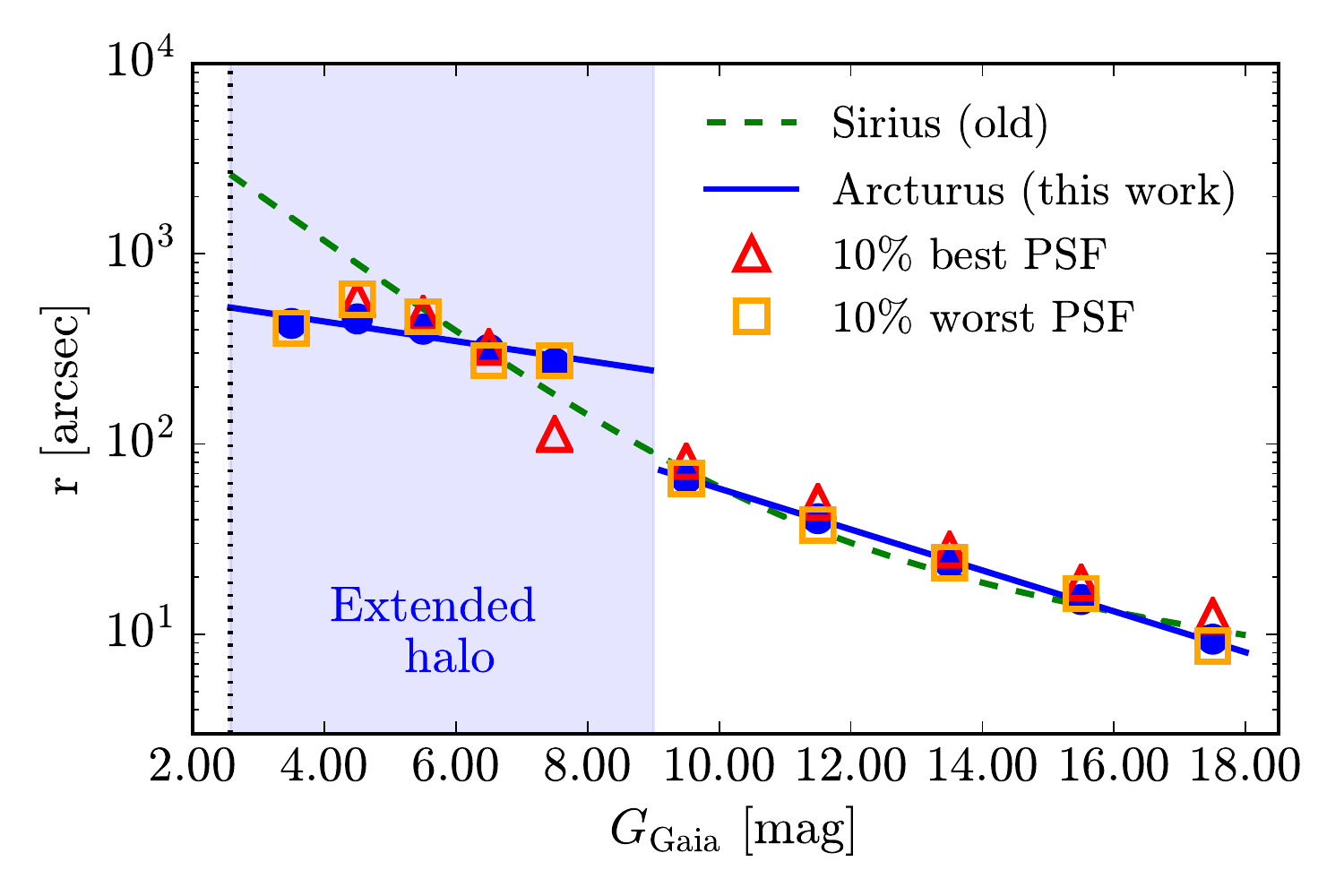}
 \end{center}
   \caption{Mask radius as a function of star magnitude. The measured radii are shown as blue dots. The magnitude dependence is modelled by two exponential functions displayed as the two blue solid curves (below and above $G_\mathrm{Gaia} = 9$). { The red triangles and orange squares} are the measured radii around the bright stars for the 10\% best-PSF data and 10\% worst-PSF data, respectively. The green dashed line shows the previously adopted relation (Sirius) for the S15B, S16A, and DR1 releases. The vertical black dotted line on the left marks the brightest star in the HSC-SSP footprint ($G_\mathrm{Gaia} = 2.57$). }
    \label{fig:maskRadius}
\end{figure}

Figure~\ref{fig:maskRadius} shows the mask radius as a function of star magnitude. The measured radii are shown as blue dots. The former relation (Sirius version) is also shown for comparison as the green dashed line: one can see the diverging radius at magnitudes brighter than $\sim5$ which explains the very large masks seen in S15B, S16A and DR1 releases.

To build a circular mask for each bright star as a function of magnitude, we model the magnitude dependence of the radius by fitting two exponential functions below and above $G_\mathrm{Gaia} = 9$, finding the following relations:
\begin{equation}
\label{eq:maskSize}
\label{eq:rvsmag}
  r\,\mathrm{[arcsec]}=\left\{
                \begin{array}{ll}
                  708.9\,\times \exp(-G_\mathrm{Gaia}/8.41), \,\, \mathrm{Gaia} < 9\\
                  694.7\,\times \exp(-G_\mathrm{Gaia}/4.04), \,\, \mathrm{Gaia} \ge 9
                \end{array}
              \right.
\end{equation}
The best-fit relation is shown as the blue solid line on Figure~\ref{fig:maskRadius}, and the resulting circular masks are shown as white circles in Figures~\ref{fig:stackedImages1} and \ref{fig:stackedImages2}.

\subsection{Accounting for the anisotropic effects}

Next, we account for the saturated bleed trails that appear as horizontal black lines on the images in all filters. We mask these with horizontal rectangles centred on the star position, and whose length and width are $1.5$ and $0.15$ times the radius of the circular mask, respectively. Similarly, to mask out the $Y$-band vertical spikes, we use vertical rectangles whose length and width are also $1.5$ and $0.15$ times the radius of the circular mask, respectively. The choice for the rectangle sizes are based on visual inspection. Although the vertical spikes are observed only on the $Y$-band images, we replicate the vertical rectangles for the other filter masks, for simplicity. 

These two additional mask components are used for stars fainter than $9$ and shown as the horizontal and vertical rectangles in Figure~\ref{fig:stackedImages2}. Around stars brighter than $9$, the extent of the luminous halo goes beyond the typical extent of the bleed trails and vertical spikes, so we do not add any rectangle.

\subsection{The reflection ghosts}

Finally, we note that the reflection ghosts 
for very bright stars (one can see an example of a ghost in the left part of the top $g$-band image in Figure~\ref{fig:stackedImages1}) 
will be directly removed from the images in a future version of the HSC pipeline.


{ In brief, ghosts are caused by internal reflections inside the optics system and their shape strongly depends on the position of the star on the focal plane. Since the surface brightness of the circular ghost is almost uniform within the circle, it can be subtracted from the image once the boundary of the circular ghost is detected. By fitting the boundary with two ellipses, one corresponding to the pupil image and the other to the vignetting, \citet{Yagi:2017} successfully fitted the circular ghosts featuring a high signal-to-noise ratio. Then, they subtracted them from the Suprime-Cam images. The method will be implemented in a future version of the HSC pipeline as an extension of the ghost subtraction procedure which is already dealing with the cometary-shape and arc-shape ghosts (see Figure~18 of \cite{Aihara:2017}). If a perfect subtraction of the circular ghosts is carried out, the atmospheric component of the PSF remains. Hence, the mask size for the brightest stars would follow the same relation as for fainter stars ($G_{\rm gaia}>9$, as seen in Figure~\ref{fig:maskRadius}), resulting in a reduction in size by a factor of $\sim$3.}

\subsection{Handling the masks in the HSC pipeline}

In practice the masks are recorded in DS9 region (``\texttt{.reg}'') format. A file is produced for each individual patch (the smaller subdivision of the HSC-SSP survey). The mask file is read by the HSC pipeline, during data processing. Then  the footprint of the mask is converted into pixels and the corresponding bit value is set to the general mask plane (see \cite{Bosch:2017}, for more details). When extracting the source properties, the pipeline records if the centre or any pixel of a detected source overlaps with the bright-star mask (but the masks are not used \emph{during} data processing).

\section{Masks validation}
\label{sec:valid}

In this section we assess the quality of the bright-star catalogue and the masks using the data from the S16A release. We perform visual inspection on the images, we address the potential issues with seeing and color dependence of the bright-star artefacts, and we run a number of validation tests. We end this section with an estimate of the masked fraction.

\subsection{The bright-star sample purity}
\label{sec:purity}

For many science cases, the purity of the star sample is essential to avoid masking bright galaxies. To test the robustness of our method in selecting a pure star sample (as described in Section~\ref{sec:pureSample}), we match the bright-star catalogue with the HSC-SSP sources from the S16A release, assuming that extended galaxies do not saturate on the images. With image quality superior to SDSS, the HSC-SSP images are more robust to identify extended sources towards faint magnitudes.

\begin{figure*}
\begin{center}
  \includegraphics[height=0.211\textwidth]{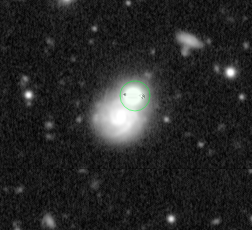}
  \includegraphics[height=0.211\textwidth]{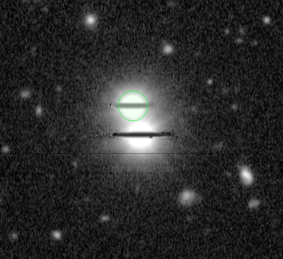}
  \includegraphics[height=0.211\textwidth]{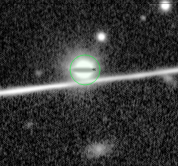} \\
  \includegraphics[height=0.20\textwidth]{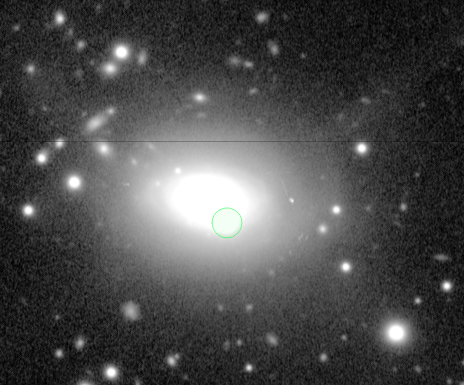}
  \includegraphics[height=0.20\textwidth]{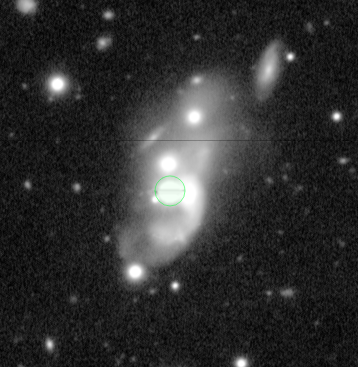}
  \includegraphics[height=0.20\textwidth]{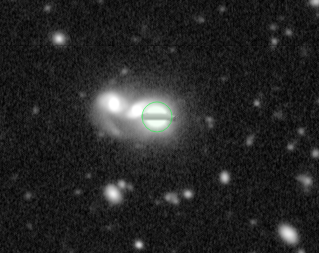}
 \end{center}
   \caption{Top row: examples of stars randomly aligned with an extended source (left), binary stars not resolved in SDSS but in HSC (middle), or stars near artefacts in HSC and flagged as not extended in the SDSS (right). The green circles are the Gaia positions. These cases can be safely masked, as they affect the photometry of the nearby galaxy and occur randomly. Bottom row: extended sources that are masked as bright stars but are probably galaxies with a bright concentrated bulge or hosting a quasar. In the entire S16A data, we find only 13 cases like this.}
    \label{fig:HSCextended}
\end{figure*}

As seen from the bottom panel of Figure~\ref{fig:extended}, the SDSS-based criterium tends to flag as extended a higher number of sources than the HSC-SSP-based one, but it does not necessarily means that \emph{all} truly extended sources are safely removed. Therefore, we select the sources (from the parent sources samples) that are flagged as extended in HSC-SSP but not in SDSS. Out of a total of $325\,028$ sources in the S16A HSC-SSP footprint, extended sources in HSC-SSP amount to $2\,609$ ($0.8\%$). We visually inspect all of the $2\,609$ objects and find that most of them are:
\begin{itemize}
\item binary stars resolved in HSC-SSP (unlike in SDSS),
\item stars randomly aligned with other objects (we remind that our HSC criterium takes only \emph{non}-deblended sources),
\item or stars near artefacts.
\end{itemize}
Those represent a nuisance for nearby galaxies and do not need to be removed from the star sample. We show some examples in the top row of Figure~\ref{fig:HSCextended}.

The galaxies that we do not want to mask are shown in the bottom row of Figure~\ref{fig:HSCextended}. We found a total of 13 cases ($0.004\%$). Among those, a few are most probably galaxies hosting a bright quasar. The need for masking those sources depends on the science case: the photometry (hence the photometric redshift and physical properties) is likely affected, but identifying galaxies hosting a quasar is obviously a very important goal. Although the number of such cases is small (most are safely identified as extended in the SDSS), it may be a concern for studies aiming at finding quasars in detected galaxies. One must also keep in mind that bright quasars with no detected host are most likely in the bright-star sample and therefore masked, so that quasar studies must use the masks with caution.

\subsection{Filter dependence of star brightness}
\label{sec:flterDepence}

For each individual star, we set identical size and shape for all five HSC filters, based on its Gaia $G$-band brightness, which covers the wide optical range from $4\,000$ to $10\,000$~\AA{}. Although this range covers the HSC bands from $g$ to $z$ (and slightly into the $Y$ band), one may ask the question whether a star significantly brighter in one band could impact the nearby sources beyond the mask limits, that are computed from the source density averaged over the five bands.

\begin{figure*}
\begin{center}
  \includegraphics[width=1.0\textwidth]{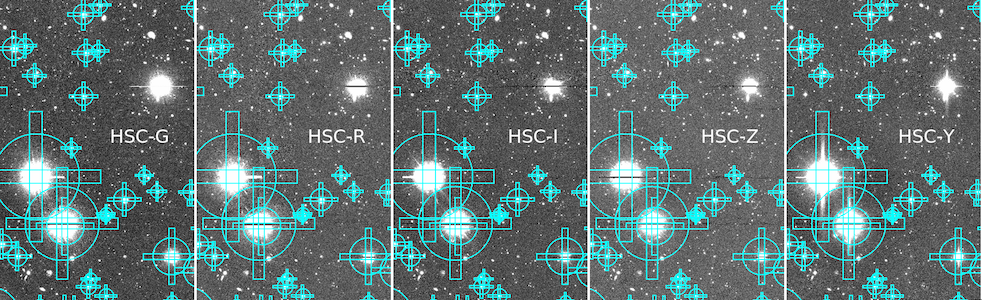}
\end{center}
\caption{Example of the bright-star masks (in blue) in the five HSC filters $grizY$, in the most star-crowded field in the S16A release, near the Galactic plane. The mask sizes properly account for the stars brightness in all bands. { Interestingly, we can see one bright star which is not masked, but this is most likely one case among the sources flagged as extended in the SDSS and hence removed from the bright-star sample.}}
    \label{fig:example}
\end{figure*}

Fortunately, the HSC pipeline first detects sources in all bands independently, hence if some artefacts artificially increase the number of detected/deblended sources in one band and not the others, these fake sources will appear in the catalogue, and hence will be caught by our estimate based on the increase in source density. So we believe that our approach properly accounts for such variations in brightness from one filter to another. We illustrate this in Figure~\ref{fig:example}, which shows the bright-star masks on top of the five-band HSC images. This field is located in the currently most star-crowded region in S16A (tract 9799 and patch 6,5), located in the ``GAMA09H'' area, close to the Galactic plane. We see that the masks safely enclose the bright stars in all filters.

However, some stars may be significantly brighter in one filter but may not pass the Gaia magnitude cut ($G_\mathrm{Gaia}<18$), especially NIR-bright stars. This means that some saturated stars in the $Y$ band may not be properly masked. This caveat must be kept in mind in the studies extensively relying on the $Y$-band photometry (e.g. when selecting high redshift dropout galaxies).

\subsection{PSF-size dependence}

Similarly, bad seeing (which increases the size of the PSF luminous pattern) or sharp seeing (which will saturate fainter stars and enlarge the seeing pattern due to the brighter-fatter effect) may require enlarging the required size of the masks. To check the impact of seeing on the masks, we recompute our estimates (see Section~\ref{ref:maskSize}) for the $i$-band $10\%$ best- and $10\%$ worst-PSF data.

Table~\ref{tab:bestworstseeing} shows the mean PSF size values for the full star sample, the $10\%$ best- and $10\%$ worst-PSF samples and the results are shown in Figure~\ref{fig:maskRadius}.
\begin{table}
  \tbl{Mean PSF sizes in the $i$-band for the full, $10\%$ best-, and $10\%$ worst-PSF star samples. For the the brightest stars, the PSF estimate often fails due to the nearby bright star.}{%
\begin{tabular}{ lccc }
\hline
 $G_{\rm Gaia}$  & All &  10\% best PSF  & 10\% worst PSF  \\
\hline
3.50 & 0.60 & -- & 0.64 \\ 
4.50 & 0.67 & 0.52 & 0.85 \\ 
5.50 & 0.71 & 0.57 & 0.89 \\ 
6.50 & 0.68 & 0.53 & 0.91 \\ 
7.50 & 0.69 & 0.53 & 0.87 \\ 
9.50 & 0.70 & 0.52 & 0.91 \\ 
11.50 & 0.70 & 0.54 & 0.90 \\ 
13.50 & 0.70 & 0.55 & 0.87 \\ 
15.50 & 0.69 & 0.53 & 0.88 \\ 
17.50 & 0.69 & 0.53 & 0.89 \\ 
\hline
\end{tabular}}
\label{tab:bestworstseeing}
\end{table}
The red triangles represent the $10\%$ best-PSF results, whereas the orange squares represent the $10\%$ worst-PSF results. For both cases, we observe no significant change over the full magnitude range. 


Overall, we find that the PSF-size dependence has little impact on the measured radius. This is partly due to the fact that all five bands are used in the source detection (as discussed in the previous section) and the effect of the PSF is averaged over the five bands. We note however that these tests are made with the $i$ band which has a constraint on good seeing for shape measurement. 


\subsection{Number counts}

We now perform a number of comparisons with the CFHTLenS dataset \citep{Heymans:2012}. The CFHTLenS project is a cosmic-shear driven data reduction of the Canada-France-Hawaii Telescope Legacy Survey (CFHTLS), a medium-deep large-scale cosmological survey covering about $150$~deg$^2$ observed in five optical ($ugriz$) bands. Results from the CFHTLenS are now a reference in a number of extragalactic and cosmological topics (see e.g. \cite{Kilbinger:2013}). One remarkable feature of the CFHTLenS is that the masking was carefully done and refined by hand by the team to meet the stringent requirements on shape measurement for cosmic shear. It is therefore the ideal dataset to assess the quality of our masks. To do a fair comparison between the two surveys, we select all sources brighter than $i<22.5$ in both the CFHTLenS and HSC-SSP surveys, in a common area covering about 20~deg$^2$, located in the CFHTLenS ``W4'' field. Both $i$-band magnitudes are total magnitudes (\texttt{MAG\_AUTO} for  CFHTLenS and \texttt{cmodel} for HSC-SSP), and the $i$-band filter response is similar for both.

\begin{figure}
\begin{center}
 	\includegraphics[width=0.49\textwidth]{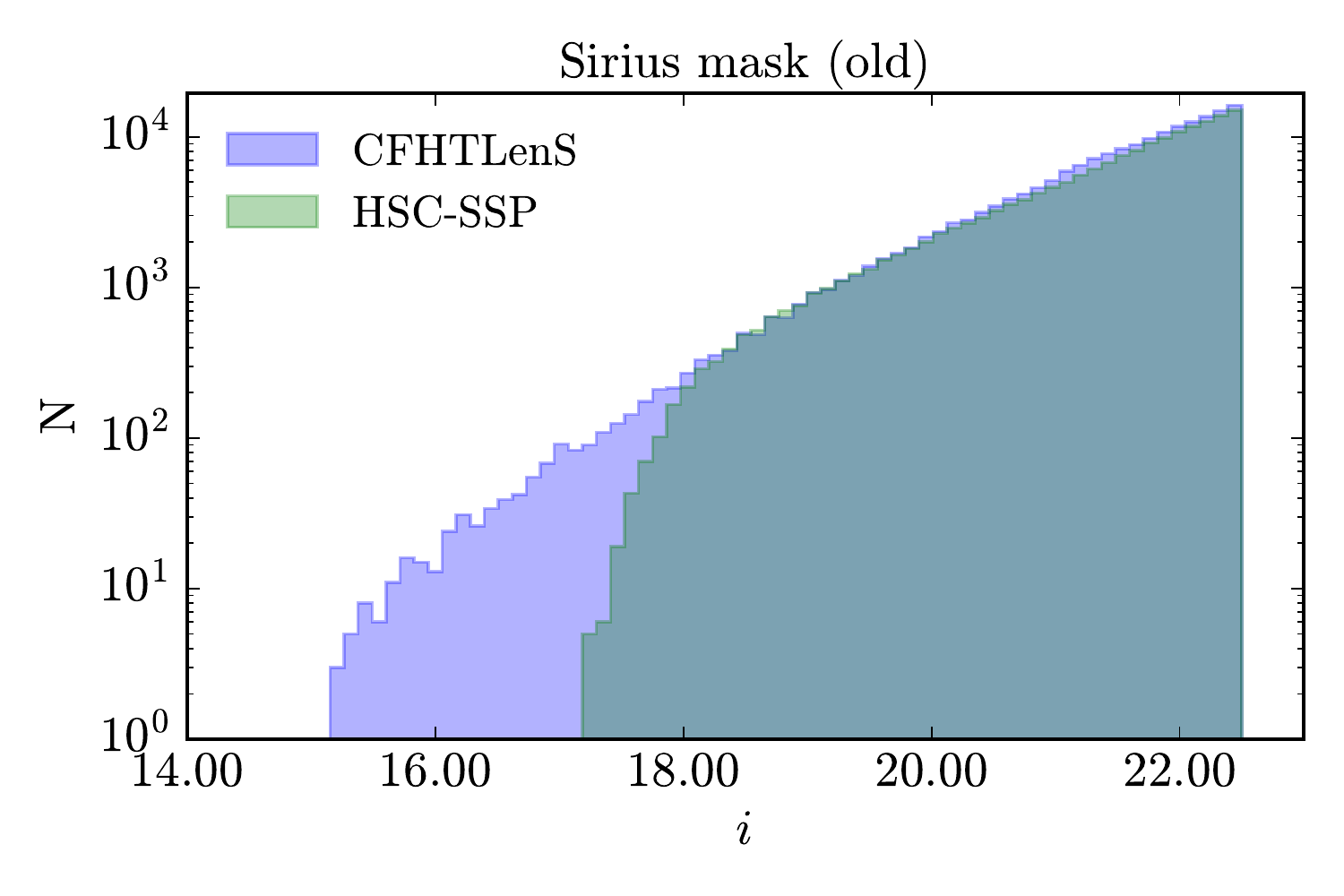}\\
 	\includegraphics[width=0.49\textwidth]{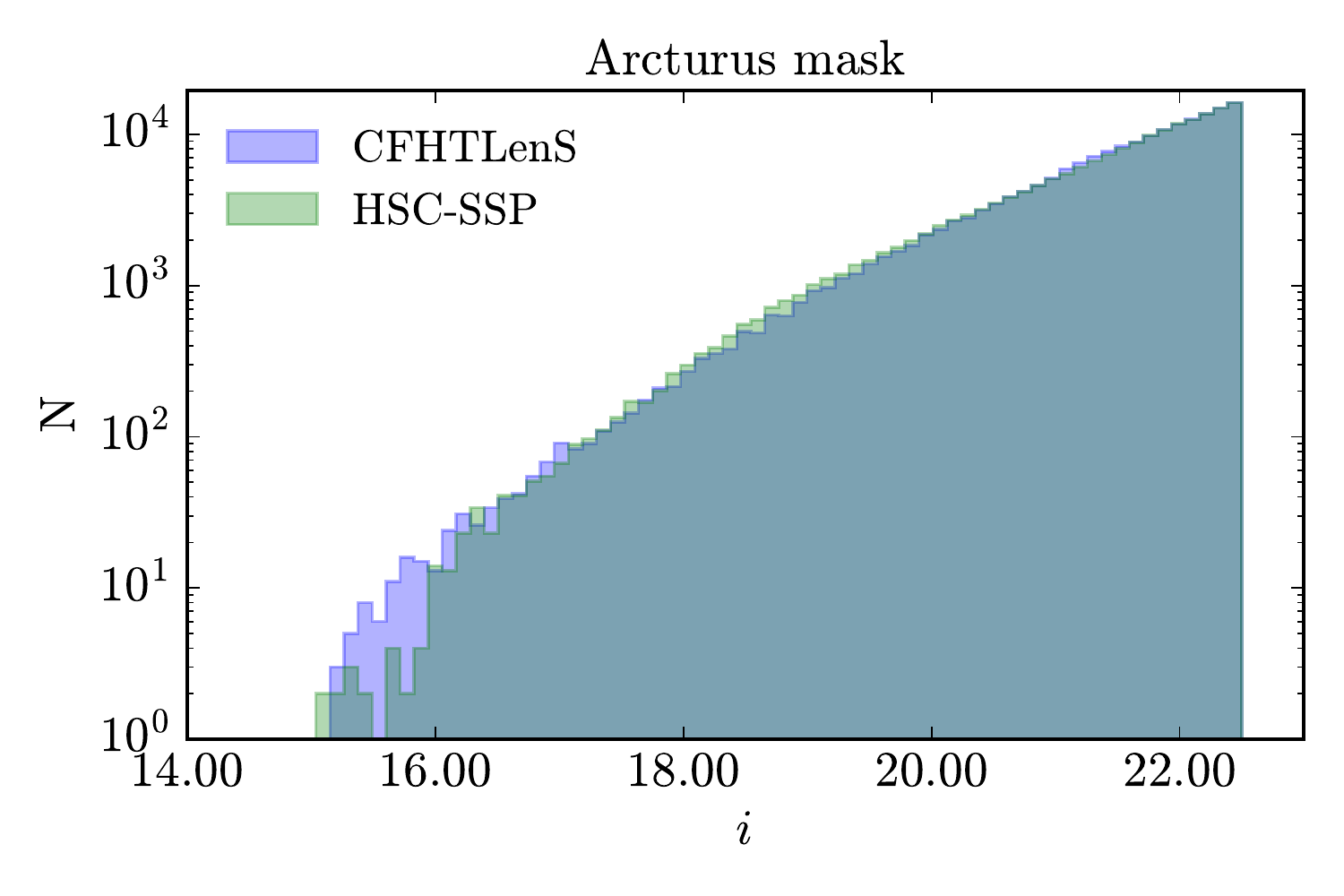}
 \end{center}
   \caption{Galaxy number counts in HSC-SSP (green) and CFHTLenS (blue) as a function of $i$-band total magnitude. The top panel shows the number counts after filtering out galaxies with the old version of the masks (Sirius), whereas the bottom panel shows the same sample after filtering with the new version of the masks (Arcturus).}
    \label{fig:numberCounts}
\end{figure}

We begin the comparison with the galaxy number counts in both catalogues. The star/galaxy separation is done using color and morphology for CFHTLS \citep{Coupon:2015}, whereas it is done using the \texttt{extendedness} estimate in HSC-SSP \citep{Bosch:2017}, based on PSF vs \texttt{cmodel} flux comparison. Results are shown in Figure~\ref{fig:numberCounts}.
The top panel shows the old mask version (Sirius), whereas the bottom panel shows the new mask version (Arcturus) developed in this work. In the former case, many bright galaxies are missing due to a number of extended sources that were wrongly included in the bright-star sample (see Appendix~\ref{sec:sirius}). In the latter case, the HSC-SSP number counts are in remarkable agreement with the CFHTLenS in the magnitude range $16<i<22.5$. We note however a few missing sources brighter than $i=16$, which are due to \texttt{cmodel} measurement failures for large radius galaxies \citep{Huang:2017}.

\subsection{The two-point correlation function}

Next, we compare the galaxy angular two-point correlation function (TPCF). The TPCF is an observable sensitive to the photometric artefacts, as it captures the local changes in density. The spurious detections in the luminous halos around very bright stars or the drop in source density observed near less-bright stars may cause significant biases in the TPCF.

\begin{figure*}
\begin{center}
 	\includegraphics[width=0.49\textwidth]{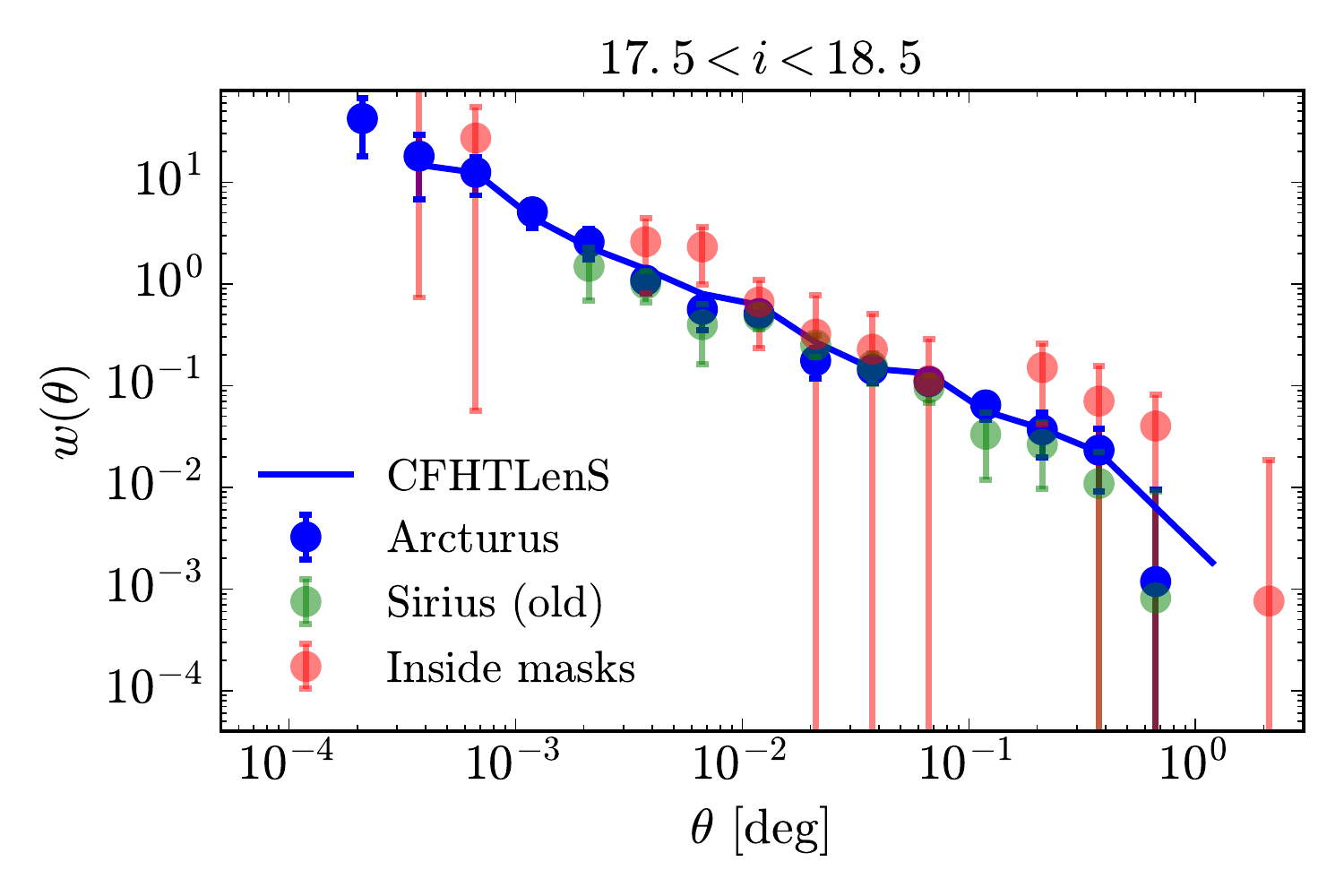}
 	\includegraphics[width=0.49\textwidth]{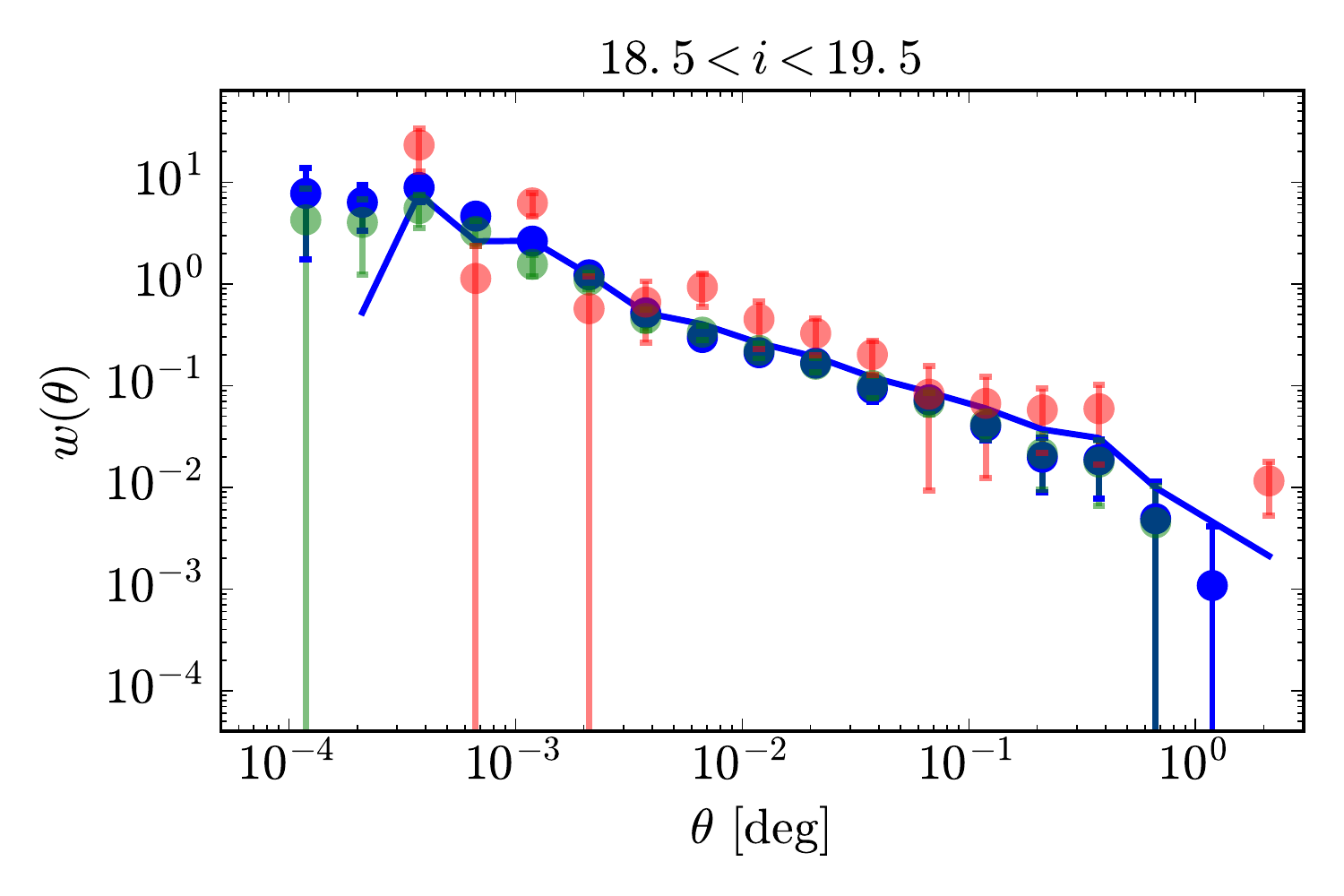}
 	\includegraphics[width=0.49\textwidth]{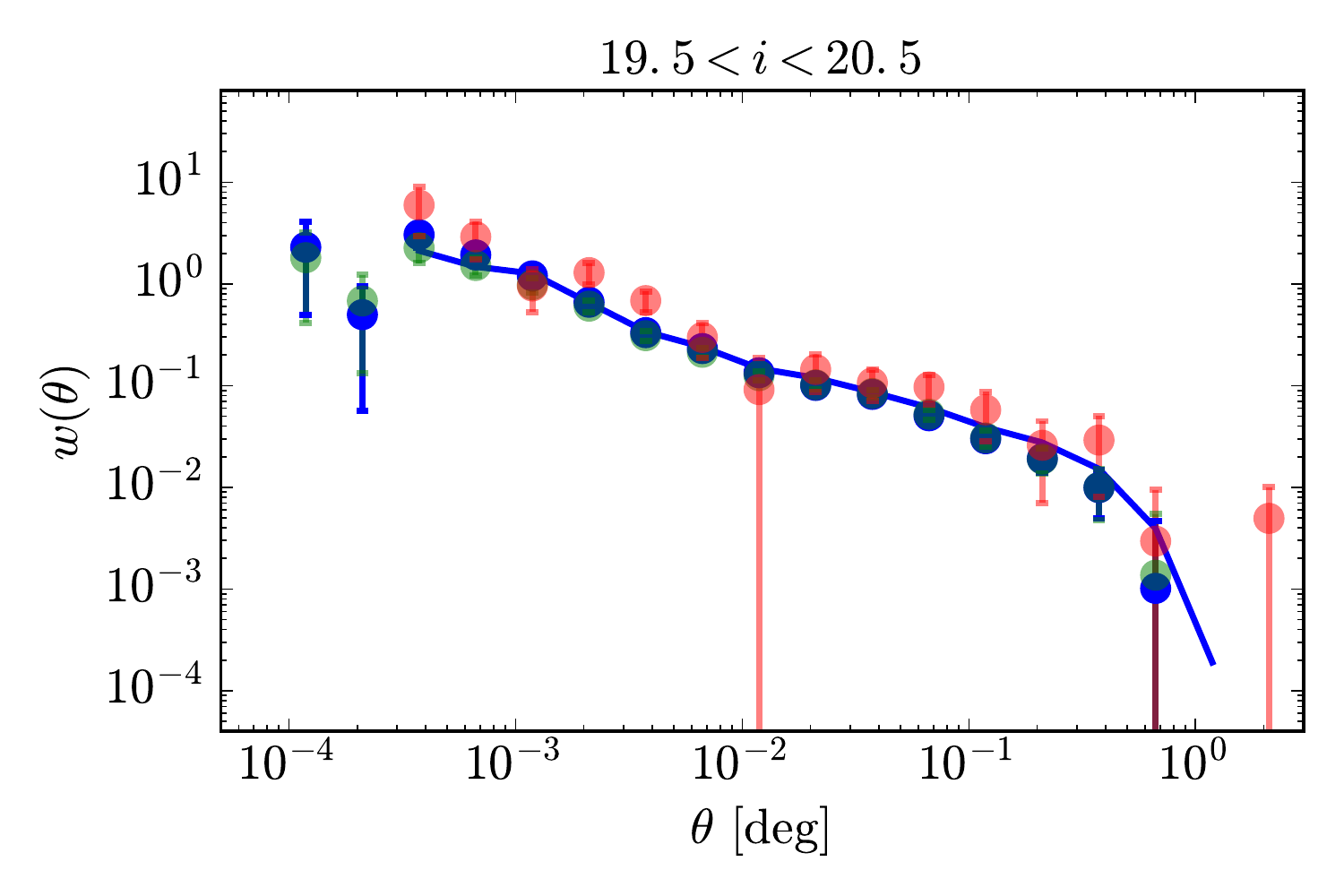}
 	\includegraphics[width=0.49\textwidth]{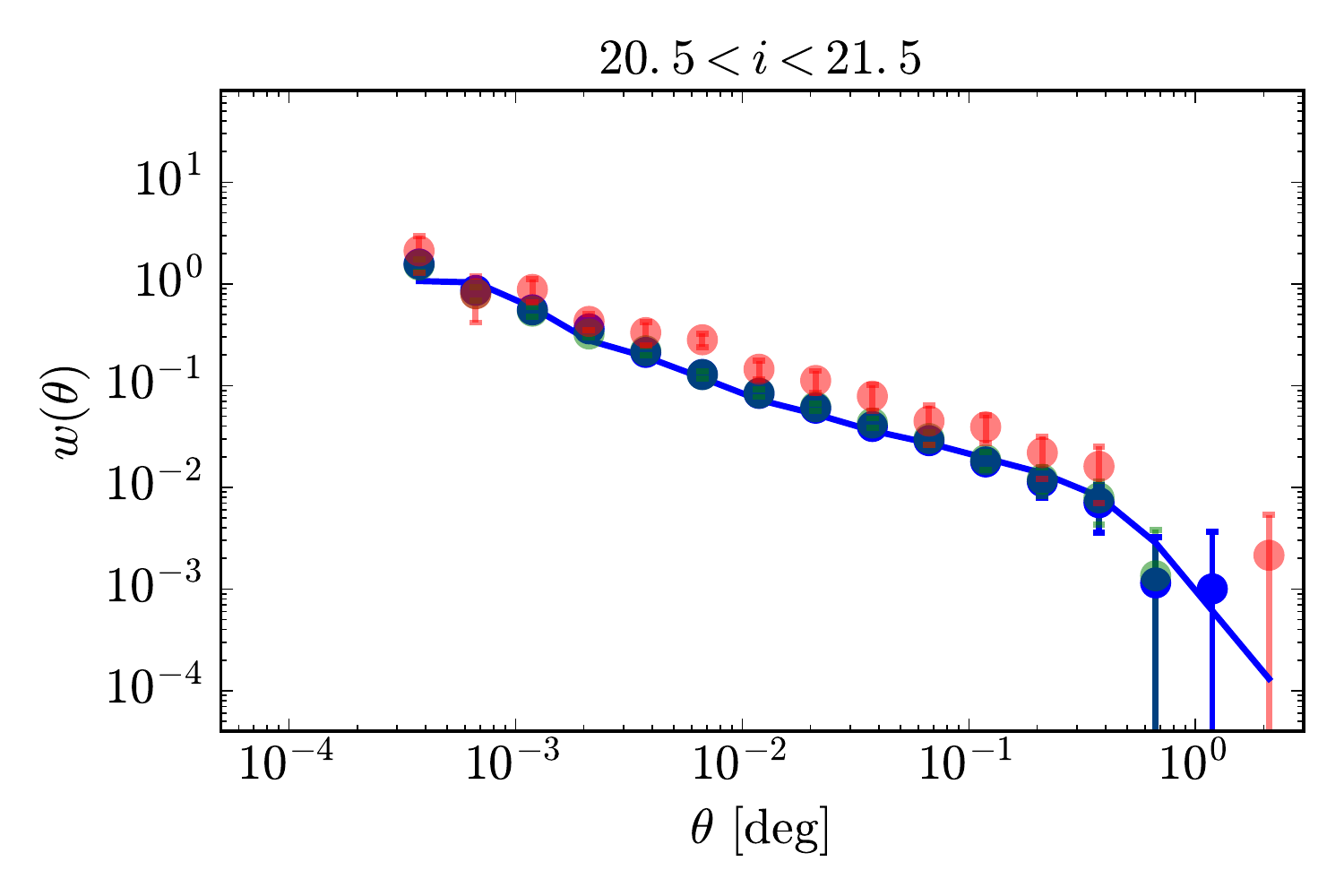}
	\includegraphics[width=0.49\textwidth]{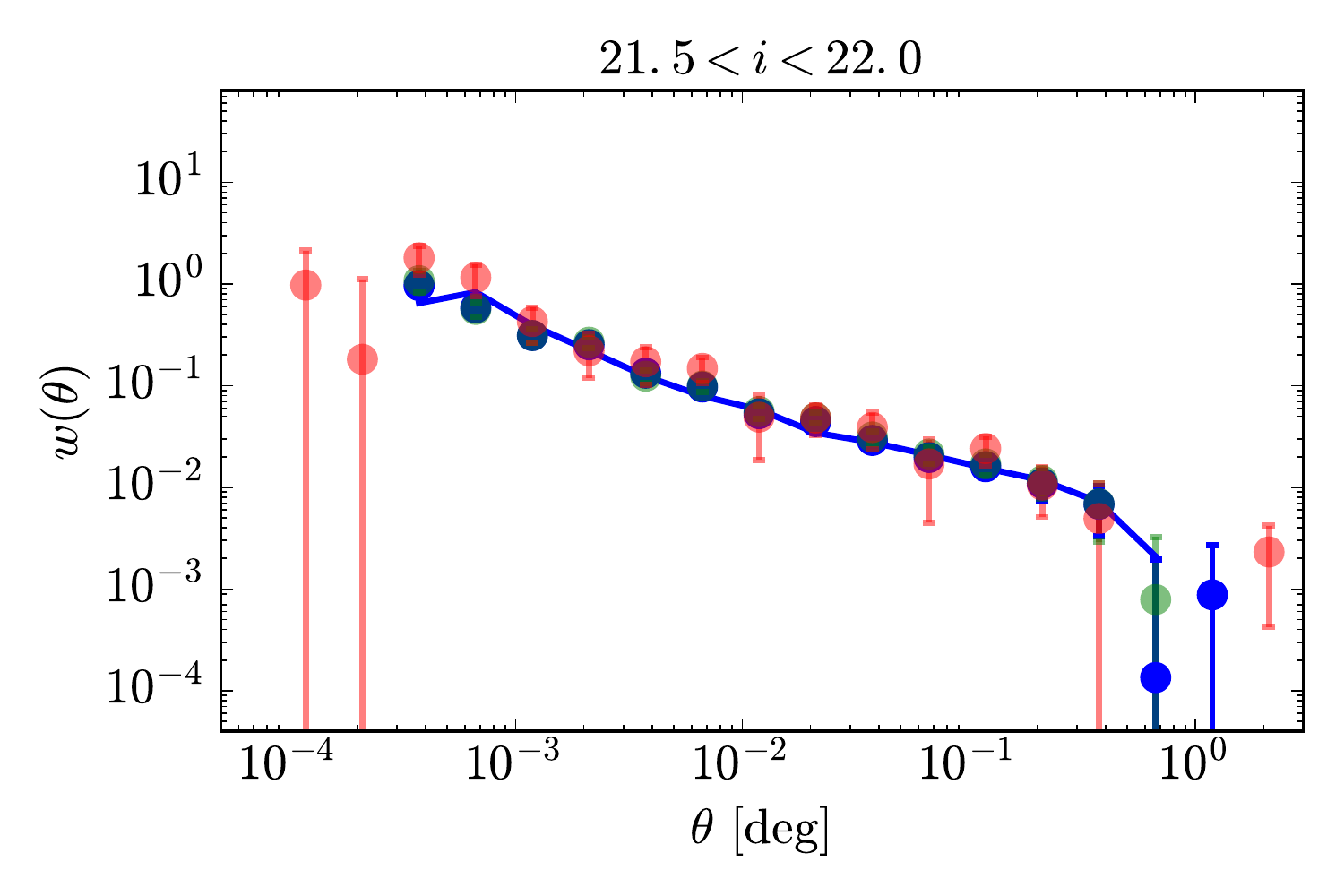}
 	\includegraphics[width=0.49\textwidth]{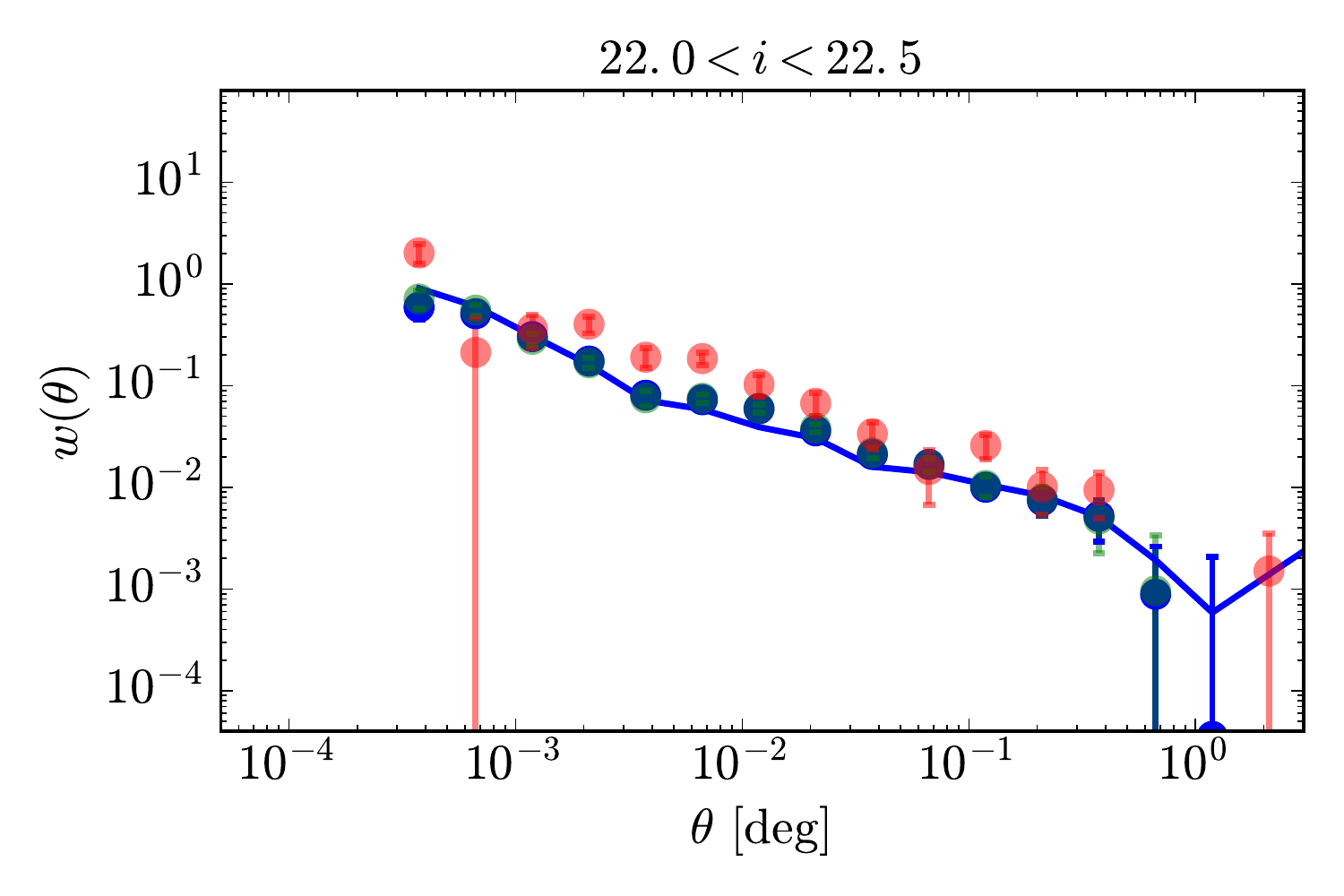}
 \end{center}
   \caption{Angular two-point correlation functions in six $i$-band magnitude bins (increasing brightness from left to right and top to bottom). The blue solid curve is the estimate in CFHTLenS. The blue and green dots are the Arcturus and Sirius (old mask) estimates, respectively, and the red dots are the estimate inside the mask.}
    \label{fig:wtheta}
\end{figure*}

We measure the TCPFs in six $i$-band magnitude bins in the range $17.5 < i < 22.5$, filtering sources using the new (Arcturus) and the old masks (Sirius), displayed in Figure~\ref{fig:wtheta} as blue and green dots, respectively.
We compare these with the CFHTLenS TPCF, shown as a blue solid line. The agreement is remarkably good in the case where the new masks are applied (in blue). This result shows the robustness of the masks, both in its ability to remove problematic data around bright stars, but without wrongly removing bright galaxies.

In the case of the Sirius masks (shown in green), we observe a drop in density at small scales, especially at bright magnitude, due to the removal of bright galaxies.  We also show the TPCF measured \emph{inside} the bright-star masks, to illustrate the effects of the nearby bright stars, that lead to an estimate with significant systematics.

\subsection{CAMIRA clusters}

\begin{figure}
\begin{center}
 	\includegraphics[width=0.49\textwidth]{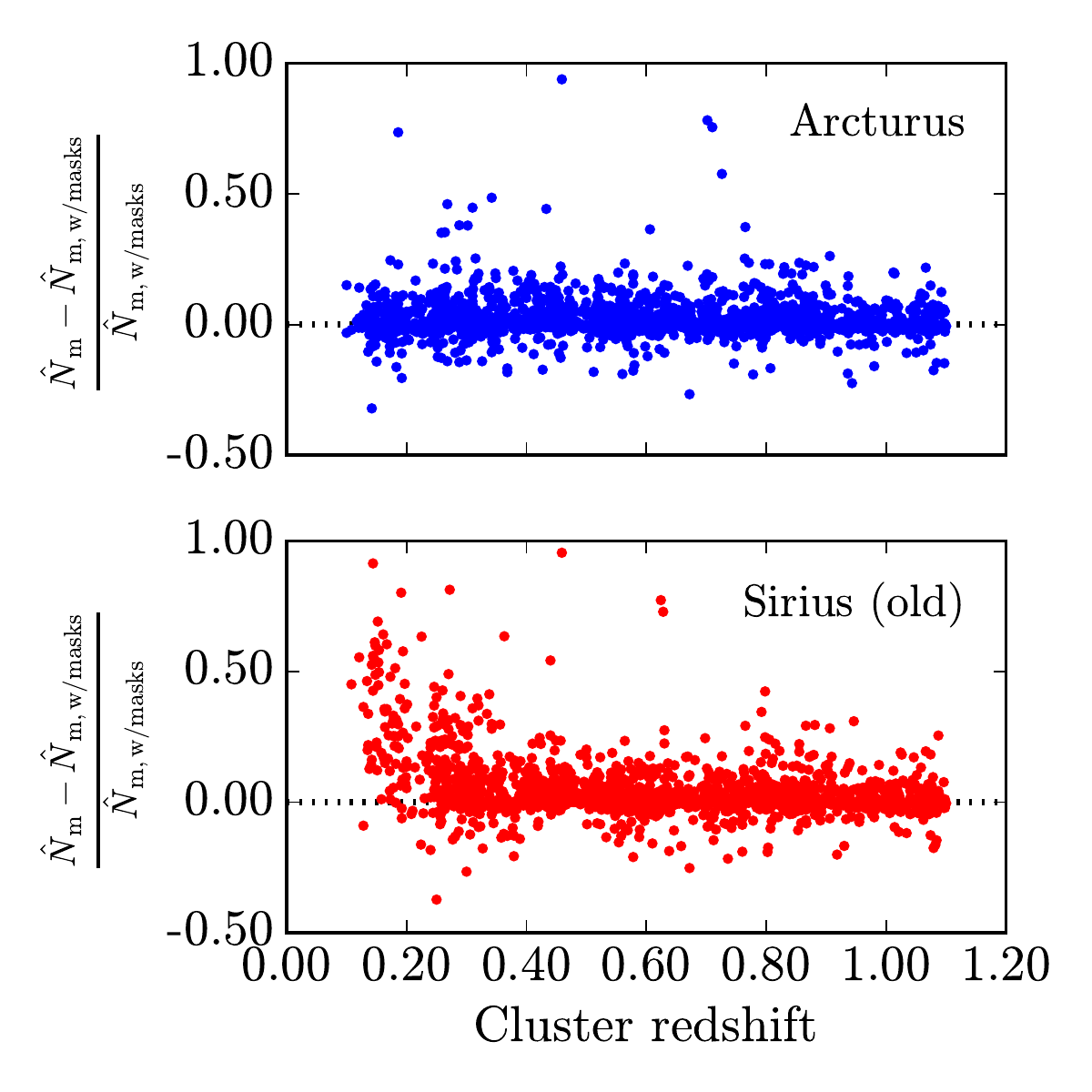}
 \end{center}
    \caption{Relative cluster member estimates before ($\hat{N}_\mathrm{m}$) and after ($\hat{N}_\mathrm{m, w/masks}$) applying the masks. The top panels is for the Arcturus version of the mask, whereas the bottom panel is for the Siriurs version.}
    \label{fig:CAMIRA}
\end{figure}

Finally, we check the cluster member estimates in the cluster catalogue generated by the CAMIRA cluster finder \citep{Oguri:2017}. Figure~\ref{fig:CAMIRA} shows the relative fraction of cluster members after and before applying the masks. The top and bottom panels show the Arcturus and Sirius versions, respectively.

After applying the Sirius version mask, low-redshift clusters feature a significant lower number of cluster members at low redshift, due to the masking of bright galaxies. This trend disappears with the (new) Arcturus version, showing that the latter is well suited for cluster detection and cluster member identification.

\subsection{The masked fraction}

\begin{figure*}
 \begin{center}
  \includegraphics[width=0.9\textwidth]{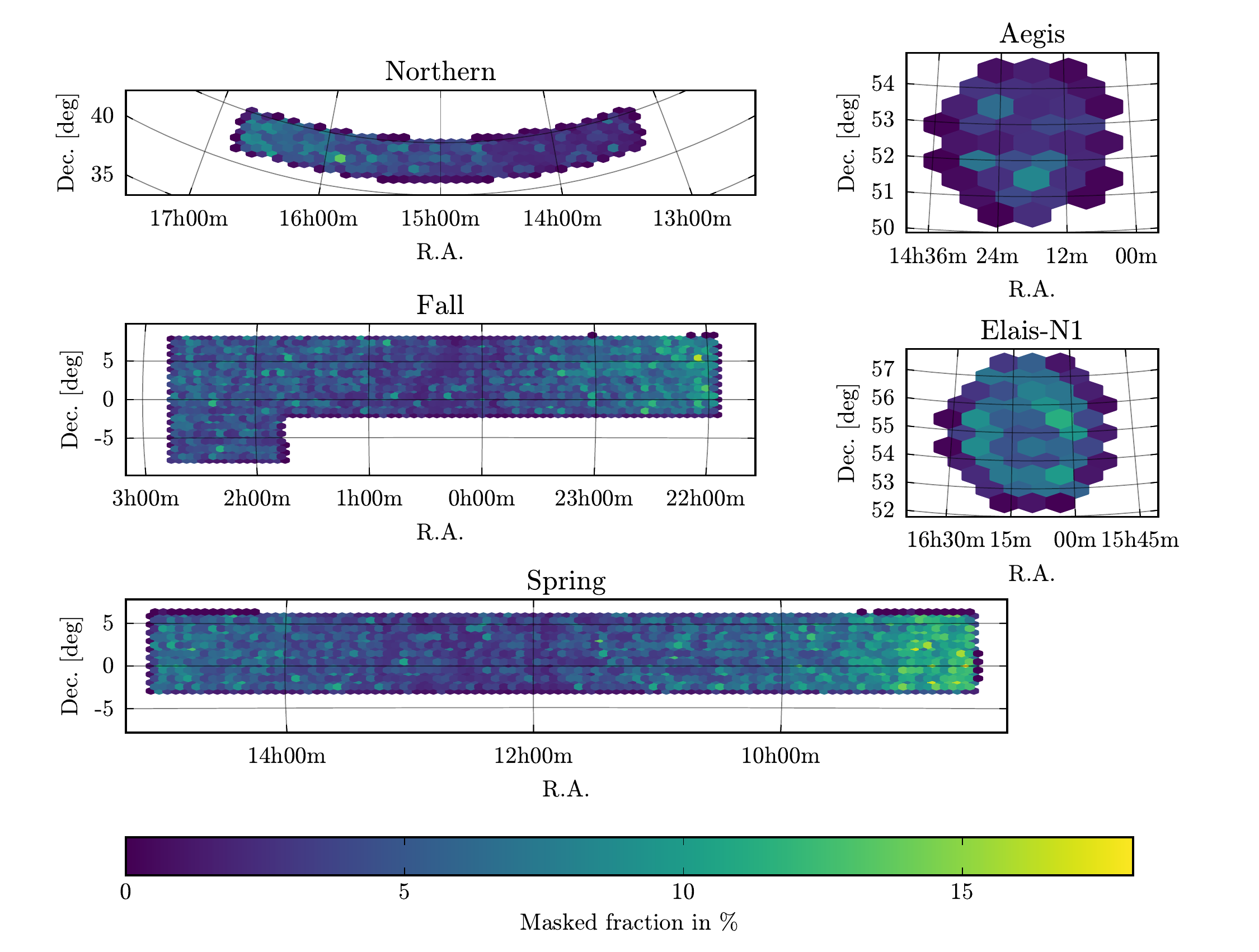}
  \end{center}
 \caption{Masked fraction in percent as a function of sky position in the HSC-SSP footprint.}
    \label{fig:maskedFraction}
\end{figure*}

Finally, we compute the masked fraction as a function of position on the sky. We generate a sample of random points and compute the fraction of points that fall inside the bright-star masks, compared to the total number of random points in a given area. The result for the full planned HSC-SSP footprint is shown in Figure~\ref{fig:maskedFraction}.
The fraction ranges from a few percents to nearly 18\% in the most star-crowded regions.

\section{Conclusions}
\label{sec:conclusions}

We gathered a pure star sample used to build photometric masks around saturated stars in the full planned HSC-SSP survey footprint. { The masks were validated on the S16A internal release data. We performed visual inspection on the HSC images and we measured a number of observables that we compared with the literature.}


Given the high sensitivity of the HSC camera, the typical { good-seeing conditions  at the telescope location (e.g., a median value of \timeform{0.6''} for the $i$ band in the Wide layer)} and the exposure times adopted in the HSC-SSP survey, we found that the bright-star saturation limit magnitudes, in best-seeing conditions, are as faint as $19.2,20.0,20.7,18.5,18.2$, for the $g,r,i,z,Y$ bands, respectively, and depend on the PSF. The effect of saturated stars translates into a PSF-shaped luminous pattern (whose extent increases as a function of star brightness due to the increasing contrast), extended luminous haloes and ghosts, linear horizontal bleed trails, and vertical spikes (in the $Y$ band).

We built a sample of bright stars from the Gaia DR1 release, completed at bright magnitude ($<12$) with the Tycho-2 star catalogue. The $G_\mathrm{Gaia}$ magnitude, measured in a broad-band filter that extends from the HSC $g$ band to $z/Y$ band, is our primary star brightness estimate. For the Tycho-2 stars with no Gaia observations, we computed the $G_\mathrm{Gaia}$ using an analytical colour transformation applied to SDSS filter-emulated magnitudes from the literature. We matched the sample with the SDSS source catalogue and measured the fraction of sources that appear extended in the SDSS. For Gaia sources brighter than $G_\mathrm{Gaia}=18$, we found that about $1.5\%$ appear as extended in the SDSS, which we removed from the star sample. Using the HSC-SSP S16A data we confirmed that our cleaning strategy leads to a star sample which is $99.2\%$ pure and, from visual inspection, the remaining sources that appear extended in HSC are blended binary stars, stars near artefacts, and overlapped with extended sources. Beyond $G_\mathrm{Gaia}>18$ where we do not use Gaia stars, the fraction of extended sources lies between $10$ and $20\%$, depending on the HSC-SSP or SDSS criterium. Due to our conservative approach to { identify and withdraw from the star catalogue all sources that appear as extended in the SDSS}, and the incomplete Gaia observations between $14<G_\mathrm{Gaia}<18$ in a few small areas in the HSC-SSP footprint, our bright-star sample suffer from some incompleteness towards faint magnitudes.

To set up the mask sizes as a function of star brightness, we first matched the bright-star positions with all the sources detected by the HSC pipeline. Second, we computed the mean source density as a function of distance from the star position, and measured two kinds of estimates: the parent source density and the primary source density. The former indicator is used to capture the drop in density caused by the PSF-shaped luminous pattern that screens the foreground sources. The latter indicator reveals the typical extent of the luminous circular halo. We set the circular mask radius as the largest value between the two. We found that for stars brighter than $G_\mathrm{Gaia}<9$, nearby sources are primarily affected by the luminous halo, whereas for fainter stars they are primarily affected by the screening effect from the PSF-shaped luminous pattern. To automate the process of building individual masks, we parametrised the required mask radius as a function of Gaia magnitude, using two exponential functions below and above $G_\mathrm{Gaia}=9$. To mask the additional anisotropic effects caused by the bleed trails and the $Y$-band vertical spikes, we added two horizontal and vertical rectangles scaled to the size of the circular mask.

After applying the masks to the latest data release from HSC-SSP (S16A), we performed a number of validation checks to make sure of (1) the purity of the bright-star sample and (2) the robustness of the masks. We found no indication of a significant loss of bright galaxies due to the masks and we measured good agreements, within the error bars, between HSC-SSP (after masking) and CFHTLenS, for the galaxy number counts and the two-point correlation functions.

These procedures are well suited for wide-field surveys in which building (and validating) all individual masks by hand is too time consuming given the large amount of imaging data. In this work we have shown that the great quality of the Gaia star sample and the remarkable data homogeneity of HSC allowed to build robust star masks  to be used for a large number of science cases, which is promising in the context of the future LSST \citep{Ivezic:2008} and Euclid \citep{Laureijs:2017} surveys.

\begin{ack}

The Hyper Suprime-Cam (HSC) collaboration includes the astronomical communities of Japan and Taiwan, and Princeton University. The HSC instrumentation and software were developed by the National Astronomical Observatory of Japan (NAOJ), the Kavli Institute for the Physics and Mathematics of the Universe (Kavli IPMU), the University of Tokyo, the High Energy Accelerator Research Organization (KEK), the Academia Sinica Institute for Astronomy and Astrophysics in Taiwan (ASIAA), and Princeton University. Funding was contributed by the FIRST program from Japanese Cabinet Office, the Ministry of Education, Culture, Sports, Science and Technology (MEXT), the Japan Society for the Promotion of Science (JSPS), Japan Science and Technology Agency (JST), the Toray Science Foundation, NAOJ, Kavli IPMU, KEK, ASIAA, and Princeton University. 

This paper makes use of software developed for the Large Synoptic Survey Telescope. We thank the LSST Project for making their code available as free software at \url{http://dm.lsst.org}.

The Pan-STARRS1 Surveys (PS1) have been made possible through contributions of the Institute for Astronomy, the University of Hawaii, the Pan-STARRS Project Office, the Max-Planck Society and its participating institutes, the Max Planck Institute for Astronomy, Heidelberg and the Max Planck Institute for Extraterrestrial Physics, Garching, The Johns Hopkins University, Durham University, the University of Edinburgh, Queen's University Belfast, the Harvard-Smithsonian Center for Astrophysics, the Las Cumbres Observatory Global Telescope Network Incorporated, the National Central University of Taiwan, the Space Telescope Science Institute, the National Aeronautics and Space Administration under Grant No. NNX08AR22G issued through the Planetary Science Division of the NASA Science Mission Directorate, the National Science Foundation under Grant No. AST-1238877, the University of Maryland, and Eotvos Lorand University (ELTE) and the Los Alamos National Laboratory.

Based in part on data collected at the Subaru Telescope and retrieved from the HSC data archive system, which is operated by Subaru Telescope and Astronomy Data Center at National Astronomical Observatory of Japan.

This work has made use of data from the European Space Agency (ESA) mission Gaia (\url{https://www.cosmos.esa.int/gaia}), processed by the Gaia Data Processing and Analysis Consortium (DPAC, \url{https://www.cosmos.esa.int/web/gaia/dpac/consortium}). Funding for the DPAC has been provided by national institutions, in particular the institutions participating in the Gaia Multilateral Agreement.

We thank Laurent Eyer for useful discussions about the Gaia DR1 and the referee for her/his useful comments.

\end{ack} 

\appendix

\section{Emulated Gaia magnitude in Tycho-2}
\label{sec:magComp}

\begin{figure}
 \begin{center}
  \includegraphics[width=0.49\textwidth]{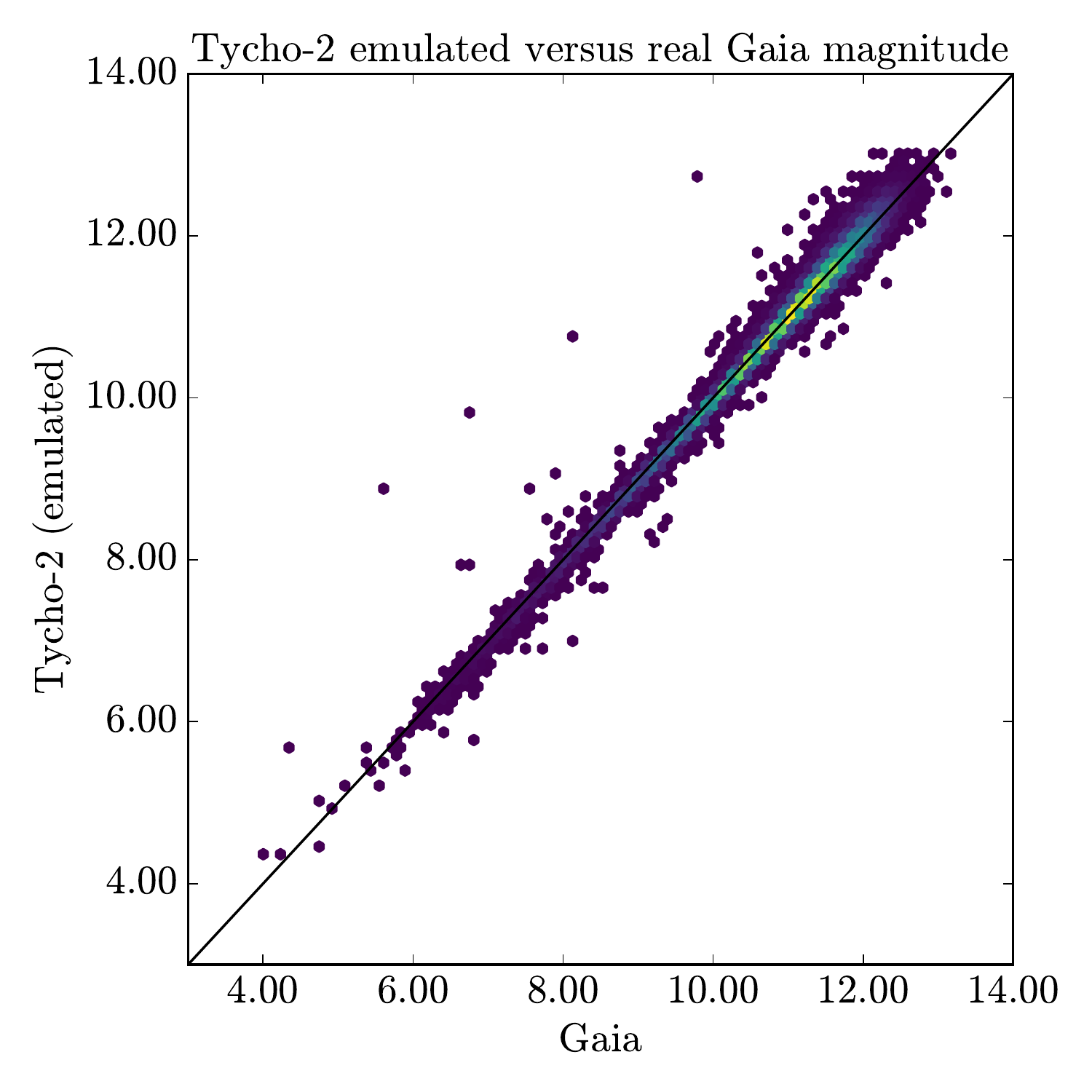}
   \end{center}
 \caption{Emulated $G_{\rm gaia}$ magnitude for Tycho-2 stars versus the true $G_{\rm gaia}$ value for the stars in common.}
    \label{fig:magComp}
\end{figure}

We show in Figure~\ref{fig:magComp} the emulated magnitude for Tycho-2 stars into $G_{\rm gaia}$ magnitude using Equation~\ref{eq:emulGaia}, versus the true $G_{\rm gaia}$ magnitude for the stars in common. The transformation gives unbiased $G_{\rm gaia}$-like magnitudes.

\section{Content of the released bright-star catalogue and masks}

The bright-star catalogue, the masks, and the tools to flag sources are distributed as a unique archive, available at this location\footnote{Also available at \url{http://jeancoupon.com/brightStarMasks}.}: \url{ftp://obsftp.unige.ch/pub/coupon/brightStarMasks/HSC-SSP/}. The repository contains a permanent link to the latest version (\texttt{HSC-SSP\_brightStarMask\_latest.tgz}) and a \texttt{README} file.

\subsection{Version history}

The current mask version described in this work is Arcturus, which differs from previous versions as follows:
\begin{itemize}
\item Arcturus (April 21th, 2017): identical to Canopus, with tract- and patch-based region files included.
\item Canopus (early April 2017): Gaia DR1, Tycho-2 and SDSS, pure star sample, a few areas with lower Gaia-star density due to low-scanned areas.
\item Sirius (March 2016, used in S15B, S16A and DR1 releases): ``old'' masks, 9\% of bright galaxies are masked, the size of a dozen masks is over conservative for magnitude $\sim5$ stars.
\end{itemize}

\subsection{Uncompressing the archive}

To uncompress the archive, run:
\begin{verbatim}
$ tar xzvf HSC-SSP_brightStarMask_VERSION.tgz
\end{verbatim}
(where \texttt{VERSION} is the latest version).

\subsection{Installing venice}

\texttt{venice} is a mask utility program that reads a mask file (DS9 or fits type) and a catalogue of objects to:
\begin{itemize}
\item create a pixelized mask,
\item find objects inside/outside a mask,
\item or generate a random catalogue of objects inside/outside a mask.
\end{itemize}
{ The code is optimised to deal with a large number of objects and input masks. Currently, \texttt{venice} allows to flag 150M objects with 5M-region mask file on a desktop machine in less than an hour.}

The code source is in \\
\texttt{HSC-SSP\_brightStarMask\_VERSION/venice-V.V.V/} \\
(where \texttt{V.V.V} is the latest stable version of \texttt{venice}\footnote{For most recent versions of \texttt{venice}, see \url{https://github.com/jcoupon/venice}.}). To compile it, you first need to install the \texttt{gsl} and \texttt{cfitsio} libraries (\url{http://www.gnu.org/software/gsl/}, \url{http://heasarc.gsfc.nasa.gov/fitsio/}). Then, go to the \texttt{venice} directory:
\begin{verbatim}
$ cd HSC-SSP_brightStarMask_VERSION/venice-V.V.V/
\end{verbatim}
and run:
\begin{verbatim}
$ make
\end{verbatim}
or, if \texttt{gsl} and \texttt{cfisio} libraries are installed in a different directory than \texttt{/usr/local}, run:
\begin{verbatim}
$ make PREFIX_GSL=DIRECTORY_NAME \
    PREFIX_CFITSIO=DIRECTORY_NAME
\end{verbatim}

\texttt{gcc} is the default compile. If you wish to use a different compiler than \texttt{gcc}, run:
\begin{verbatim}
$ make CC=COMPILER_NAME
\end{verbatim}

The compiled program is automatically installed in \texttt{HSC-SSP\_brightStarMask\_VERSION/venice-V.V.V/bin/}.

\subsection{The bright-star catalogue}

The bright-star catalogue is located in \texttt{HSC-SSP\_brightStarMask\_VERSION/star}. It contains the following columns:
\begin{verbatim}
 1: source_id(Long)
 2: ra(Double)/Angle[deg]
 3: dec(Double)/Angle[deg]
 4: G_Gaia(Double)/Magnitude[mag]
 5: origin(String)/Gaia or Tycho-2
 6: G_Gaia_SDSS(Double)/Magnitude[mag]
\end{verbatim}

The source identification for the Gaia sources is the released \texttt{source\_id} column, however, for Tycho-2 stars, it is a new running number independent from the released catalogue and the original Tycho-2 sample. So, to get a unique identification number from this catalogue, it is necessary to select both the \texttt{source\_id} and \texttt{origin} identifiers.

\subsection{flagging a catalogue using \texttt{venice}}

To flag a catalogue using \texttt{venice}, run:
\begin{verbatim}
$ venice-V.V.V/bin/venice \
    -m reg/masks_all.reg -f all \
    -cat MY_INPUT_CAT \
    -xcol RA_COLUMN_NAME -ycol DEC_COLUMN_NAME \
    -o MY_OUTPUT_CAT
\end{verbatim}
where \texttt{reg/masks\_all.reg} is the master region mask file.

Note: \texttt{venice} can read both fits files (default) and ascii files. For input and output ascii files, set \texttt{-ifmt ascii} and \texttt{-ofmt ascii}, respectively.

\subsection{Tract and patch region files}

The \texttt{reg/patches.tgz} and \texttt{reg/tracts.tgz} archives contain the region mask files split in tracts and patches for the five filters. To untar it, run:
\begin{verbatim}
$ cd HSC-SSP_brightStarMask_VERSION/reg
$ tar xzvf tracts.tgz
$ tar xzvf patches.tgz
\end{verbatim}
The tracts archive expands into $5\,715$ files and the patches archive expands into $356\,550$ files.

The paths and names for the patches and tracts files are:
\begin{verbatim}
tract: HSC-SSP_brightStarMask_VERSION/reg/tracts/
BrightStarMask-TRACT-FILTER.reg
patch: HSC-SSP_brightStarMask_VERSION/reg/tracts/
TRACT/BrightStarMask-TRACT-PATCH-FILTER.reg
\end{verbatim}

Currently, the mask is identical for each filter, so $g$, $r$, $z$, and $Y$ filter masks are symbolic links pointing to the $i$ filter mask.

\subsection{A concrete example: flagging objects in tract 9376}
\label{sec:flagging}

First, create a catalogue of random points to be used for this example:
\begin{verbatim}
$ venice-V.V.V/bin/venice \
    -r -xmin 221.476 -xmax 222.967 \
    -ymin -1.5014 -ymax 0.0442215 \
    -coord spher -o tract_9376.fits
\end{verbatim}

Then, run the catalogue through the mask file:
\begin{verbatim}
$ venice-V.V.V/bin/venice \
    -m tracts/9376/BrightStarMask-9376-HSC-I.reg \
    -cat tract_9376.fits -xcol ra -ycol dec \
    -f all -flagName isOutsideMask \
    -o tract_9376_flagged.fits
\end{verbatim}

Options:
\begin{itemize}
\item \texttt{-m tracts/9376/BrightStarMask-9376-HSC-I.reg}: masks in region format
\item \texttt{-cat tract\_9376.fits}: catalogue to flag
\item \texttt{-xcol ra -ycol dec}: names of the input coordinates columns
\item \texttt{-f all}: keep all objects, 1: outside the mask, 0: inside the mask
\item \texttt{-flagName isOutsideMask}: name of the flag column
\item \texttt{-o tract\_9376\_flagged.fits}: the output file
\end{itemize}

\section{Galaxy contamination in the (previously released) Sirius bright-object catalogue}
\label{sec:sirius}

An earlier version of the bright-star mask (Sirius version) was used in the S15B, S16A and DR1 releases. The description of the star catalogue and the construction of the masks are detailed in 
\citet{Mandelbaum:2017}. The code source used for the Sirius version can be retrieved at \url{https://bitbucket.org/czakon/hsc-data-analysis}. 

\begin{figure*}
\begin{center}
\begin{minipage}{0.55\linewidth}
\begin{center}
  \includegraphics[width=1\textwidth]{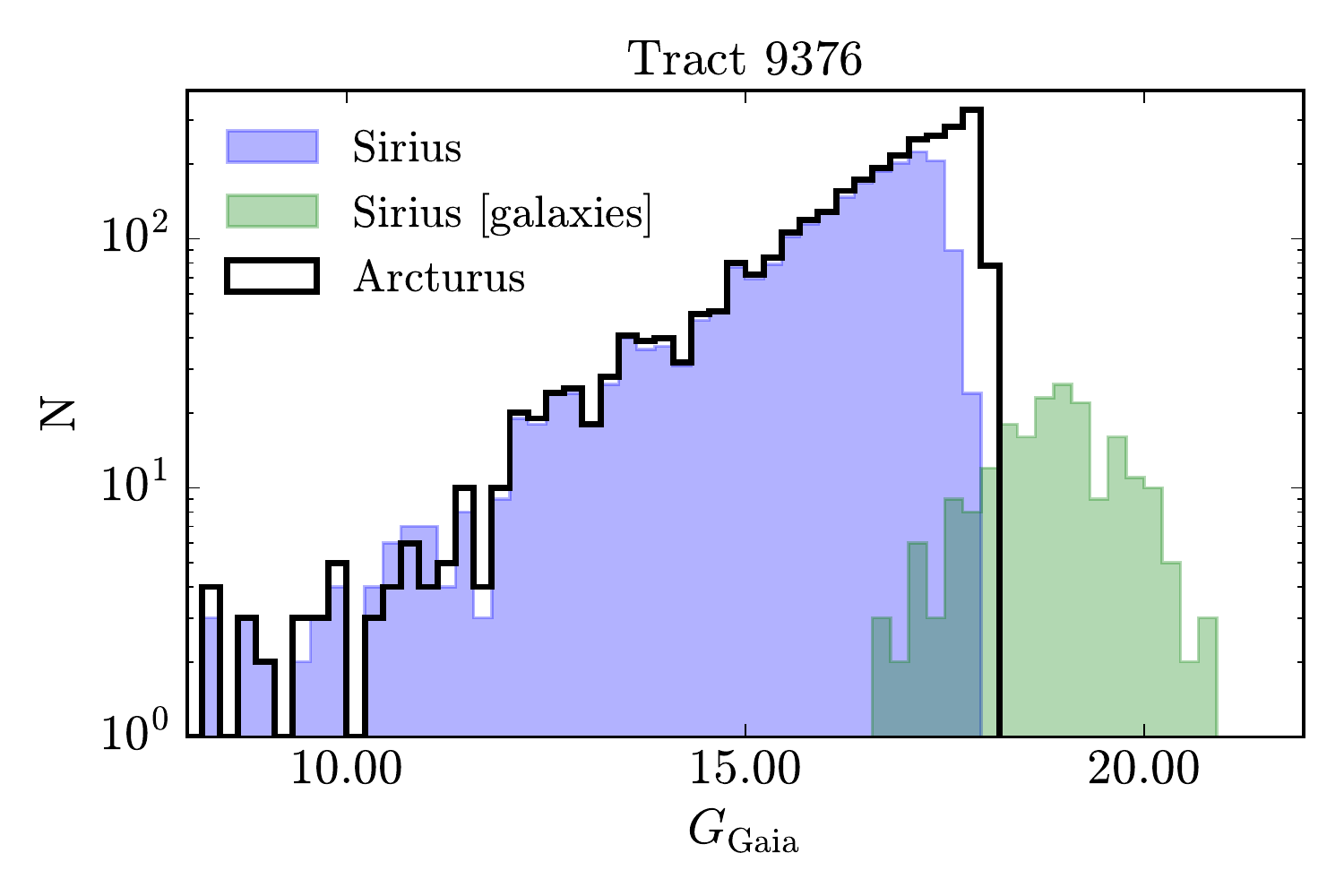}\\
\end{center}
\end{minipage}%
\begin{minipage}{0.45\linewidth}
\begin{center}
  \includegraphics[height=0.275\textwidth]{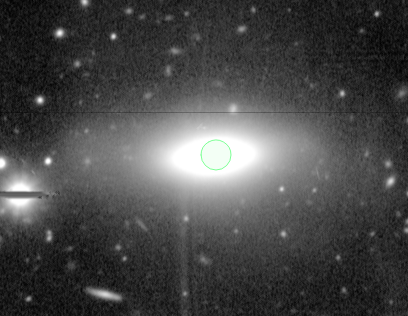}
    \includegraphics[height=0.275\textwidth]{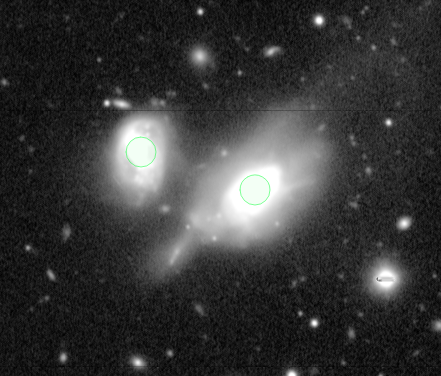}\\
  \includegraphics[height=0.30\textwidth]{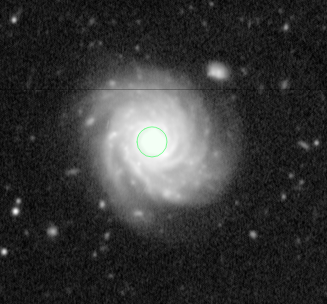}
  \includegraphics[height=0.30\textwidth]{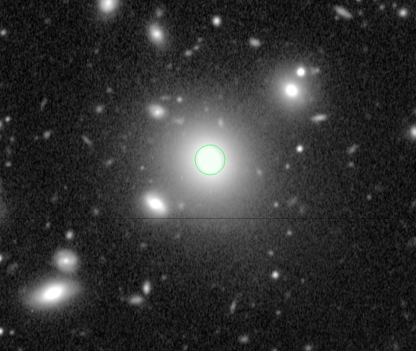}  
\end{center}
\end{minipage}%
\end{center}
  \caption{Left: magnitude distribution of sources in the Sirius catalogue, matched to the SDSS catalogue. In green we show the distribution of sources identified as galaxies in the SDSS (note that the bright part is completed with the bright-star catalogue, since SDSS stars saturate at $G_{\rm Gaia} < 14$), and in blue the distribution of the point-like sources. The black solid curve shows the magnitude distribution of the Arcturus star catalogue. Right: examples of galaxies in the Sirius catalogue wrongly identified as bright stars, but properly removed in the new Arcturus version.}
\label{fig:S16AMagDist}
\end{figure*}

To evaluate the fraction of galaxies in the Sirius catalogue we match it to the SDSS catalogue, in a single HSC-SSP tract (9376). First, we compute the emulated $G_\mathrm{Gaia}$ magnitudes using the SDSS PSF magnitudes, and we identify the sources flagged as extended in the SDSS. In the left panel of Figure~\ref{fig:S16AMagDist}, we show the magnitude distribution of point-like and extended sources, as the blue and green curves, respectively. We also show the distribution of the Arcturus sample as the black solid curve. 
One can see that for extended sources, the emulated $G_{\rm Gaia}$ magnitudes extend to the fainter part, as it is computed from the SDSS PSF magnitudes (hence in the central, fainter, part of the galaxies). Visual inspection reveals that most of the extended sources are bright galaxies with a prominent bulge, as illustrated in the right panel of Figure~\ref{fig:S16AMagDist}.

Out of $3\,255$ objects (in the HSC-SSP tract 9376) from the Sirius catalogue, we find that $20\%$ are not matched with none of the $G_{\rm Gaia} < 18$ or the bright SDSS sample. This is mostly due to some sources from the Sirius catalogue that scatter to fainter magnitude. $2\,609$ sources are matched with at least Gaia or SDSS, among which $211$ (8.9\%) are flagged as extended. Note that this number is an underestimate since we did not measure the fraction of galaxies beyond the limiting magnitude of the bright-star Gaia and SDSS catalogues.



\begin{thebibliography}{}

\bibitem[Aihara et al.(2017)]{Aihara:2017} Aihara H., et al.\ 2017, preprint, (ArXiv:1702.08449)
\bibitem[Albareti et al.(2016)]{Albareti:2016} Albareti F. D., et al.\ 2016, preprint, (arXiv:1608.02013)
\bibitem[Antilogus et al.(2014)]{Antilogus:2014} Antilogus, P., et al.\ 2014, Journal of Instrumentation, 9, C03048
\bibitem[Bertin \& Arnouts(1996)]{Bertin:1996} Bertin, E. \& Arnouts, S.\ 1996, \aaps, 117, 393
\bibitem[Chambers et al.(2016)]{Chambers:2016} Chambers, K. C., et al.\ 2016, preprint, (arXiv:1612.05560)
\bibitem[Bosch et al.(in\ prep.)]{Bosch:2017} Bocsh, J., et al.\ in prep.
\bibitem[Coupon et al.(2012)]{Coupon:2012} Coupon, J., et al.\ 2012, \aap, 500, 981
\bibitem[Coupon et al.(2015)]{Coupon:2015} Coupon, J., et al.\ 2015, \mnras, 449, 1352
\bibitem[Crocce et al.(2016)]{Crocce:2016} Crocce, M., et al.\ 2016, \mnras, 455, 4301
\bibitem[Dark Energy Survey Collaboration et al.(2005)]{DES:2005} Dark Energy Survey Collaboration, et al.\ 2016, preprint, (arXiv:astro-ph/0510346)
\bibitem[Dark Energy Survey Collaboration et al.(2016)]{DES:2016} Dark Energy Survey Collaboration, et al.\ 2016, \mnras, 460, 1270
\bibitem[de Jong et al.(2015)]{dejong:2015} de Jong, J. T. A, et al.\ 2015, \aap, 582, A62
\bibitem[de Jong et al.(2013)]{dejong:2013} de Jong, J. T. A, et al.\ 2013, ExA, 35, 25
\bibitem[Erben et al.(2009)]{Erben:2009} Erben, T., et al.\ 2009, \aap, 493, 1197
\bibitem[Erben et al.(2013)]{Erben:2013} Erben, T., et al.\ 2013, \mnras, 433, 2545
\bibitem[Gaia Collaboration et al.(2016a)]{Gaia-Collaboration:2016ab} Gaia Collaboration et al.\ 2016a, \aap, 595, A1
\bibitem[Gaia Collaboration et al.(2016b)]{Gaia-Collaboration:2016aa} Gaia Collaboration et al.\ 2016b, \aap, 595, A2
\bibitem[Granett et al.(2012)]{Granett:2012} Granett, B. R., et al.\ 2012, \mnras, 421, 251
\bibitem[H\o g et al.(2000)]{Hog:2000aa} H\o g E., et al.\ 2000, \aap, 355, L27
\bibitem[Guzzo et al.(2014)]{Guzzo:2014} Guzzo, L., et al.\ 2014, \aap, 566, 108
\bibitem[Heymans et al.(2012)]{Heymans:2012} Heymans, C., et al.\ 2012, \mnras, 427, 146
\bibitem[HSC collaboration et al.(in\ prep)]{HSC:2017} HSC collaboration et al., in prep.
\bibitem[Huang et al.(in\ prep.)]{Huang:2017} Huang, S., et al.\ 2017, in prep.
\bibitem[Ivezi\'c et al.(2004)]{Ivezic:2004} Ivezi\'c, \`Z., et al.\ 2004, Astron. Nachr., 325, 583
\bibitem[Ivezi\'c et al.(2008)]{Ivezic:2008} Ivezi\'c, \`Z., et al.\ 2008, preprint, (arXiv:0805.2366)
\bibitem[Jordi et al.(2010)]{Jordi:2010aa}  Jordi C., et al.\ 2010, \aap, 523, A48
\bibitem[Juri\'c et al.(2015)]{Juric:2015} Juri\'c, M.\ 2015, preprint, (arXiv:1512.07914)
\bibitem[Kilbinger et al.(2013)]{Kilbinger:2013} Kilbinger, M., et al.\ 2013, \mnras, 430, 2200
\bibitem[Kuijken et al.(2015)]{Kuijken:2015} Kuijken, K., et al.\ 2015, \mnras, 454, 3500
\bibitem[Lang et al.(2016)]{Lang:2016} Lang, D., et al.\ 2016, \aj, 151, 36
\bibitem[Laureijs et al.(2011)]{Laureijs:2017} Laureijs, R.\ 2011, preprint, (arXiv:1110.3193)
\bibitem[Magnier et al.(2004)]{Magnier:2004} Magnier, E. A., et al.\ 2004, \pasp, 116, 449
\bibitem[Magnier et al.(2013)]{Magnier:2013} Magnier, E. A., et al.\ 2013, \apjs, 205, 20
\bibitem[Mandelbaum et al.(in\ prep.)]{Mandelbaum:2017} Mandelbaum, R., et al., in prep.
\bibitem[Miyazaki et al.(2012)]{Miyazaki:2012} Miyazaki, S., et al.\ 2012, SPIE, 8446E, 0ZM
\bibitem[Miyazaki et al.(in\ prep.)]{Miyazaki:2017} Miyazaki, S., et al.\ 2017, in prep.
\bibitem[Oguri et al.(2017)]{Oguri:2017} Oguri, M.\ 2017, preprint, (arXiv:1701.00818)
\bibitem[Pickles \& Depagne (2010)]{Pickles:2010aa} Pickles A., Depagne \'E.\ 2010, \pasp, 122, 1437
\bibitem[Racine(1996)]{Racine:1996} Racine, R.\ 1996, \pasp, 108, 699
\bibitem[Schirmer(2013)]{Schirmer:2013} Schirmer, M.\ 2013, \apjs, ApJS, 209, 21
\bibitem[Schlafly et al.(2012)]{Schlafly:2012} Schlafly, E. F.\ 2012, \apj, 756, 158
\bibitem[Tonry et al.(2012)]{Tonry:2012} Tonry, J.L.\ 2012, \apj, 750, 99
\bibitem[Tanaka et al.(2017)]{Tanaka:2017} Tanaka, M.\ 2017, preprint, (arXiv:1704.05988)
\bibitem[Tonry et al.(2012)]{Tonry:2012} Tonry, J.L.\ 2012, \apj, 750, 99
\bibitem[Yagi, Utsumi \& Komiyama (in\ press)]{Yagi:2017} Yagi, M., Utsumi, Y., \& Komiyama, Y. \ 2015b, adass, in press
\bibitem[York et al.(2000)]{York:2000aa}  York D. G., et al.\ 2000, \aj, 120, 1579
\end{thebibliography}
\end{document}